\documentclass[iop,apj]{emulateapj}
\usepackage{amssymb}
\usepackage{amsmath}
\usepackage{graphicx}
\usepackage{lscape} 
\graphicspath{{./figures/}}

\newcommand\kms{km~s$^{-1}$}

\newcommand\mjb{mJy~beam$^{-1}$}

\newcommand\pp{$^{\prime\prime}$}
\newcommand\um{$\mu$m}

\newcommand\q{$\sim$}

\newcommand\h{H$_{2}$}
\newcommand\msun{M$_{\odot}$}  

\newcommand\lsun{L$_{\odot}$}

\newcommand{\ammonia}{NH$_3$}

\newcommand\water{H$_{2}$O}

\newcommand\ujb{$\mu$Jy~beam$^{-1}$}
\newcommand{\meth}{CH$_3$OH}

\newcommand{\vlsr}{$v_{LSR}$}

\newcommand{\noprint}[1]{}

\defcitealias{egocat}{C08} 	
\defcitealias{maserpap}{C09} 	
\defcitealias{C11}{C11a} 
\defcitealias{C11vla}{C11b} 		
\defcitealias{Chen11}{CE11} 

\begin{document}
\clearpage

\shortauthors{Cyganowski et al.}

\title{A Water Maser and \ammonia\/ Survey of GLIMPSE Extended Green Objects (EGOs)}
\author{C.J. Cyganowski\altaffilmark{1,11}, J. Koda\altaffilmark{2}, E. Rosolowsky\altaffilmark{3}, S. Towers\altaffilmark{2,4}, J. Donovan Meyer\altaffilmark{2}, F. Egusa\altaffilmark{5,6}, R. Momose\altaffilmark{7,8,9}, T. P. Robitaille\altaffilmark{10}}

\email{ccyganowski@cfa.harvard.edu}

\altaffiltext{1}{Harvard-Smithsonian Center for Astrophysics, Cambridge, MA 02138, USA}
\altaffiltext{2}{Department of Physics and Astronomy, Stony Brook University, Stony Brook, NY 11794, USA} 
\altaffiltext{3}{Department of Physics and Astronomy, University of British Columbia, Okanagan, Kelowna BC V1V 1V7, Canada }
\altaffiltext{4}{Department of Physics, Western Michigan University, Kalamazoo, MI 49008, USA}
\altaffiltext{5}{Institute of Space and Astronautical Science, Japan Aerospace Exploration Agency, Chuo-ku, Sagamihara, Kanagawa 252-5210, Japan}
\altaffiltext{6}{Department of Astronomy, California Institute of Technology, Pasadena, CA 91125, USA}
\altaffiltext{7}{Department of Astronomy, University of Tokyo, Hongo, Bunkyo-ku, Tokyo 113-0033, Japan}
\altaffiltext{8}{National Astronomical Observatory of Japan, Mitaka, Tokyo 181-8588, Japan }
\altaffiltext{9}{Institute for Cosmic Ray Research, University of Tokyo, 5-1-5 Kashiwa-no-Ha, Kashiwa City, Chiba, 277-8582, Japan}
\altaffiltext{10}{Max Planck Institute for Astronomy, Heidelberg, Germany}
\altaffiltext{11}{NSF Astronomy and Astrophysics Postdoctoral Fellow}

\begin{abstract}

We present the results of a Nobeyama 45-m \water\/ maser and
\ammonia\/ survey of all 94 northern GLIMPSE Extended Green Objects
(EGOs), a sample of massive young stellar objects (MYSOs) identified
based on their extended 4.5 \um\/ emission.  We observed the
\ammonia(1,1), (2,2), and (3,3) inversion lines, and detect emission
towards 97\% , 63\%, and 46\% of our sample, respectively (median rms
$\sim$ 50 mK).  The \water\/ maser detection rate is 68\% (median rms
$\sim$ 0.11 Jy).  The derived \water\/ maser and clump-scale gas
properties are consistent with the identification of EGOs as young
MYSOs.  To explore the degree of variation among EGOs, we analyze
subsamples defined based on MIR properties or maser associations.
\water\/ masers and warm dense gas, as indicated by emission in the
higher-excitation \ammonia\/ transitions, are most frequently detected
towards EGOs also associated with both Class I and II \meth\/ masers.
95\% (81\%) of such EGOs are detected in \water\/ (\ammonia(3,3)),
compared to only 33\% (7\%) of EGOs without either \meth\/ maser type.
As populations, EGOs associated with Class I and/or II \meth\/ masers have significantly higher
\ammonia\/ linewidths, column densities, and kinetic temperatures than
EGOs undetected in \meth\/ maser surveys.  However, we find no
evidence for statistically significant differences in \water\/ maser
properties (such as maser luminosity) among any EGO subsamples.
Combining our data with the 1.1 mm continuum Bolocam Galactic Plane
Survey, we find no correlation between isotropic \water\/ maser
luminosity and clump number density.  \water\/ maser luminosity is
weakly correlated with clump (gas) temperature and clump mass.

\end{abstract}


\section{Introduction}\label{intro}

The early stages of massive star formation remain poorly understood,
due in part to the difficulty of identifying young massive young
stellar objects (MYSOs)\footnote{We define MYSOs as young stellar
objects (YSOs) that will become O or early B type main sequence stars
(M$_{ZAMS}>$ 8 \msun)} that are actively accreting and driving
outflows.
Large-scale \emph{Spitzer Space Telescope} surveys of the Galactic
Plane have recently yielded a promising new sample of candidates:
Extended Green Objects \citep[EGOs;][]{egocat,maserpap}, selected
based on extended 4.5 \um\/ emission, and named for the common coding
of three-color InfraRed Array Camera \citep[IRAC;][]{Fazio04} images
(RGB: 8.0, 4.5, 3.6 \um).  Modeling, mid-infrared (MIR) spectroscopy,
and narrowband near-infrared (NIR) imaging have shown that
shock-excited molecular line emission, predominantly from \h, can
dominate the 4.5 \um\/ broadband flux in active protostellar outflows
\citep[e.g.][]{SmithRosen05,Smith06,Davis07,Ybarra09,Ybarra10,DeBuizer10}.
While all the IRAC filters include \h\/ lines, only the 4.5 \um\/ band
\emph{lacks} Polycyclic Aromatic Hydrocarbon (PAH) emission features
\citep[e.g. Fig. 1 of][]{Reach06}, which are readily excited in
massive star forming regions (MSFRs).  Morphologically distinct
extended 4.5 \um\/ emission is thus a common feature of well-known
MSFRs \citep[e.g. DR21, S255N, NGC6334I(N), G34.4+0.23, IRAS
18566+0408:][]{Davis07,Cyganowski07,Hunter06,Shepherd07,Araya07}, and a
means of identifying candidate MYSOs with active outflows.

Cyganowski et al. \citepalias[2008, hereafter][]{egocat} cataloged
over 300 EGOs in the Galactic Legacy Infrared Mid-Plane Survey
Extraordinaire survey area \citep[GLIMPSE-I;][]{Churchwell09}.  At the
time, the only data available for most EGOs were IR surveys.
Using the GLIMPSE images, \citetalias{egocat} divided cataloged EGOs
into ``likely'' and ``possible'' outflow candidates based on the
morphology and angular extent of their extended excess 4.5 \um\/ emission.
As detailed by \citetalias{egocat}, two phenomena in the IRAC images
have the potential to be confused with moderately extended 4.5 \um\/
emission: multiple nearby point sources and image artifacts near
bright IRAC sources.  To categorize the \citetalias{egocat} EGOs, two
observers independently reviewed three-color IRAC images:
if either observer thought the MIR morphology could be attributable to
one of these phenomena, the EGO was considered a ``possible'' outflow
candidate.  Of the 302 EGOs in the \citetalias{egocat} catalog, 133
(44\%) were classified as ``likely'' outflow candidates, 165 (55\%) as
``possible'' outflow candidates, and 4 (1\%) as ``outflow-only''
sources (in which the extended outflow emission could be readily
separated from the central source).
\citetalias{egocat} also tabulated
whether each EGO was or was not associated with an Infrared Dark Cloud
(IRDC) visible against the diffuse 8 \um\/ background.  A majority
(67\%) of GLIMPSE EGOs are associated with IRDCs, which are thought to
be sites of the earliest stages of massive star and cluster formation
\citep[e.g.][]{Rathborne06,Rathborne07,Chambers09,Wang11}.  A somewhat
higher fraction of EGO ``likely'' outflow candidates are found in
IRDCs: 71\% compared to 64\% of ``possible'' outflow candidates \citepalias{egocat}.
The GLIMPSE survey is too shallow to detect distant low-mass outflows; based primarily on the MIR data, \citetalias{egocat}
argued that GLIMPSE EGOs were likely outflow-driving \emph{massive} YSOs.

Testing this hypothesis required correlating extended 4.5 \um\/
emission with other massive star formation tracers at high angular resolution.  
Interferometric studies at cm-mm
wavelengths have provided much of the key evidence to date that EGOs
are indeed young, massive YSOs driving active outflows.
The first strong evidence was remarkably high detection rates for two
diagnostic types of \meth\/ masers in sensitive, high angular resolution Very
Large Array (VLA) surveys (Cyganowski et al. 2009, hereafter C09): 6.7
GHz Class II and 44 GHz Class I \meth\/ masers.
Radiatively pumped by IR emission from warm dust, Class II \meth\/
masers are excited near the (proto)star \citep[e.g.][and references
therein]{Cragg05,maserpap}, and recent work suggests that the
luminosities and relative strengths of different Class II transitions
change as the central source evolves \citep[e.g.][and references
therein]{Ellingsen11,Breen11meth}.  The 6.7 GHz transition is the
strongest and most common Class II \meth\/ maser; importantly,
numerous searches have shown that these masers are \emph{not} found towards low-mass YSOs
\citep[e.g.][]{Minier03,Bourke05,Xu08,Pandian08}.  Collisionally
excited in the presence of weak shocks, Class I \meth\/ masers are
generally associated with molecular outflows and outflow/cloud
interactions \citep[e.g.][]{PlambeckMenten90,Kurtz04,Voronkov06},
though recent work suggests Class I masers may also be excited by
shocks driven by expanding HII regions \citep{Voronkov10}.  As a
result of their association with outflows, Class I \meth\/ masers are
more spatially distributed than Class II masers, and may be found many
tens of arcseconds from the driving (proto)star
\citepalias[e.g.][]{maserpap}.

\citetalias{maserpap} detected 6.7 GHz \meth\/ masers towards
$\gtrsim$64\% of their 28 EGO targets, and 44 GHz \meth\/ masers
towards \q90\% of the subset searched for Class I emission (19 EGOs,
18 with 6.7 GHz \meth\/ masers).  Their full sample of 28 EGOs was
chosen to be visible from the northern hemisphere and to span a range
in MIR properties including presence/absence of 8 and 24 \um\/
counterparts, morphology, IRDC association and angular extent of 4.5
\um\/ emission.  The 19 sources observed with the VLA at 44 GHz were all
``likely'' outflow candidates and, in essence, a 6.7 GHz \meth\/
maser-selected subsample \citepalias[for further details see][]{maserpap}.
Subsequent
high-resolution mm-$\lambda$ observations of two of the
\citetalias{maserpap} EGOs revealed high-velocity bipolar molecular
outflows coincident with the 4.5 \um\/ lobes, driven by compact
millimeter continuum cores that exhibit hot core line emission
\citep[][hereafter C11a]{C11}.  Recently, exceptionally deep VLA 3.6
and 1.3 cm continuum observations of a sample of 14
\citetalias{maserpap} EGOs have shown that the vast majority of the
targets (12/14) are \emph{not} ultracompact (UC) HII regions \citep[][hereafter C11b]{C11vla}.  Most
(8/14) are undetected at both 3.6 and 1.3 cm ($\sigma \sim$ 30 and 250
\ujb, respectively); four sources are associated with weak
($\lesssim$1 mJy) cm-$\lambda$ emission consistent with hypercompact
(HC) HII regions or ionized winds or jets.  Based on their cm survey
results and complementary multiwavelength data, \citetalias{C11vla}
argued that these EGOs represent an early stage of massive star
formation, before photoionizing feedback from the central MYSO becomes
significant.

Detailed, high-resolution followup studies have, of necessity, been
limited to relatively small EGO subsamples, and have generally focused
on \citetalias{egocat} ``likely'' outflow candidates \citepalias[see
also][]{maserpap}.  Assessing the variation within the
\citetalias{egocat} catalog and the significance of their MIR
classifications requires large, uniform surveys in tracers of dense
gas and star formation activity.  Few such surveys have been conducted to date.
\citet{Chen10} searched 88 (of 94) northern ($\delta \gtrsim
-$20$^{\circ}$) EGOs for 3 mm HCO$^{+}$, $^{12}$CO, $^{13}$CO, and
C$^{18}$O emission, with the primary goal of detecting infall
signatures.  They found a larger ``blue excess'' towards EGOs
associated with IRDCs compared to those not associated with IRDCs, and
towards ``possible'' compared to ``likely'' outflow candidates;
however, the interpretation of these results was complicated by the
likelihood that multiple sources/dynamical phenomena were present
within their large (\q60-80\pp) beam.  Recently, \citet{He12}
conducted a 1 mm line survey, covering \q 251.5-252.5 GHz and \q
260.2-261.2 GHz, towards 89 northern EGOs (resolution \q 29\pp).  \citet{He12} focus on
linewidth and line luminosity correlations, however, and do not analyze
EGO subsamples.  Chen et al.\ (2011, hereafter CE11) searched for 95
GHz Class I \meth\/ masers towards 192 northern and southern EGOs (of
302 total) with the MOPRA telescope ($\theta_{\rm FWHP}$\q36\pp,
3$\sigma$\q1.6 Jy).  They found a higher 95 GHz \meth\/ maser
detection rate towards ``likely'' than towards ``possible''
\citetalias{egocat} EGOs (62\% and 49\%, respectively), and very
similar detection rates towards EGOs associated/not associated with
IRDCs (55\%/53\%).  Their Class I \meth\/ maser detection rate is also
much higher towards EGOs associated with Class II \meth\/ masers
(80\%) than towards those without (38\%), consistent with the very
high Class I maser detection rate of \citetalias{maserpap}.

Like Class I \meth\/ masers, \water\/ masers are collisionally pumped
\citep[e.g.][]{Elitzur89} and associated with protostellar outflows;
notoriously variable, \water\/ masers also often exhibit high-velocity
emission features, offset by 30 \kms\/ or more from the systemic
velocity \citep[e.g.][]{Breen10water,CB10}.  While Class I \meth\/
masers are excited under moderate conditions \citep[T\q80
K, n(H$_2$)\q10$^{5}$-10$^{6}$ cm$^{-3}$, e.g.][]{Leurinithesis} and associated
with outflow-cloud interfaces, \water\/ masers require more extreme
conditions \citep[T\q400
K, n(H$_2$)\q10$^{8}$-10$^{10}$ cm$^{-3}$,][]{Elitzur89} and are thought
to originate behind fast shocks in the inner regions of the outflow
base. 
Numerous correlations have been reported
between the properties of \water\/ masers and those of the driving
source or surrounding clump, including recent evidence that
$L_{H_{2}O} \propto L_{bol}$ over many orders of magnitude
\citep[e.g.][]{Urquhart11,Bae11}.  
This suggests that \water\/ masers
may be used to investigate the properties of their driving sources, at least
in a statistical sense for different subsamples--a possibility of
interest for EGOs, since their bolometric luminosities are in
most cases poorly constrained by available data \citepalias[see
also][]{C11vla}.

Large \water\/ maser and \ammonia\/ surveys with single-dish
telescopes have long been recognized as powerful tools for
characterizing massive star forming regions
\citep[e.g.][]{Churchwell90,Anglada96,Sridharan02}, and continue to be
applied to new samples \citep[e.g.][]{Urquhart11,nh3-innergal}.
\ammonia\/ traces high-density gas \citep[\q 10$^{4}$ cm$^{-3}$,
e.g.][]{Evans99,SP05}, and provides a wealth of information about
clump kinematics and physical properties; notably, it is an excellent
``thermometer.''  This paper presents the results of a \water\/ maser
and \ammonia\/ survey of the 94 northern ($\delta \gtrsim
-$20$^{\circ}$) EGOs from the \citetalias{egocat} catalog with the
Nobeyama Radio Observatory 45-m telescope.  The motivation for this
survey was to characterize the properties of the \citetalias{egocat}
EGO sample as a whole, the main goals being to evaluate the
significance of the MIR classifications from \citetalias{egocat} and
to place EGOs in the context of other large MYSO samples.  We also
compare the \water\/ maser and \ammonia\/ properties of EGO subsamples
associated with Class I and/or II \meth\/ masers and explore
correlations between \water\/ maser and clump properties.
Evolutionary interpretations have been suggested for both \meth\/
masers and \water\/ maser properties
\citep[e.g.][]{Ellingsen06,Ellingsen07,Breen10evol,BreenEllingsen11},
and our survey, in conjunction with the 1.1 mm Bolocam Galactic Plane
Survey \citep[BGPS;][]{Aguirre11,Rosolowsky10}, provides the necessary
data to test these scenarios.

\section{Observations and Data Analysis}\label{obs_overall}

\subsection{Nobeyama 45 m Observations}\label{obs}

We targeted all 94 EGOs in the \citetalias{egocat} catalog visible
from Nobeyama (those in the northern Galactic Plane, $\delta \gtrsim
-$20$^{\circ}$).  Our sample sources are listed in
Table~\ref{sample_table}, along with information from the literature
on their MIR properties and \meth\/ maser associations. 
The NH$_3$ ($J$,
$K$)=(1,1), (2,2), and (3,3) inversion transitions and the 22.235 GHz
H$_2$O maser line were observed simultaneously with the Nobeyama Radio
Observatory 45-m telescope (NRO45)\footnote{The 45-m radio telescope
is operated by the Nobeyama Radio Observatory, a branch of the
National Astronomical Observatory of Japan, National Institutes of
Natural Sciences.} in 2008-2010.  During our winter (January/February)
observing sessions, the system temperature was typically \q 100-160 K.
The beamsize and main-beam efficiency of the NRO45 at 22 GHz are $\theta_{\rm
FWHP}=73\arcsec$ and $\eta_{\rm MB}=0.825$, respectively. 
We pointed at the EGO positions tabulated in \citetalias{egocat}, which
are the positions of the brightest 4.5 \um\/ emission associated with
each candidate outflow.  We note that these positions will not
necessarily be those of the driving sources \citepalias[which in many
cases are difficult to identify solely from the MIR data, see also][]{egocat},
though in most cases the NRO beam is large enough to encompass likely
driving sources as well as the 4.5 \um\/ extent of the EGO.

We used the H22 receiver, a cooled HEMT receiver, and eight
high-resolution acousto-optic spectrometers (AOSs) to observe both
polarizations for each line simultaneously.  The bandwidth and spectral resolution
of the AOSs are 40 MHz and 37 kHz, respectively, corresponding to
velocity coverage of \q 500 \kms\/ and resolution of \q 0.5 \kms\/ for
the observed lines.  The spectral channels were Nyquist-sampled.

The observations were conducted in position-switching mode, using
``off'' positions \q 5\arcmin\/ away.  All spectra were checked for
evidence of emission in the chosen ``off'' position, and, if
necessary, reobserved.  Initially, each target was observed for 2
minutes (on-source).  The spectra were then inspected, and weak
sources were reobserved to improve the signal-to-noise as time
permitted.  The pointing was measured and adjusted at the beginning of
each observing run using Galactic maser sources.  The absolute
pointing of the NRO45 is very accurate for 22 GHz observations, from a
few arcsec (no wind) to $\sim 10\arcsec$ in the windiest conditions in
which we observed--still a small fraction of the beamsize at 22 GHz.

\begin{figure*}
\center{\includegraphics[scale=0.45]{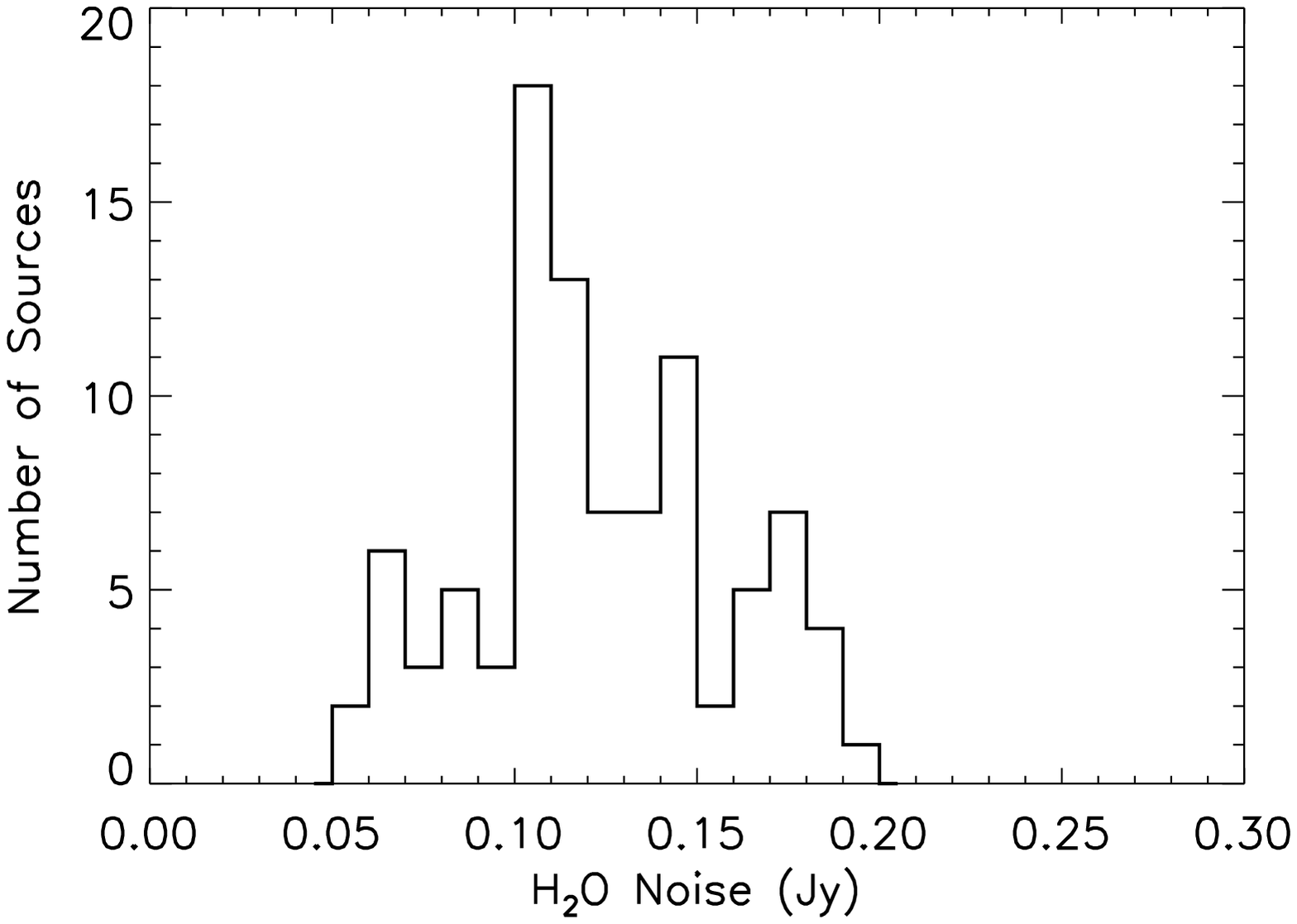}\includegraphics[scale=0.45]{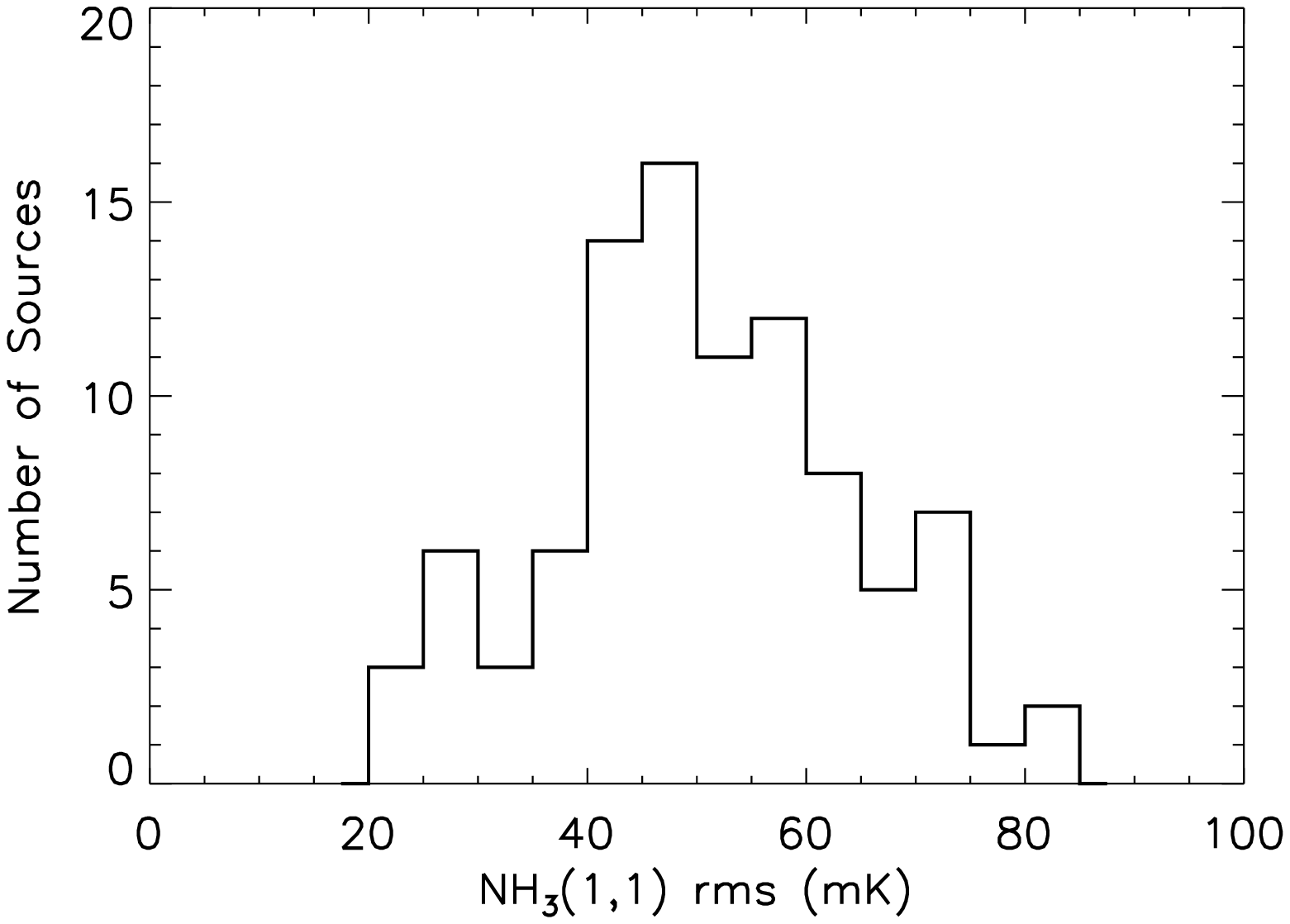}}\\
\includegraphics[scale=0.45]{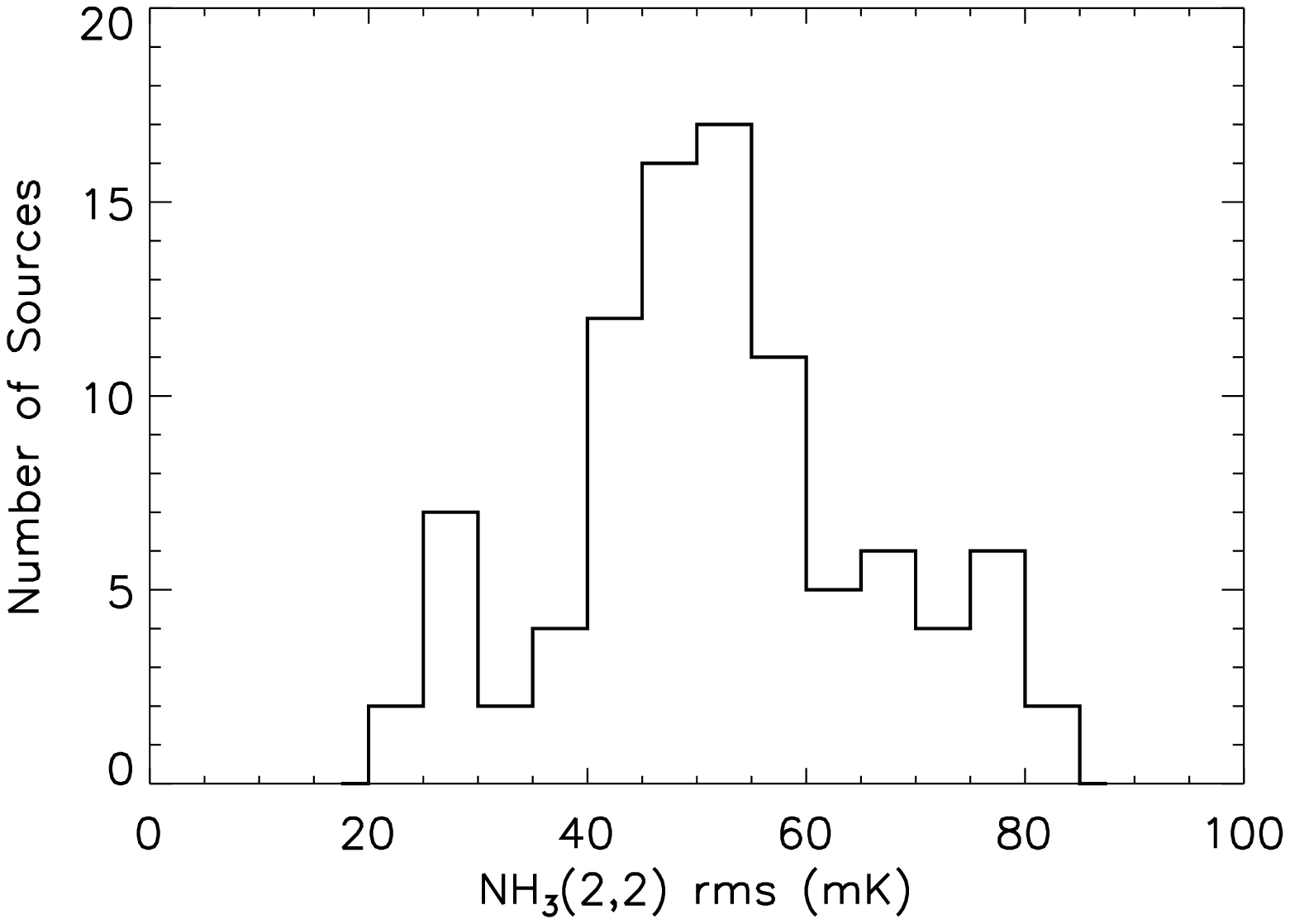}
\includegraphics[scale=0.45]{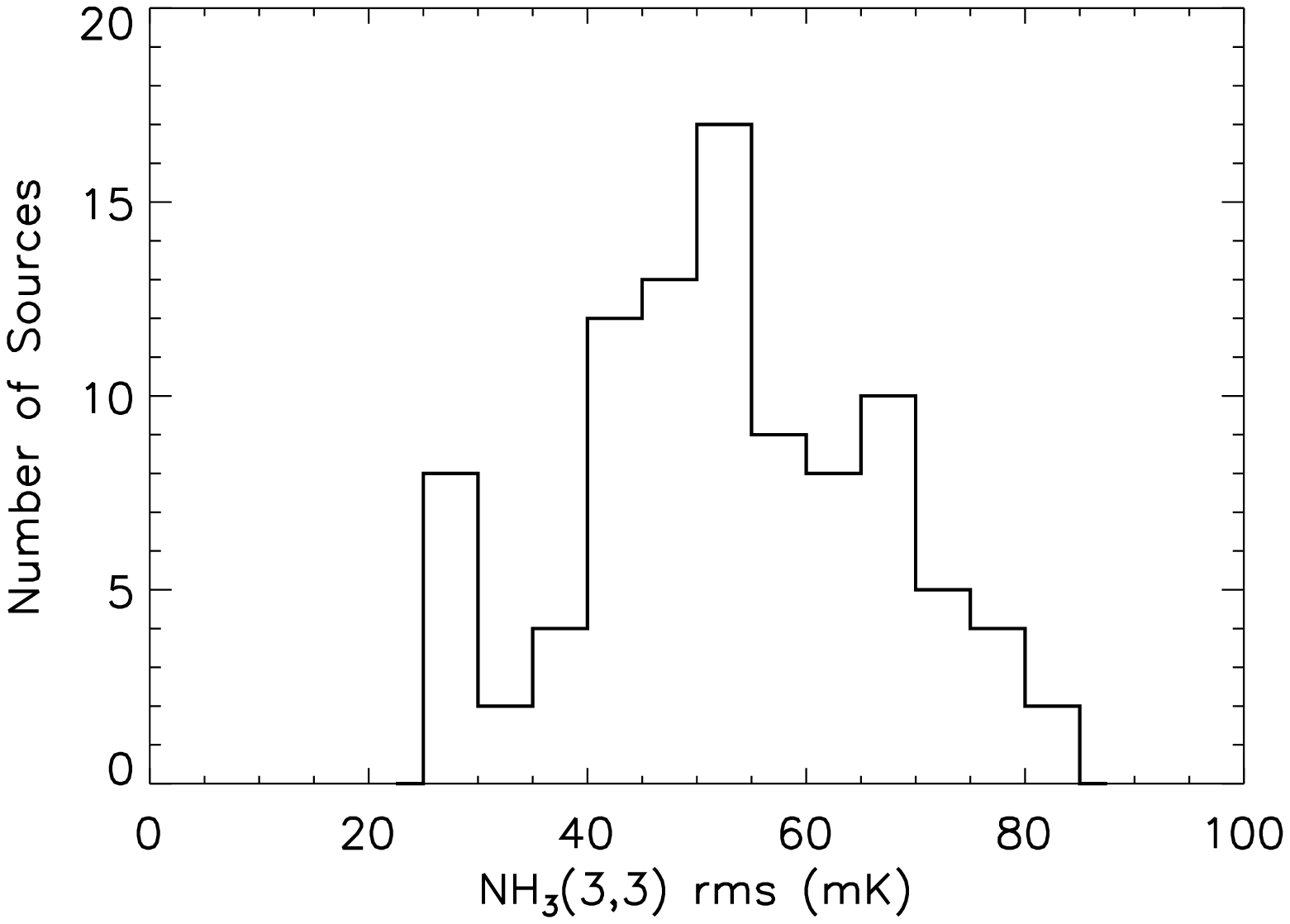}
\caption{Histograms of the distributions of rms noise for the sources in our sample for the four observed lines.}
\label{rms_fig}
\end{figure*}

The data reduction followed standard procedures using the NRO NEWSTAR
software package \citep{Ikeda01}.  For each spectrum, emission-free
channels were used to estimate and subtract a linear spectral baseline.  For
each line, the two polarizations were then co-added, weighted based on
system temperature.  The temperature scale was calibrated to the
antenna temperature ($T_{\rm A}^*$) in Kelvin with the standard
chopper-wheel method, and the main-beam temperature ($T_{\rm MB}$)
calculated as $T_{\rm MB}=T_{\rm A}^*/\eta_{\rm MB}$.  For the
\water\/ maser data, we then convert to the Jansky scale to facilitate
comparisons with other surveys.

Histograms of the rms are shown in Figure~\ref{rms_fig}.  The median
1$\sigma$ rms is \q 50, 51, and 52 mK for \ammonia(1,1), (2,2), and
(3,3), respectively.  For our \water\/ maser observations, the median
1$\sigma$ rms is \q 0.11 Jy, corresponding to a median 4$\sigma$
detection limit of \q 0.44 Jy.

\subsection{\ammonia\/ Modeling and Physical Parameter Estimation}\label{nh3_modeling}

We estimate physical properties from the observed \ammonia\/ spectra following the philosophy developed by
\citet{nh3-perseus} and adapted for use in
\citet{gemob1} and \citet{nh3-innergal}.  The emission is modeled as a
beam-filling slab of \ammonia\/ with a variable column density
($N_{\mathrm{NH3}}$), kinetic temperature ($T_{kin}$), excitation
temperature ($T_{ex}$), Gaussian line width ($\sigma_{\rm v}$), and LSR
velocity ($v_{\mathrm{LSR}}$).  The model assumes the molecules are
in thermodynamic equilibrium using an ortho-to-para ratio of 1:1,
which is the high temperature formation limit \citep{takano}.  Hence,
the ammonia molecules are partitioned among the energy levels as
\begin{tiny}
\begin{eqnarray}
Z_O&=&1+\nonumber\\
&&\sum_{J,K,i}2(2J+1)\exp\left\{-\frac{h[BJ(J+1)+(C-B)J^2]+\Delta E(J,K,i)}{kT_k}\right\}\nonumber\\
&& \mbox{for }J=K=3,6,9,\dots; i=0,1,  \label{partitionfunctionO}\\
Z_P&=&\sum_{J,K,i}(2J+1)\exp\left\{-\frac{h[BJ(J+1)+(C-B)J^2]+\Delta E(J,K,i)}{kT_k}\right\}\nonumber\\
&&\mbox{
for } J=K=1,2,4,5,\dots; i=0,1,
\label{partitionfunctionP}
\end{eqnarray}  
\end{tiny}
Here, $J$ and $K$ are the rotational quantum numbers of NH$_3$ and,
for the metastable inversion transitions, $J=K$.  The energy
difference, $\Delta E(J,K,i)$, is the splitting of the symmetric and
anti-symmetric states, representing both levels of the inversion
transition.  The antisymmetric state, $\Delta E(J,K,1)$, is $\Delta
E/k\sim 1.1$~K above the symmetric state ($\Delta E(J,K,0)=0$).  The
column density of the molecules in the $N_{\mathrm{NH3}}(J,K,i)$ ortho
state is thus $N_{\mathrm{NH3}} Z_O(J,i)/(2Z_O)$ and in the para state $N_{\mathrm{NH3}}
Z_P(J,i)/(2Z_P)$, where the factor of two arises because of the
assumption of a 1:1 ortho-to-para ratio.

The optical depths in the individual transitions are calculated from
the column densities in the individual states.  The optical depth,
hyperfine structure, velocity information and excitation conditions
are then used to model the individual spectra.  Free parameters are
optimized using the MPFIT least-squares minimization routine including
parameter bounds \citep{mpfit}.  Uncertainties in the derived
parameters are also determined from this optimization, accounting for
the covariance between the parameters.  We note that parameter
uncertainties cannot account for systematic errors stemming from the
uniform slab model being an incomplete description of the physical
system.  In all cases, derived quantities should be considered summary
properties of the system and not a complete description.  In most
cases, this simple model reproduces the emission features observed on
the large scales sampled.

\begin{figure*}
\center{\includegraphics[scale=0.5]{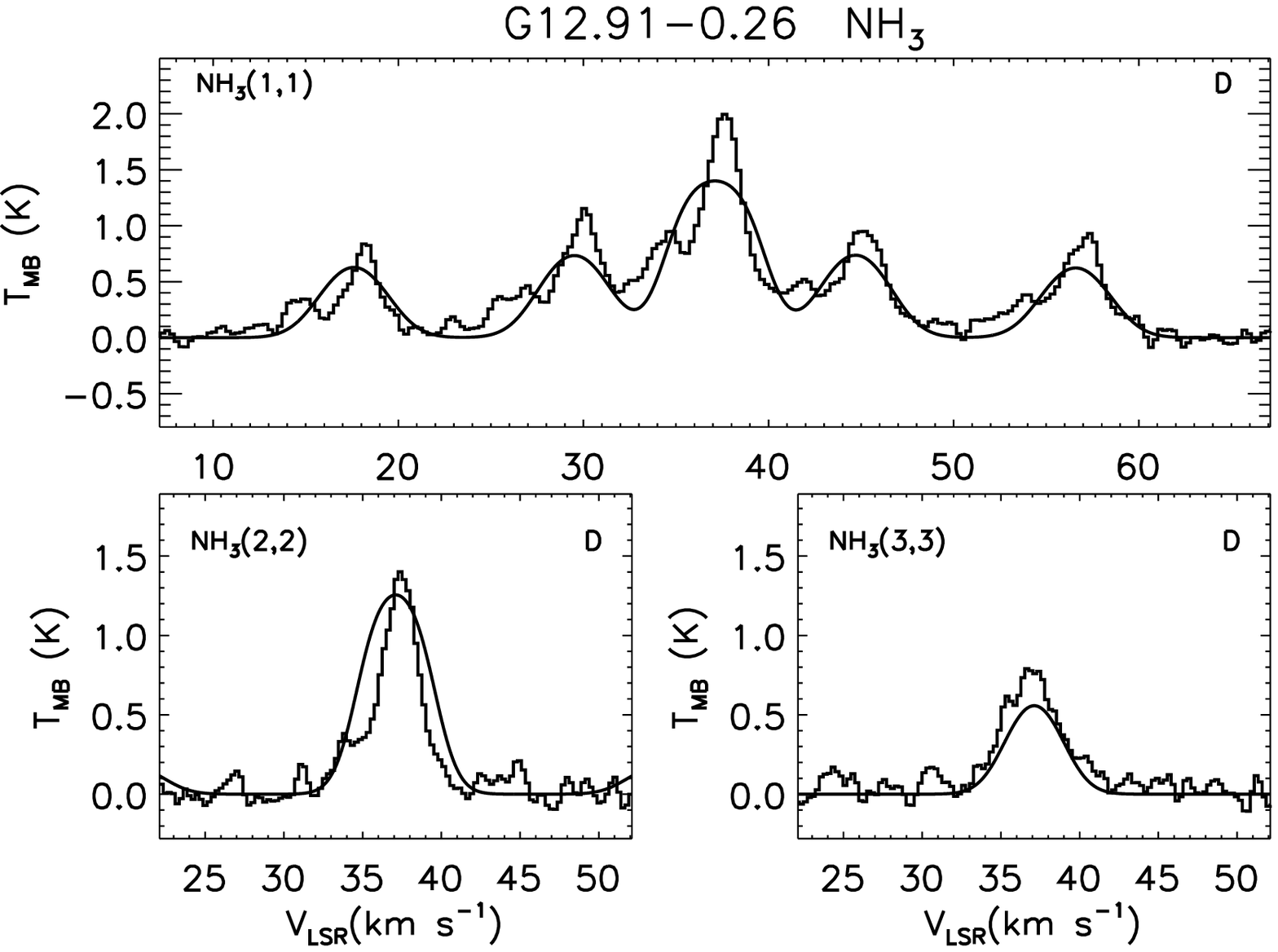} \includegraphics[scale=0.5]{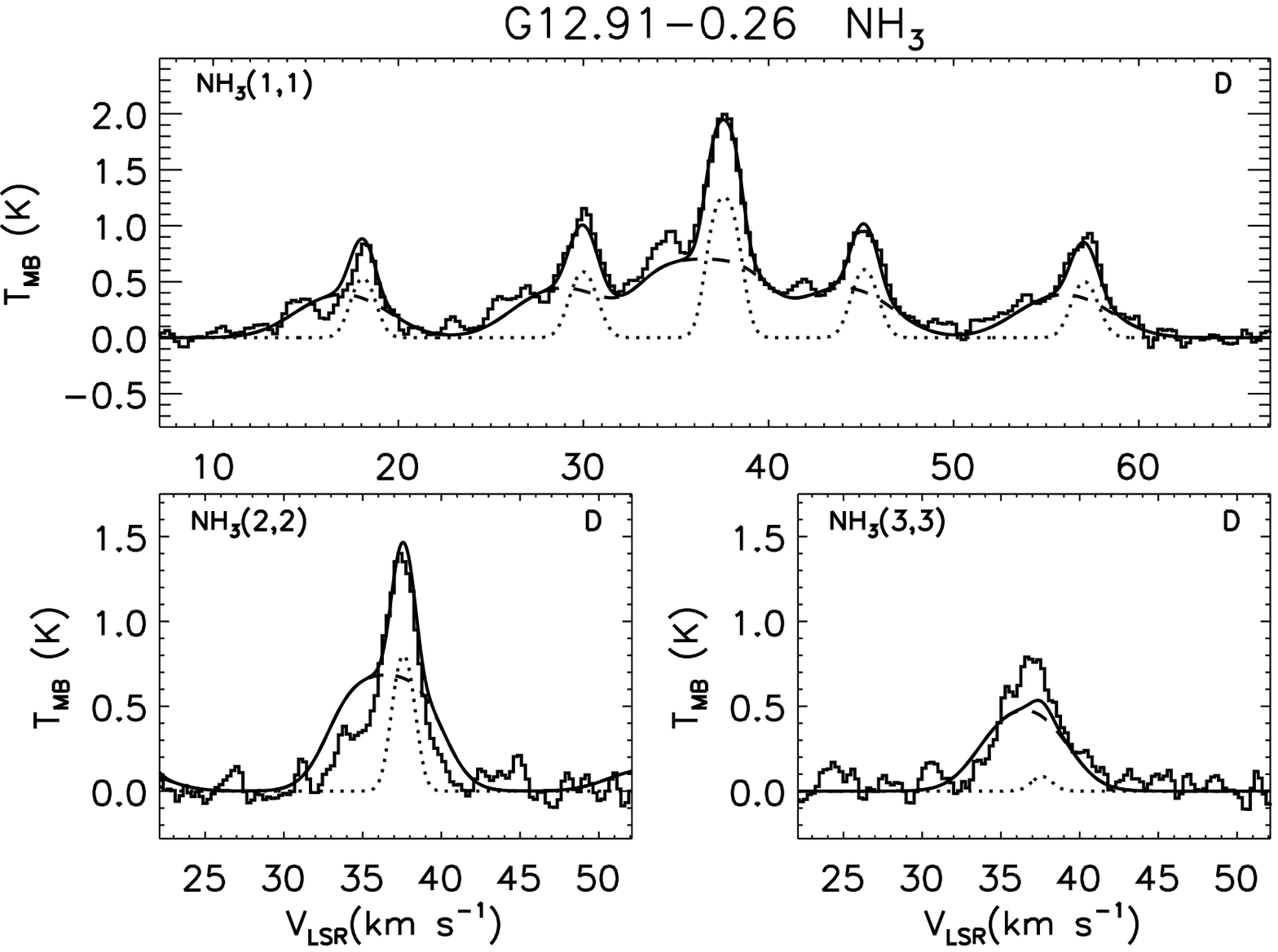}}\\
\includegraphics[scale=0.5]{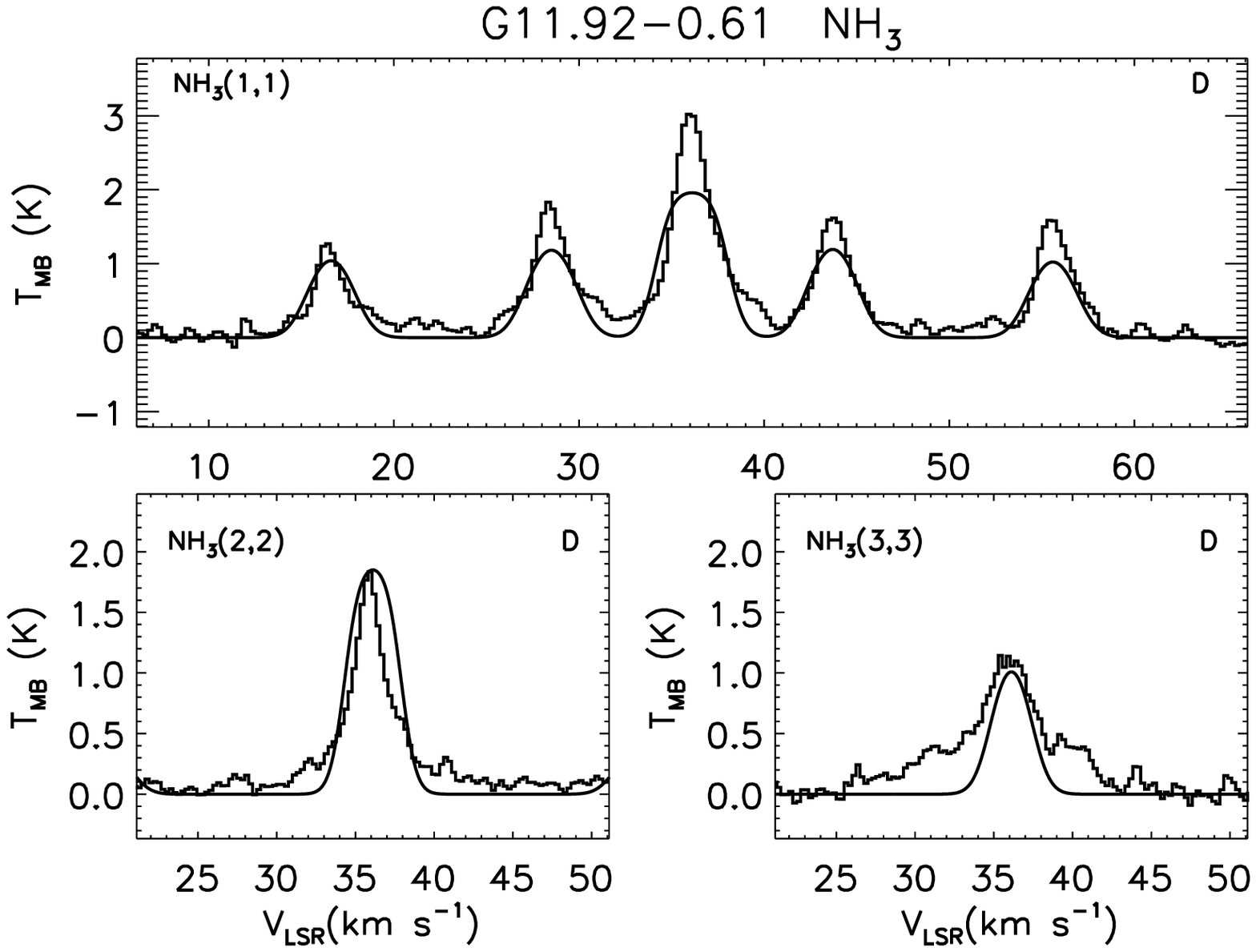}
\includegraphics[scale=0.5]{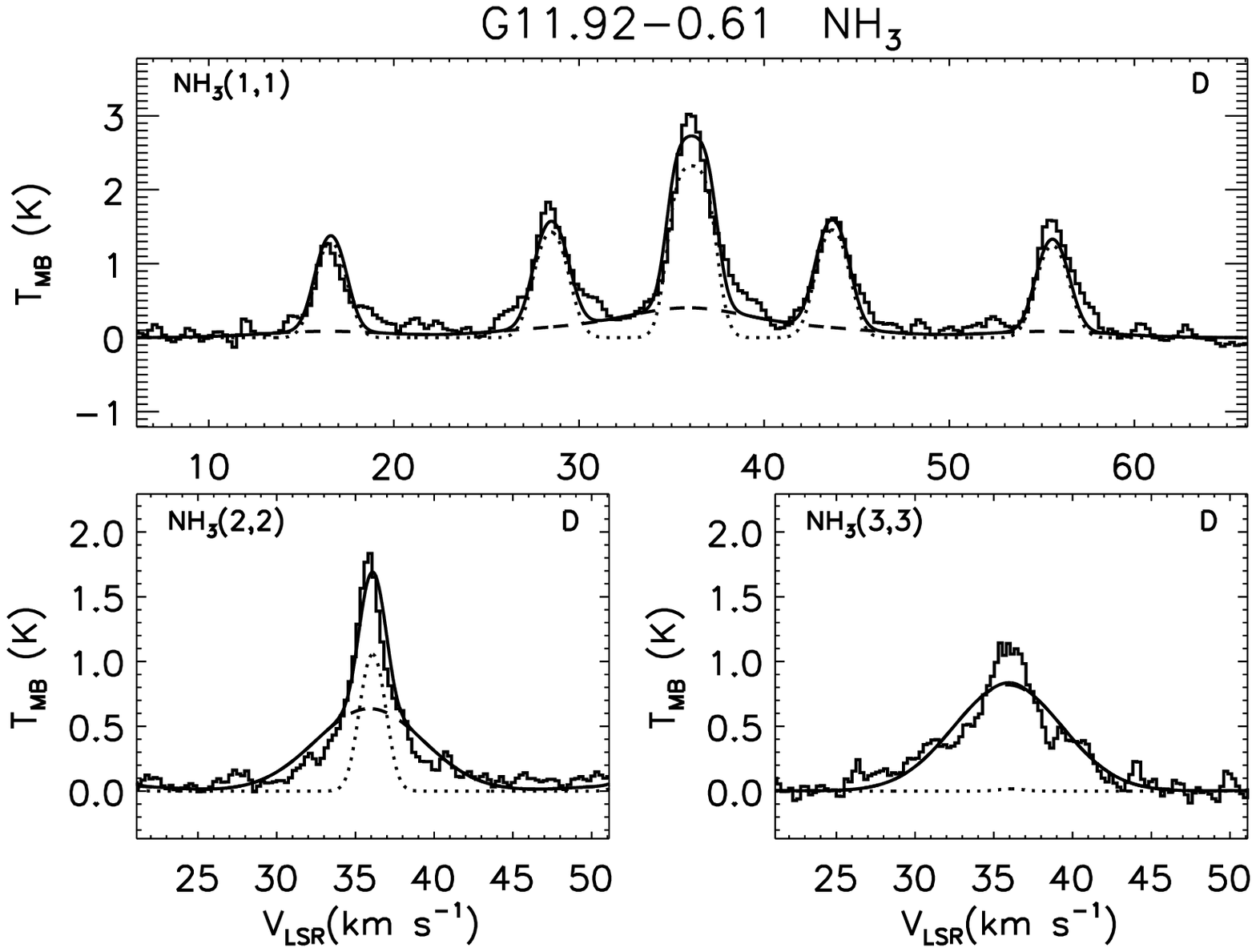}
\caption{Single component (left) and two component (right) fits to
sample \ammonia\/ spectra.  The best-fit models are overplotted on the
observed spectra.  For the two-component fits, model spectra for
each component are shown (dashed line: warmer component; dotted line:
cooler component), as well as their sum (solid line).  The 'D' at
upper right in each panel indicates that our 4$\sigma$ detection
criterion was met for that transition.  G12.91$-$0.26 (top) has two
velocity components.  For G11.92$-$0.61 (bottom), two temperature
components significantly improve the fit to the \ammonia(1,1), (2,2)
and (3,3) spectra. }
\label{2comp_fit_fig}
\end{figure*}

For some sources in our sample, however, a single-slab model does not
adequately represent the amplitudes of all three \ammonia\/
transitions.  Figure~\ref{2comp_fit_fig} shows examples of the two
cases that prompted a revision of our model: (1) spectra that showed
velocity components with different $v_{\mathrm{LSR}}$ or $\sigma_{\rm
v}$; and (2) spectra that could not be well represented by a single
temperature fit.  We found that including a second component produced
significantly better fits in these cases \citep[see][for more
details]{nh3-perseus}.  A second component was introduced for any fit
where the $\chi^2$ per degree of freedom was larger than two for any
individual inversion line (23 sources, \q 25\% of our sample).  For
two sources that met this criterion, the best-fit two component model
included a component with an unphysically low excitation temperature
($<$2.73 K).  For these sources (G14.33$-$0.64 and G19.36$-$0.03), we
retain the single component fits (leaving 21 sources with two-component fits).
In the two component model, the two slabs are nominally beam-filling,
but no radiative transfer is performed from one slab through the
other.  We see no evidence for absorption of one component through the
other in the spectra, suggesting such a treatment is not needed.  A
simple two-component fit yields a substantial improvement in the
quality of the fit for many sources, successfully identifying two
velocity/temperature components.  We again note, however, that slab
models are an incomplete description of the physical system; the
best-fit physical parameters of the two components are thus likely
representative but not definitive.  We also note that a contradiction
arises because the model takes $T_{\rm MB} = \eta_{ff}
(T_{ex}-T_{bg})(1-e^{-\tau})$ where $\eta_{ff}=1$ is the assumed beam
filling factor.  However, the parameter $\eta_{ff}$ is degenerate with
$T_{ex}$, and our assumption that $\eta_{ff}=1$ means $T_{ex}$ is a
lower limit.  Relaxing this constraint on $\eta_{ff}$ leaves $T_{ex}$
undetermined for the two components, and suggests that the success of
the simple two component fitting means the two \ammonia\/ components
are spatially distinguished on smaller scales.

\section{Results}\label{results}

\subsection{Detection Rates}\label{detection_rates}

\subsubsection{Water Masers}\label{water_detect}

We define a water maser detection as $>$4$\sigma$ emission in at least
two adjacent channels.  The overall detection rate is 68\% (64/94),
and Table~\ref{detect_rates_table} summarizes the \water\/ maser
detection rates towards various EGO subsamples.  The uncertainties
quoted in Table~\ref{detect_rates_table} were calculated using
binomial statistics.  Throughout, we treat each EGO separately, though
we note that for EGOs separated on the sky by $\lesssim$36.5\pp (half
the FWHP Nobeyama beam), our data are insufficient to determine
whether one or all are associated with \water\/ masers.  An
unavoidable limitation of single-dish surveys is the possibility that
some \water\/ maser detections are chance alignments within the single-dish beam, and not physically associated
with the target EGOs.  While this can only be definitively addressed
by future high-resolution observations of all detected EGOs, available
data suggests the effect on the sample as a whole is small.  We
searched the literature for reported \water\/ masers with
interferometric positions within 2\arcmin\/ of each EGO with a
\water\/ maser detection in our survey.  Of 27 sources with such data
available, there are only 3 cases (\q11\%) of \water\/ masers within
the Nobeyama beam and not associated with the EGO (see also
\S\ref{water_dis}).

One of the goals of this survey is to investigate whether the MIR EGO
classifications from \citetalias{egocat} correspond to differences in
\water\/ maser associations or dense gas properties.  We find a
somewhat higher \water\/ maser detection rate for EGOs classified as
'likely' MYSO outflow candidates, compared to those classified as
'possible' based on their MIR properties.  Two-tailed binomial tests
reject the null hypothesis that these two detection rates are the same
at the 5\% significance level (p-values \q 0.02).  
We also find a
slightly higher \water\/ maser detection rate towards EGOs \emph{not}
associated with IRDCs, compared to EGOs that are associated with
IRDCs.  In this case, however, two-tailed binomial tests are
consistent with the detection rates being the same, at the 5\%
significance level (p-value=0.07(0.10) adopting the non-IRDC(IRDC)
detection rate as the null hypothesis).  If, instead, EGOs are grouped
based on the \ammonia\/ transitions detected in our survey, much
larger differences in the \water\/ maser detection rates emerge.  We
detect \water\/ masers towards only 44\% of EGOs with \ammonia(1,1)
emission only, compared to 81\% of EGOs with emission in the
higher-excitation \ammonia\/ transitions: a difference of nearly a
factor of two.

There are comparably striking differences in the \water\/ maser
detection rates towards EGO subsamples defined based on \meth\/
maser associations (see Table~\ref{detect_rates_table}).  
To group EGOs by their \meth\/ maser associations, we use the data in
Table 1 of \citetalias{Chen11}.  This dataset, derived from
single-dish surveys, is the most uniform available that includes the
majority (\q 3/4) of our northern EGO targets.  \citetalias{Chen11}
searched for 95 GHz Class I \meth\/ masers towards 192 EGOs (northern
and southern) with the MOPRA telescope ($\theta_{\rm FWHP}$\q36\pp,
3$\sigma$\q1.6 Jy).  They also observed EGOs without known Class II
masers at 6.7 GHz with the University of Tasmania Mt. Pleasant
telescope ($\theta_{\rm FWHP}$\q7\arcmin, 3$\sigma$\q1.5 Jy).  This
produced a three-tiered classification for Class II maser
associations: (1) EGOs associated with Class II masers, based on
published high-resolution data (maser positions known to \q 1\pp\/ or
better); (2) EGOs for which 6.7 GHz emission was detected in the large
Mt Pleasant beam but no positional information was available (``no
information''); and (3) EGOs undetected in the Mt Pleasant
observations.  For this reason, definitive Class II maser information
is available in \citetalias{Chen11} for a smaller number of the EGOs
in our sample (51) than for Class I masers (69 EGOs).  We note one
additional caveat.  The 95 GHz Class I transition observed by
\citetalias{Chen11} is generally weaker than that at 44 GHz, and their
MOPRA observations are significantly less sensitive than the VLA
survey of \citetalias{maserpap}.  As a result, one source in the
\citetalias{maserpap} sample that has weak 44 GHz Class I masers
(G37.48$-$0.10) is listed as a Class I nondetection in
\citetalias{Chen11}.

The most dramatic difference in \water\/ maser detection rates in our
survey is between EGOs associated with both Class I and II \meth\/
masers (20/21 \q 95\%) and EGOs associated with neither type of
\meth\/ maser (5/15 \q 33\%).  The \water\/ maser detection rate is
also very high (\q90\%) towards EGOs with Class I \meth\/ masers
(considering all Class I detections, regardless of Class II
detections/information).  This correlation is consistent with both
Class I \meth\/ and \water\/ masers being associated with outflows,
though \water\/ masers are also observed towards \q46\% of EGOs
undetected at 95 GHz.  Unfortunately, comparison of ``Class I only''
and ``Class II only'' EGO subsamples is limited by the small number
statistics.

\begin{figure}
\figurenum{3}
\plotone{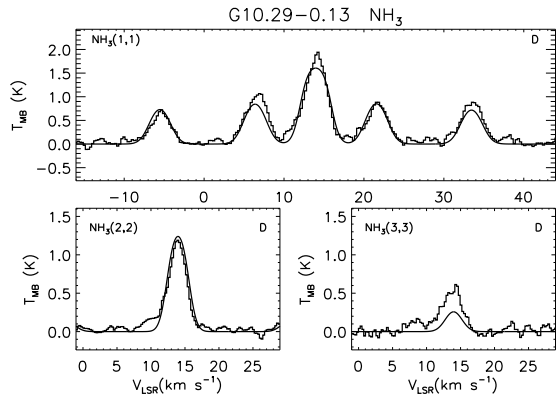}
\caption{Observed NH$_{3}$ spectra with best-fit single component model overlaid.  A 'D' in the upper right corner of a panel indicates that our 4$\sigma$ detection criterion was met for that transition. A complete figure set, including all 91 EGOs detected in \ammonia(1,1), is available in the online journal. \label{nh3_spectra_one_comp_fig} }
\end{figure}

\subsubsection{\ammonia \label{ammonia_detect}}

The vast majority (97\%) of our target EGOs are detected in
\ammonia(1,1) at the 4$\sigma$ level (peak/rms).  
For a significant fraction (34\%)
of our sample, (1,1) is the only \ammonia\/ transition detected.\footnote{One source meets our 4$\sigma$ peak/rms
detection criterion for \ammonia(1,1) and (3,3), but not (2,2).  This could be indicative of nonthermal (3,3) emission; however, 
the \ammonia\/ emission is weak and the (3,3) detection is marginal
($<$5$\sigma$).  Thus, we conservatively treat this source as an
\ammonia(1,1)-only detection in our analysis of detection rates and
in Table~\ref{detect_rates_table}.}
As shown in Table~\ref{detect_rates_table}, it is the
detection rates for the higher-energy transitions, particularly (3,3),
that show significant differences across EGO subsamples.  The
\ammonia(3,3) detection rate towards EGOs associated with IRDCs is
about twice that for non-IRDC EGOs; similarly, the detection rate
towards ``likely'' outflow candidates \citepalias[as classified
by][]{egocat} is about twice that for ``possible'' outflow candidates.
The (2,2) detection rates show the same trends.

The strongest correlation we see, however, is again with \meth\/ maser
associations.  The highest (3,3) detection rate of any subsample is
81\%, towards EGOs with both Class I and II masers, while the lowest
(7\%) is towards EGOs without either type.  The detection rate towards
EGOs with Class I masers (regardless of Class II
association/information) is similarly high, at 76\%.  The (2,2) and
(3,3) detection rates show similar trends, with (3,3) showing larger differences between subsamples.  

\begin{figure*}
\addtocounter{figure}{1}
\center{\includegraphics[scale=0.4]{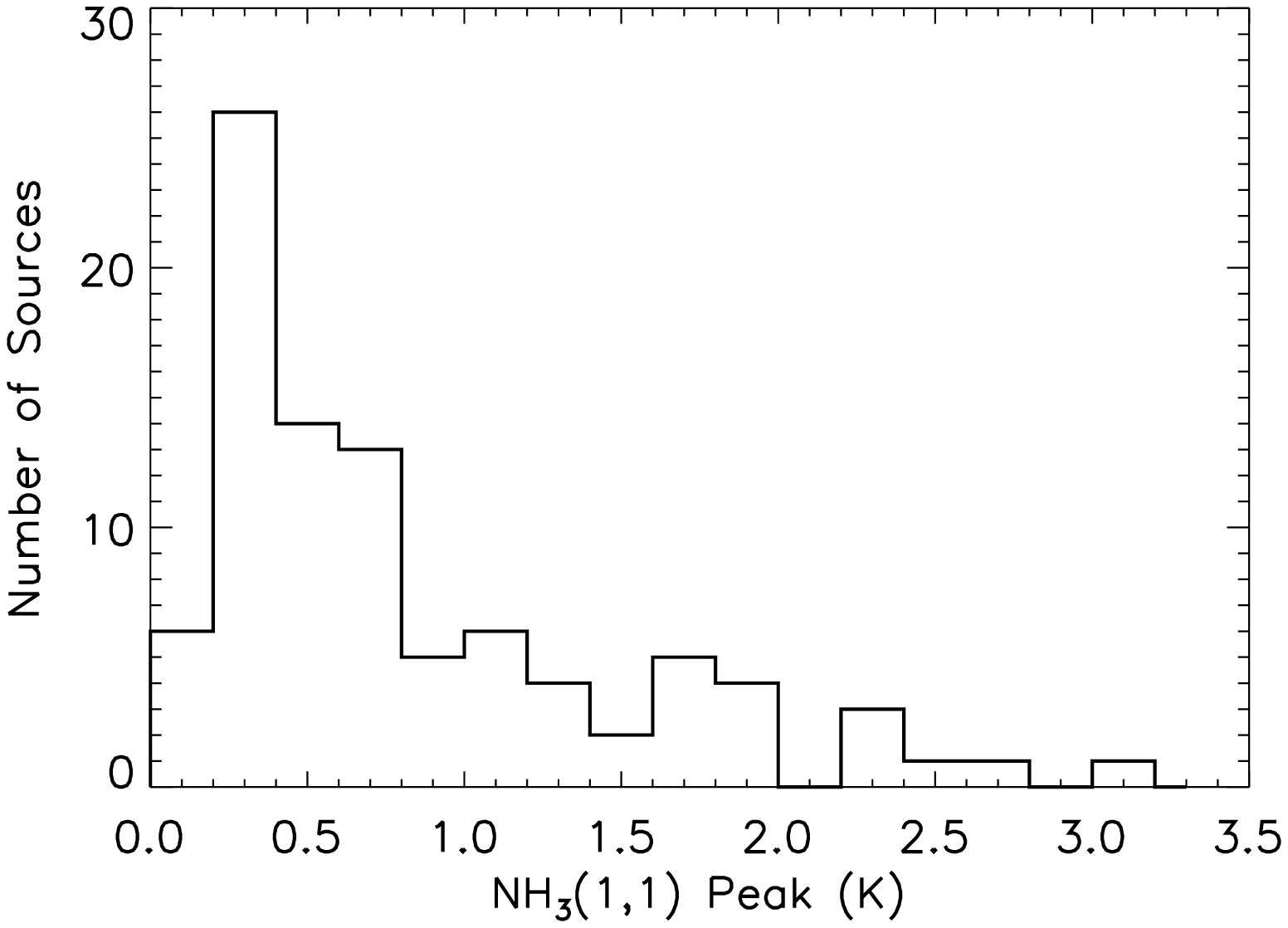}\includegraphics[scale=0.4]{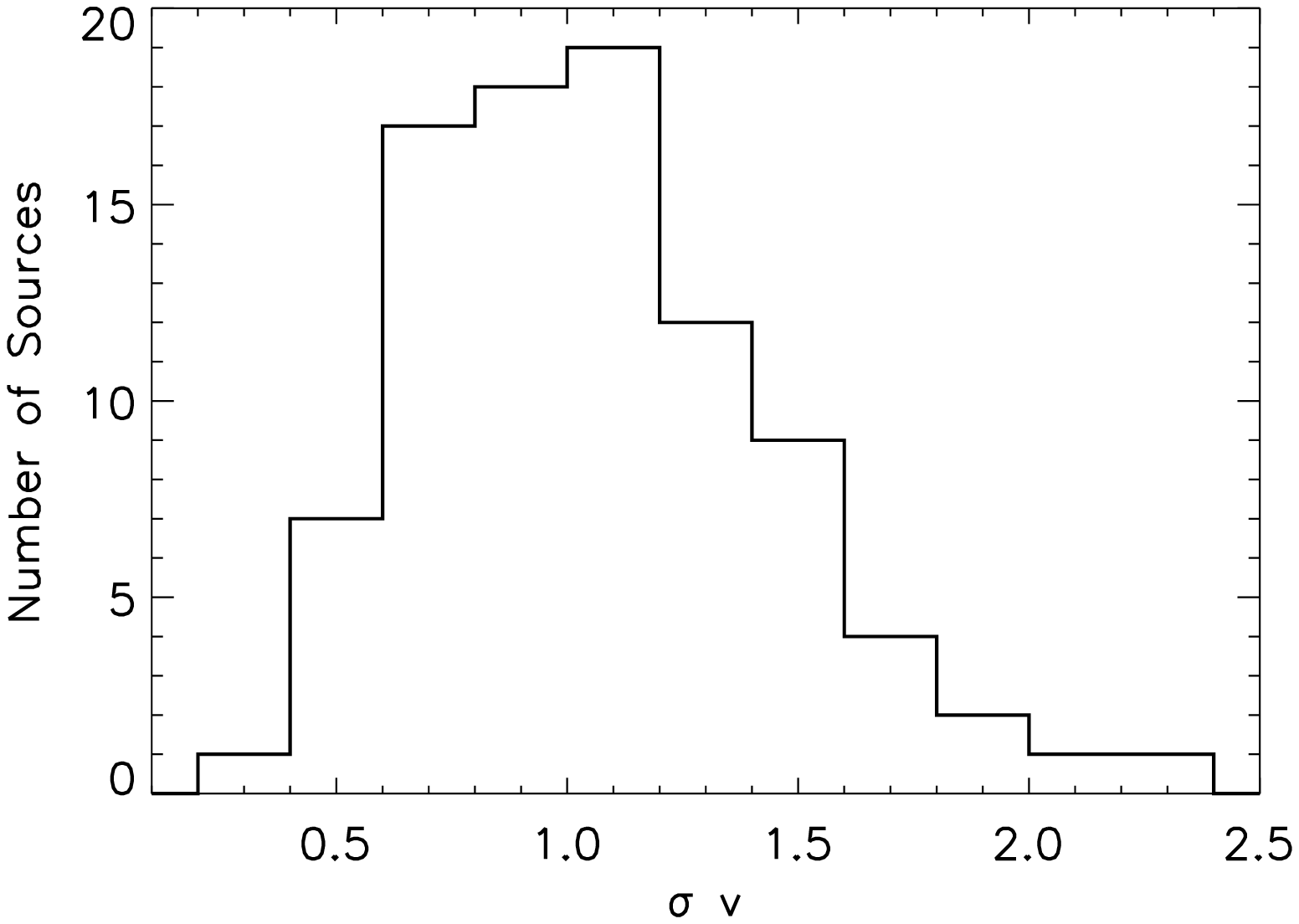}}\\
\includegraphics[scale=0.4]{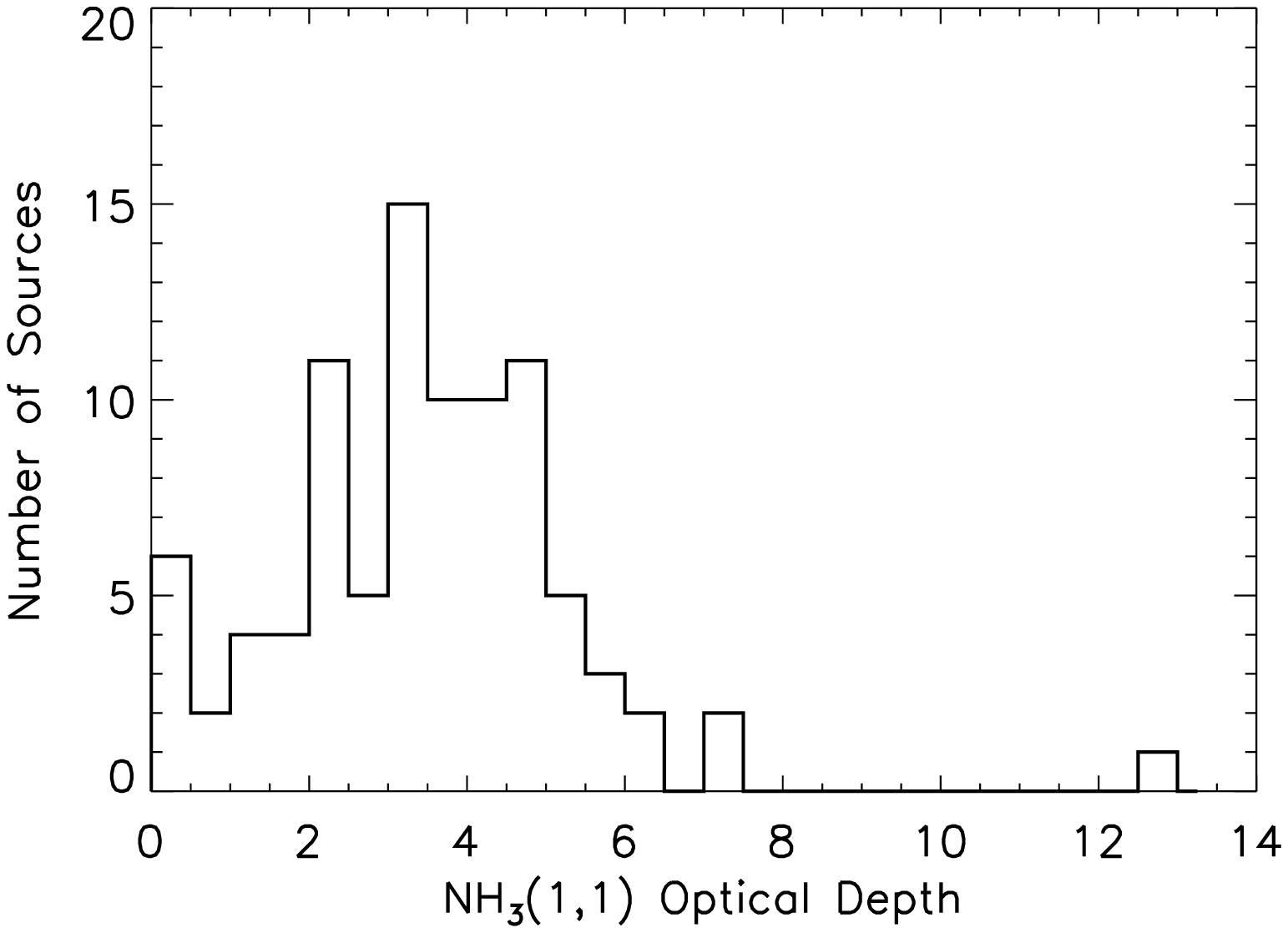}
\includegraphics[scale=0.4]{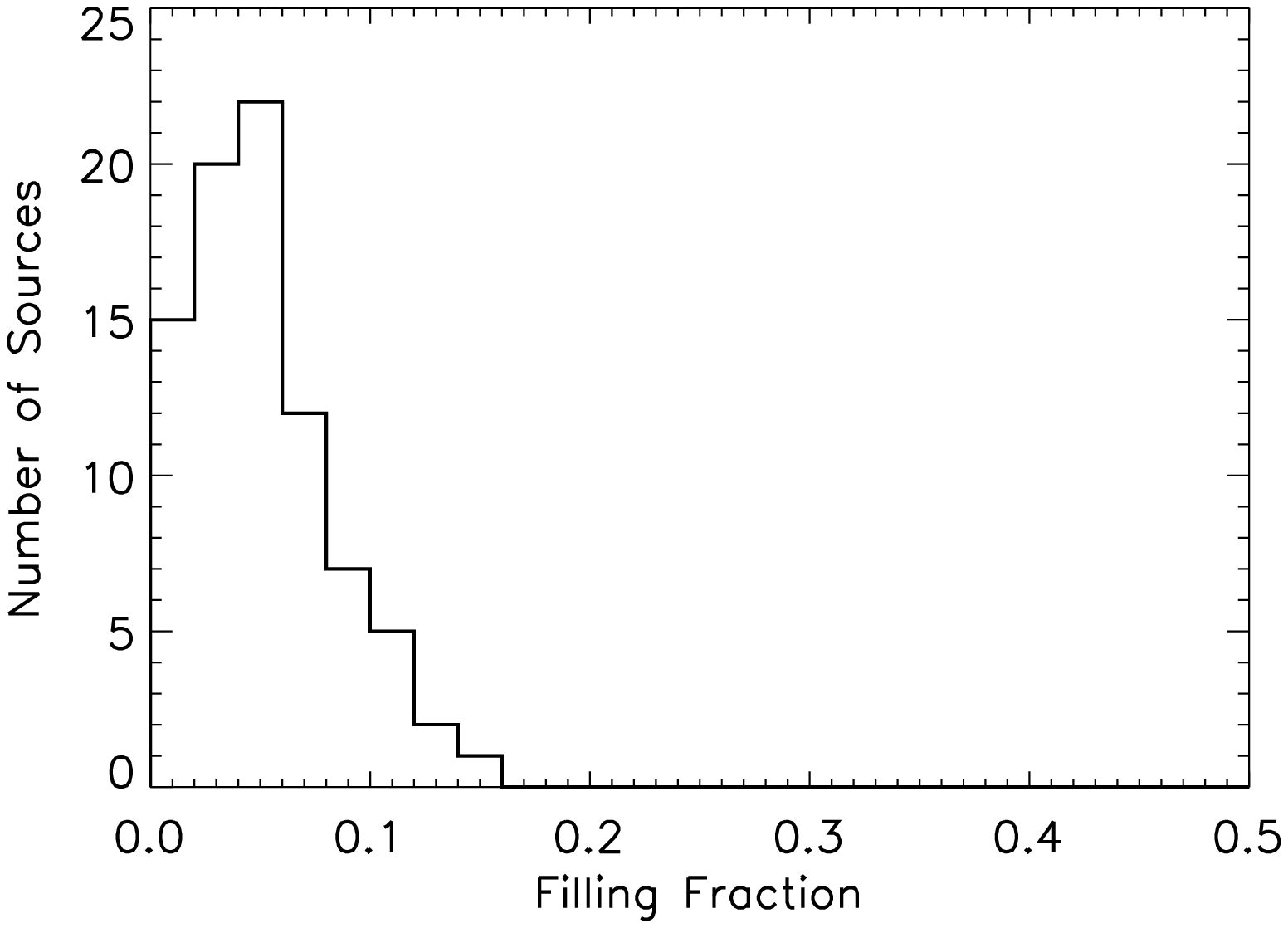}\\
\includegraphics[scale=0.4]{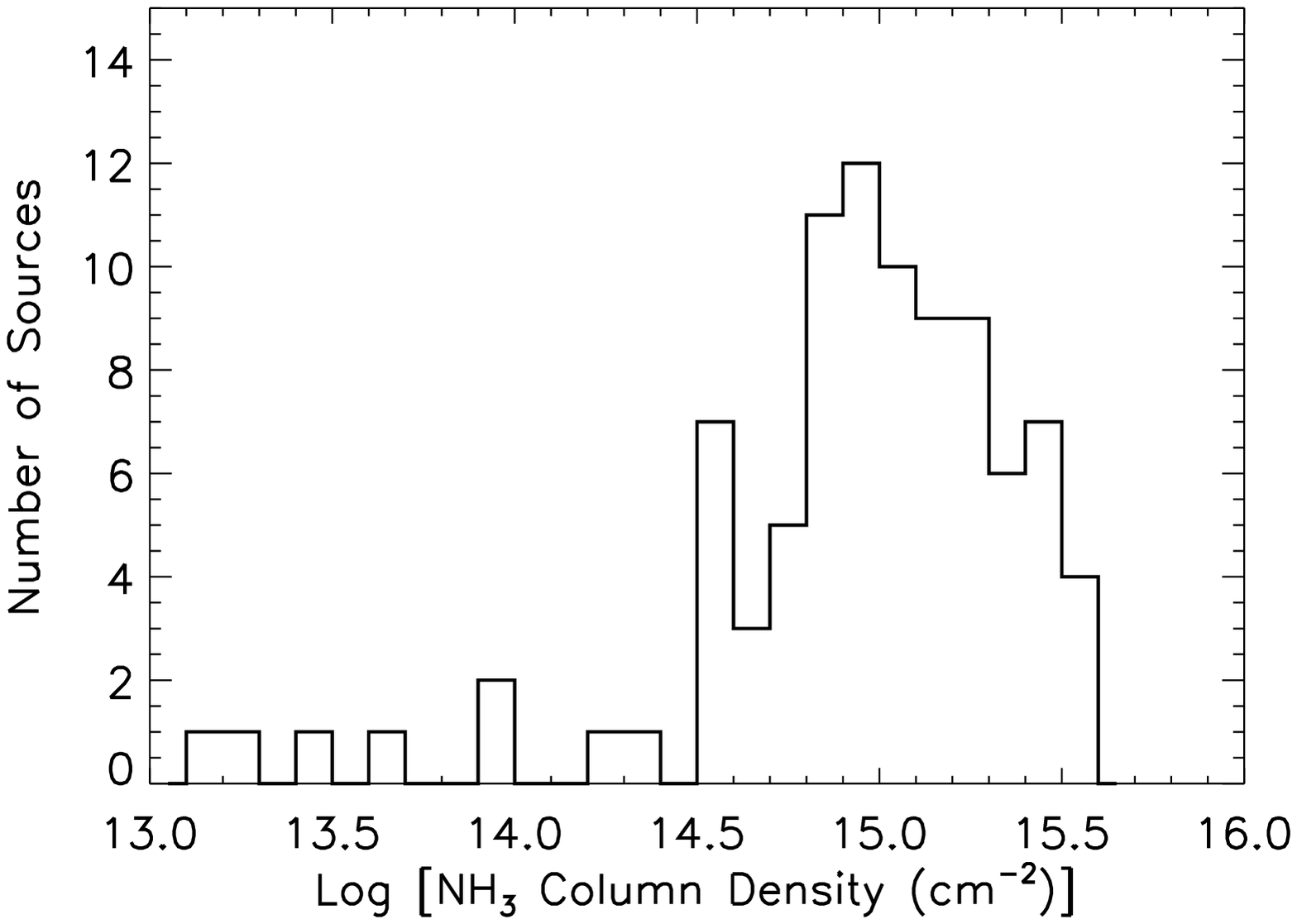}
\includegraphics[scale=0.4]{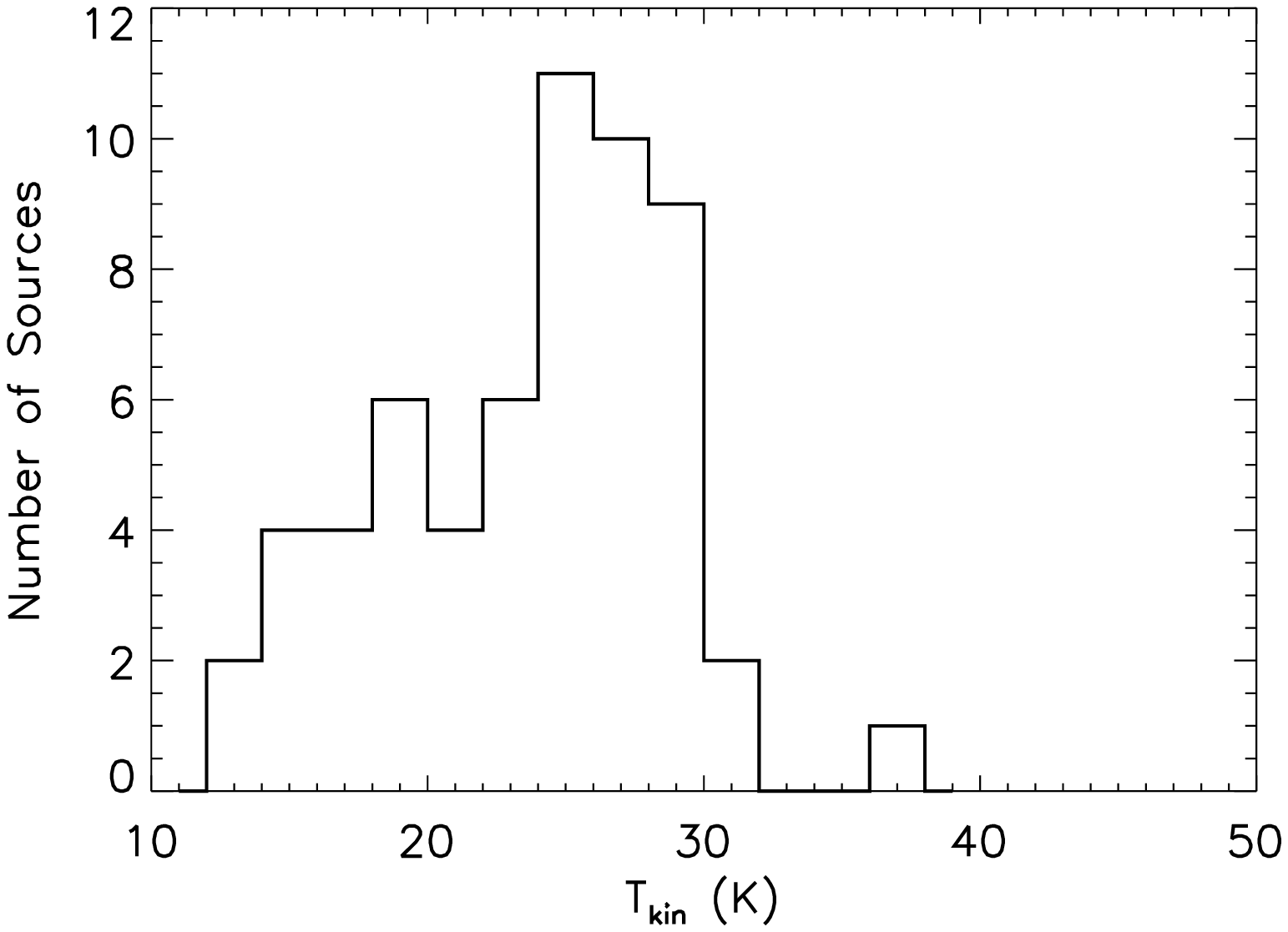}\\
\includegraphics[scale=0.4]{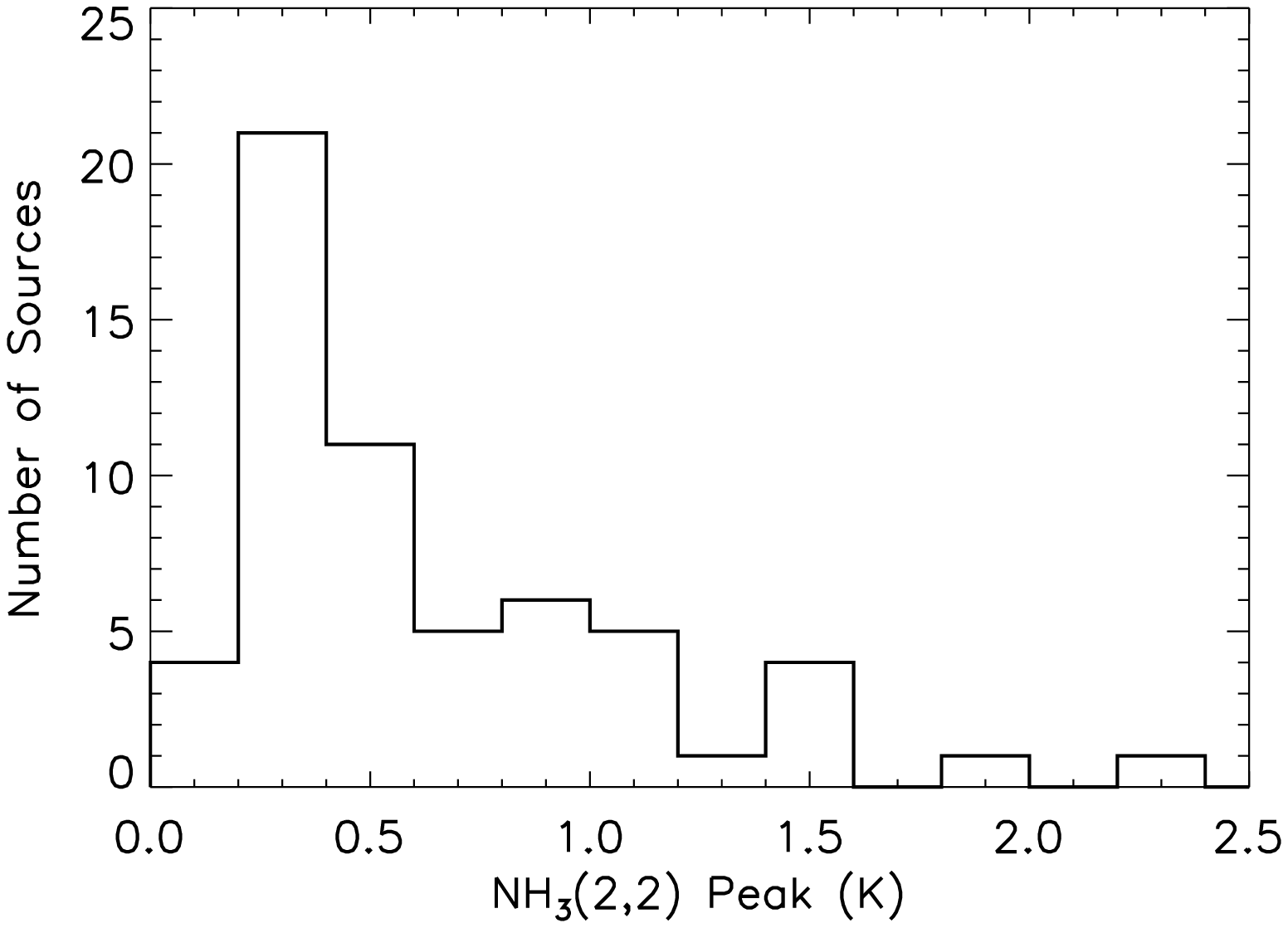}
\includegraphics[scale=0.4]{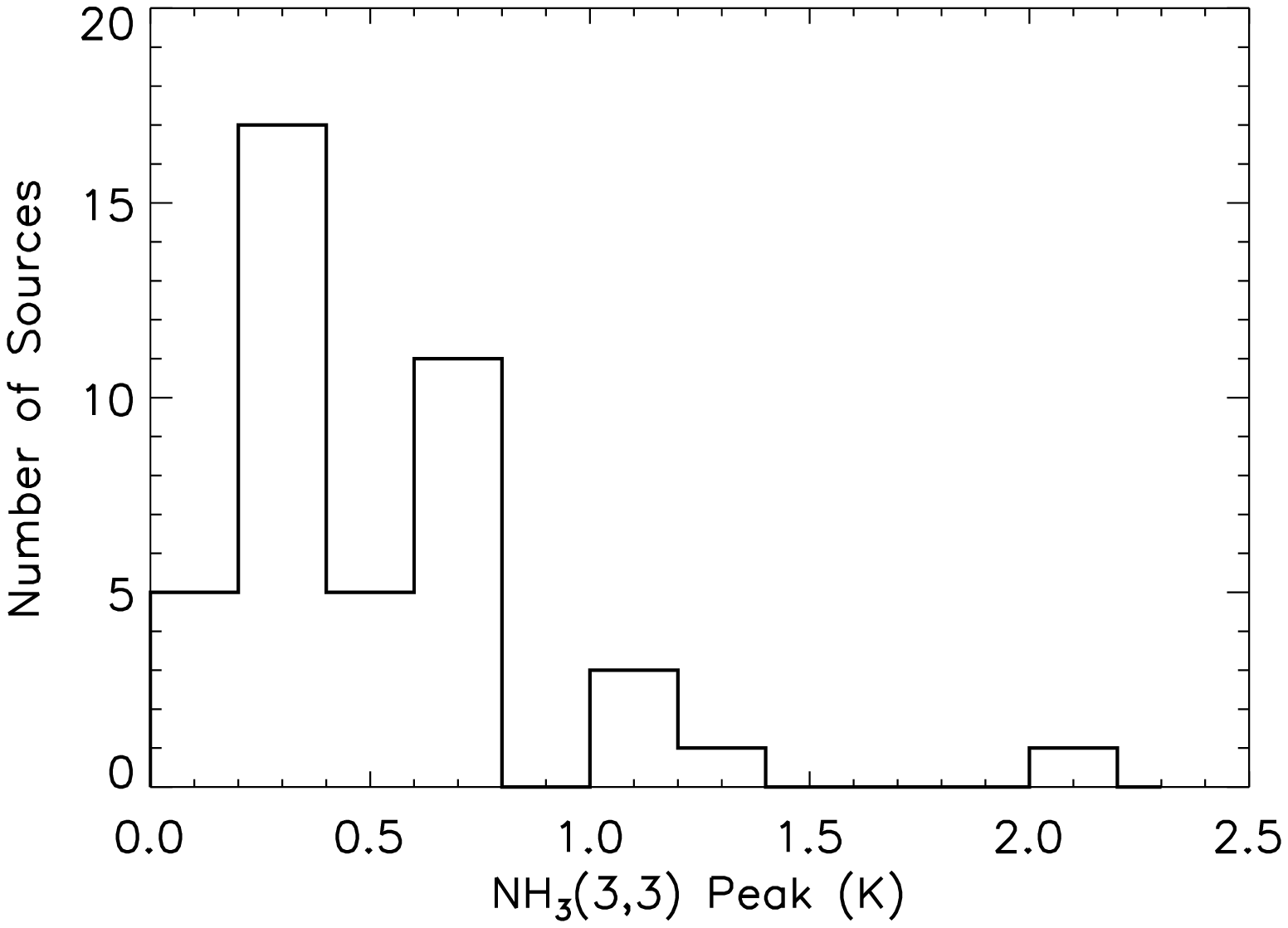}
\caption{\scriptsize Histograms showing distributions of observed \ammonia\/ properties and physical properties obtained from the \ammonia\/ modeling.  Bin sizes are 0.2 K for the \ammonia\/ peak temperatures, 0.2 \kms\/ for $\sigma_{\rm v}$, 0.5 for $\tau_{(1,1)}$, 0.02 for $\eta_{ff}$, 0.1 dex for the \ammonia\/ column density, and 2 K for T$_{kin}$.  All EGOs detected in \ammonia(1,1) are included in the first five panels ((1,1) peak, $\sigma_{\rm v}$, $\tau_{(1,1)}$, $\eta_{ff}$, and column density).  Sources for which T$_{ex}$=T$_{kin}$ (the upper limit, for $\eta_{ff}$=1) are excluded from the filling fraction plot.  EGOs detected in both \ammonia(1,1) and (2,2) are included in the T$_{kin}$ and (2,2) peak histograms, and EGOs detected in all three \ammonia\/ transitions are included in the (3,3) peak plot.  \label{nh3_prop_fig}}
\end{figure*}

\subsubsection{\ammonia\/ Nondetections}\label{nh3_nondetect}

Our extremely high \ammonia(1,1) detection rate raises the question of
whether the three nondetections are in some way unusual, or
interlopers in the EGO sample.  The (1,1) nondetections do have some
common characteristics: they are not associated with IRDCs and do not
have Class I \meth\/ masers.  Two have detected \water\/ maser
emission in our survey.  G49.42+0.33, a \citetalias{egocat} ``likely''
outflow candidate, was included in the \citetalias{maserpap} sample
and detected in thermal HCO$^{+}$(3-2), H$^{13}$CO$^{+}$(3-2), and
CH$_{3}$OH (5$_{2,3}$-4$_{1,3}$) emission with the JCMT.  Thus, there
is dense gas associated with the EGO: in combination with the
detection of Class II \meth\/ \citepalias{maserpap} and \water\/
masers (Table~\ref{water_detect_tab}), strong evidence for the presence of MYSO(s).  This
EGO is among the most distant in our sample, so our Nobeyama
\ammonia\/ nondetection may be attributable to sensitivity and/or beam
dilution.

We also detect \water\/ maser emission towards G53.92-0.07, a
\citetalias{egocat} ``possible'' outflow candidate.  Its MIR
morphology is unusual amongst the EGO sample; the ``green'' source
appears embedded in an 8 \um-bright pillar, and the 4.5 \um\/ emission
is only slightly extended.  Little is known about this source beyond
its identification as an EGO and its association with a BGPS 1.1 mm
source, but it is possible it may be a
comparatively evolved outlier in the EGO sample.

Finally, G57.61+0.02 is a ``possible'' outflow candidate located on
the edge of an 8 and 24\um-bright nebula, likely a more evolved source
(e.g. compact or UC HII region).  Formally undetected by our 4$\sigma$
criteria, we do see a weak (\q 3.9$\sigma$) \ammonia(1,1) line in our
spectra (see also \S~\ref{distances}).

\subsection{\ammonia\/  Properties}\label{nh3_prop}

Table~\ref{nh3_prop_tab} presents the physical properties obtained
from the single-component \ammonia\/ modeling for all EGOs detected in
\ammonia\/ emission in our survey.  The \ammonia(1,1), (2,2), and
(3,3) peaks (T$_{\rm MB}$) are also listed, with 4$\sigma$ upper limits
given for undetected transitions (for all sources, including
\ammonia\/ nondetections).  If \ammonia(2,2) is not detected, the
best-fit T$_{kin}$ is treated as an upper limit and is indicated as
such in Table~\ref{nh3_prop_tab}.  The observed \ammonia\/ spectra for
each detected source, overlaid with the best-fit model, are shown in
Figure~\ref{nh3_spectra_one_comp_fig} (available online in its
entirety), and the property distributions for our EGO sample are shown
in Figure~\ref{nh3_prop_fig}.  Throughout, the \ammonia(1,1) peak
(T$_{\rm MB}$), $\sigma_{\rm v}$, $\tau_{(1,1)}$, $\eta_{ff}$, and \ammonia\/
column density are presented for all EGOs detected in \ammonia(1,1)
emission.  In Figure~\ref{nh3_prop_fig}, the T$_{kin}$ and
\ammonia(2,2) peak distributions include only sources with
$>$4$\sigma$ \ammonia(2,2) detections, and the \ammonia(3,3) peak
distribution includes only sources with $>$4$\sigma$ detections in all
three \ammonia\/ transitions.  For EGOs where a two component model
provides a better fit to the observed \ammonia\/ emission
(\S\ref{nh3_modeling}), Figure~\ref{nh3_spectra_two_comp_fig}
(available online in its entirety) shows the spectra overlaid with the
best-fit two component model, and Table~\ref{nh3_prop_tab_2comp}
presents the parameters of the two-component fits.

\begin{figure}
\figurenum{5}
\plotone{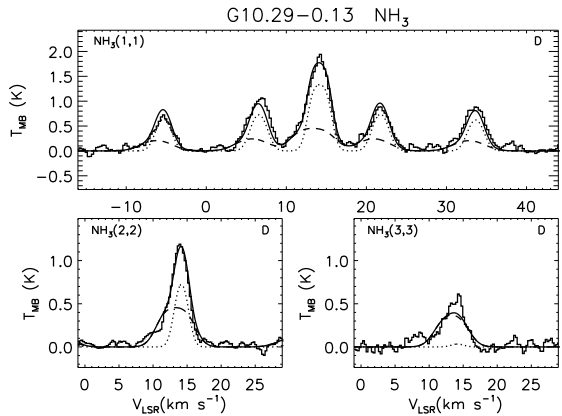}
\caption{Observed NH$_{3}$ spectra with best-fit two component model
overlaid.  Model spectra for each component are shown (dashed line:
warmer component; dotted line: cooler component), as well as their sum
(solid line).  A 'D' in the upper right corner of a panel indicates
that our 4$\sigma$ detection criterion was met for that transition.  A
complete figure set, including all 21 EGOs with two-component fits, is available in the online journal. \label{nh3_spectra_two_comp_fig}}
\end{figure}

\subsubsection{Kinematic Distances}\label{distances}

We calculate kinematic distances based on the \ammonia\/ velocities in
Table~\ref{nh3_prop_tab} and the prescription of \citet{Reid09}, using
updated input parameters (M. Reid, priv. comm., 2012; Galactic: $R_O$=
8.40 kpc, $\Theta_0$= 245.0 \kms, $d\Theta/dr$= 1.0 \kms\/ kpc$^{-1}$;
Solar: U$_0$= 10.00 \kms, V$_0$= 12.00 \kms, W$_0$= 7.20 \kms; Source
pecular motions: U$_S$= 5.00 \kms, V$_S$= -6.00 \kms, W$_S$= 0.00
\kms; and an assumed \vlsr\/ uncertainty of 7 \kms).  For sources with
distance ambiguities, the near kinematic distance is listed in Table~\ref{nh3_prop_tab}, unless otherwise noted.  The angular extent of EGOs on the sky
supports adopting the near kinematic distance, as does the association
of EGOs, as a population, with IRDCs \citepalias[see
also][]{egocat,maserpap}.  In their HI self-absorption study of 6.7
GHz \meth\/ masers, \citet{Green11} have recently suggested assigning
the far distance to masers associated with a few (eight) of our
targets.  Most of these assignments are 'Class B' in their scheme,
reflecting uncertainty in the classification.  For these sources, we
adopt the far distance calculated from the \ammonia\/ velocity.  Maser
parallax distances are adopted when available, as noted in Table~\ref{nh3_prop_tab}.
 
Three sources are undetected in \ammonia(1,1), and so present special
cases for calculating kinematic distances.  For G49.42+0.33, we use
the H$^{13}$CO$^{+}$(3-2) velocity from \citetalias{maserpap} (see also
\S~\ref{nh3_nondetect}).  For G53.92$-$0.07, the \water\/ maser
emission is very narrow ($\Delta$v=1.3 \kms), and we calculate a
kinematic distance using the \water\/ maser peak velocity (Table~\ref{water_detect_tab}).
In G57.61+0.02, we detect weak \ammonia(1,1) emission at \q
3.9$\sigma$, just below our formal detection limit.  The fitted
\vlsr\/ of 37.4$\pm$0.1 \kms\/ gives a kinematic distance of 4.50$\pm$1.96
kpc.  For completeness, we include this source in the distance
histogram shown in Figure~\ref{distance_histo}, but not in the subsequent analysis.
The mean(median) distance for our sample is 4.3 kpc(4.2 kpc). 

\begin{figure}
\addtocounter{figure}{1}
\plotone{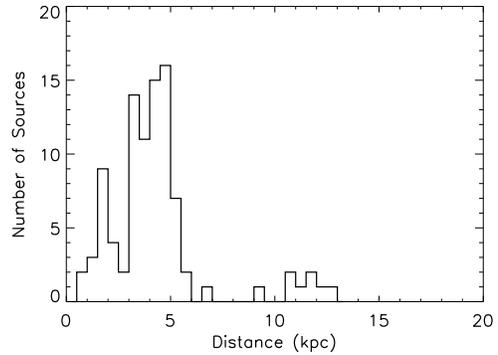}\\
\caption{Distribution of adopted distances for all sources in our sample (\ref{distances}).  The bin size is 0.5 kpc.  \label{distance_histo}}
\end{figure}

\begin{figure*}
\center{\includegraphics[scale=0.45]{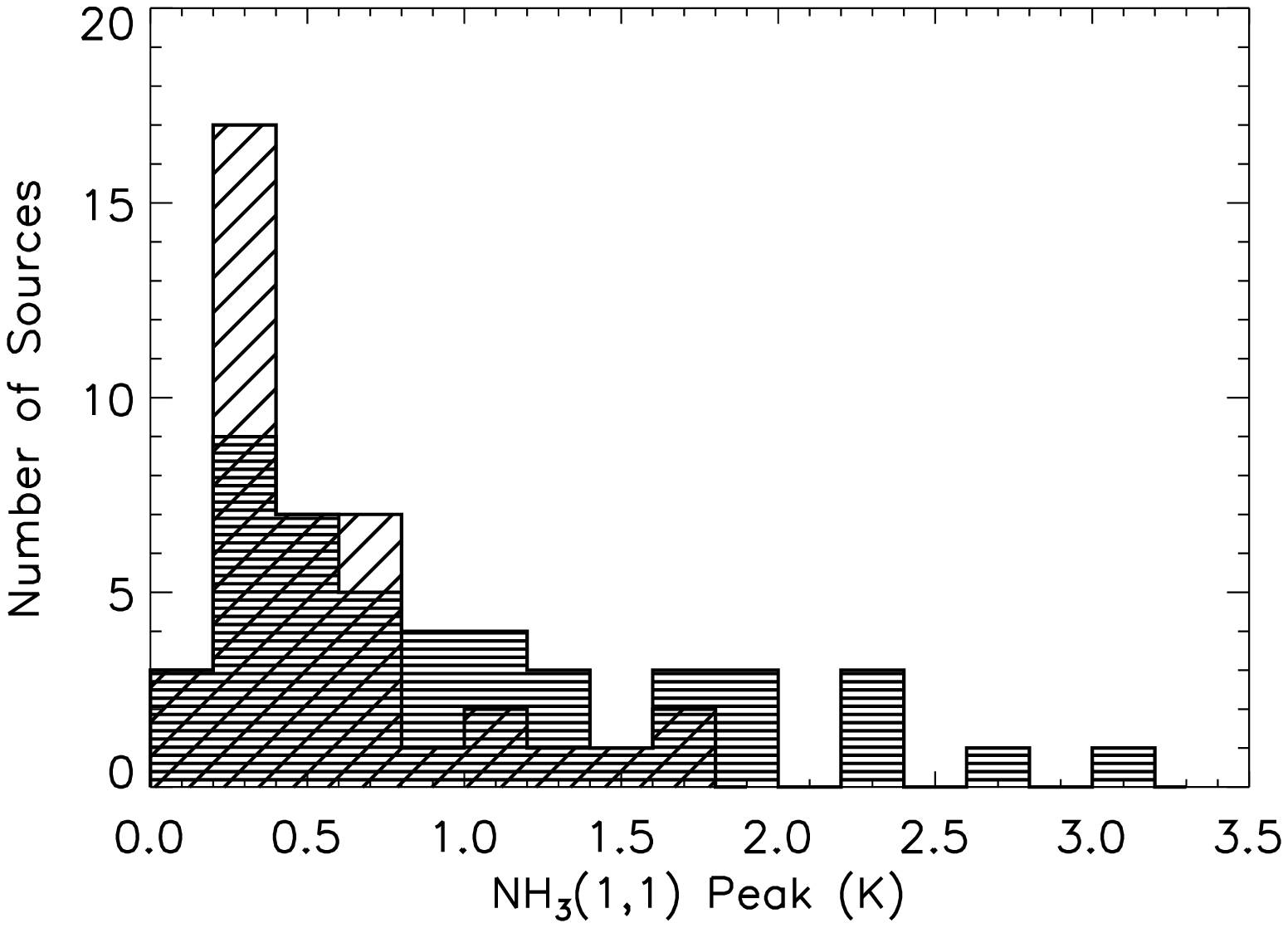}\includegraphics[scale=0.45]{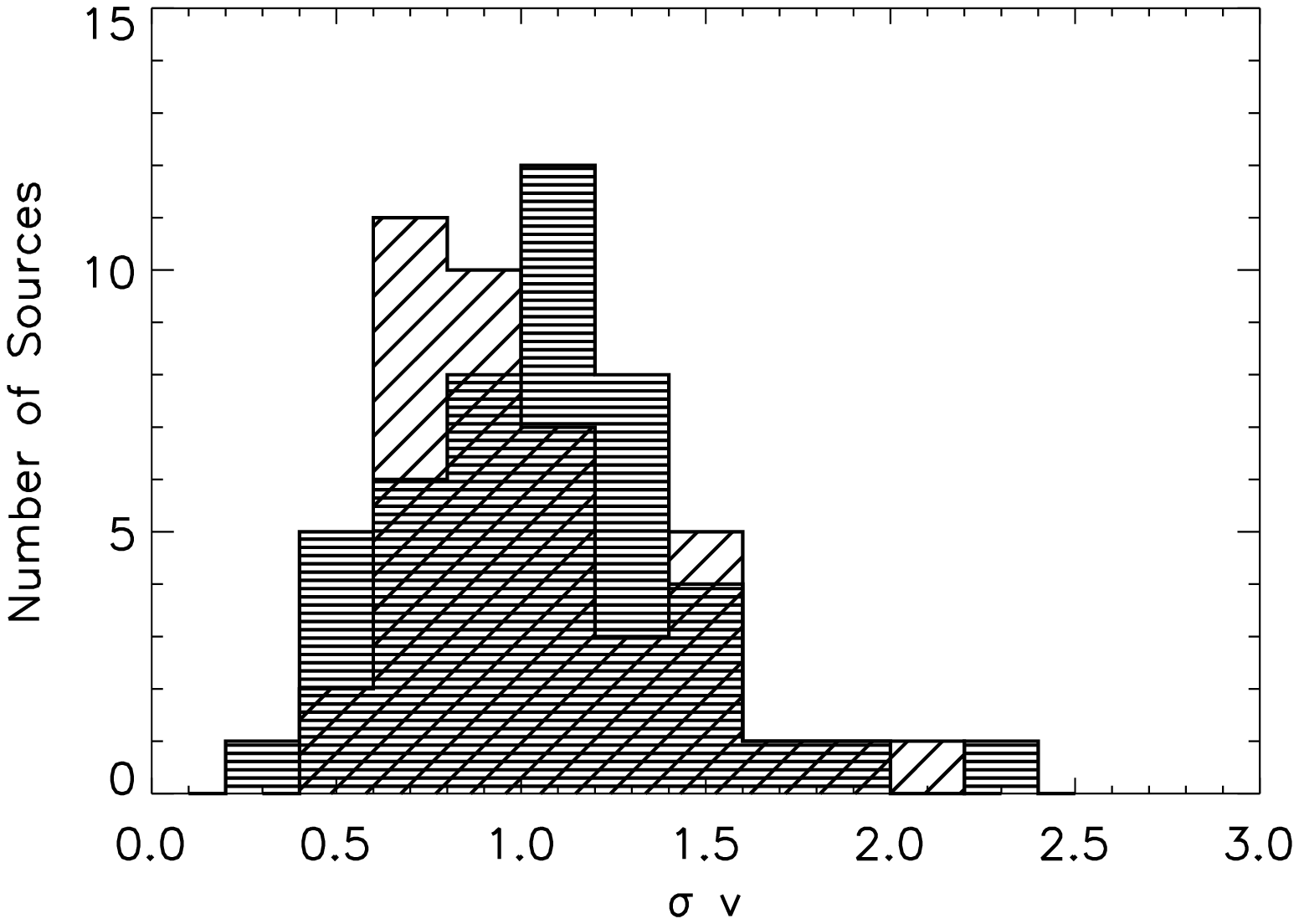}}\\
\includegraphics[scale=0.45]{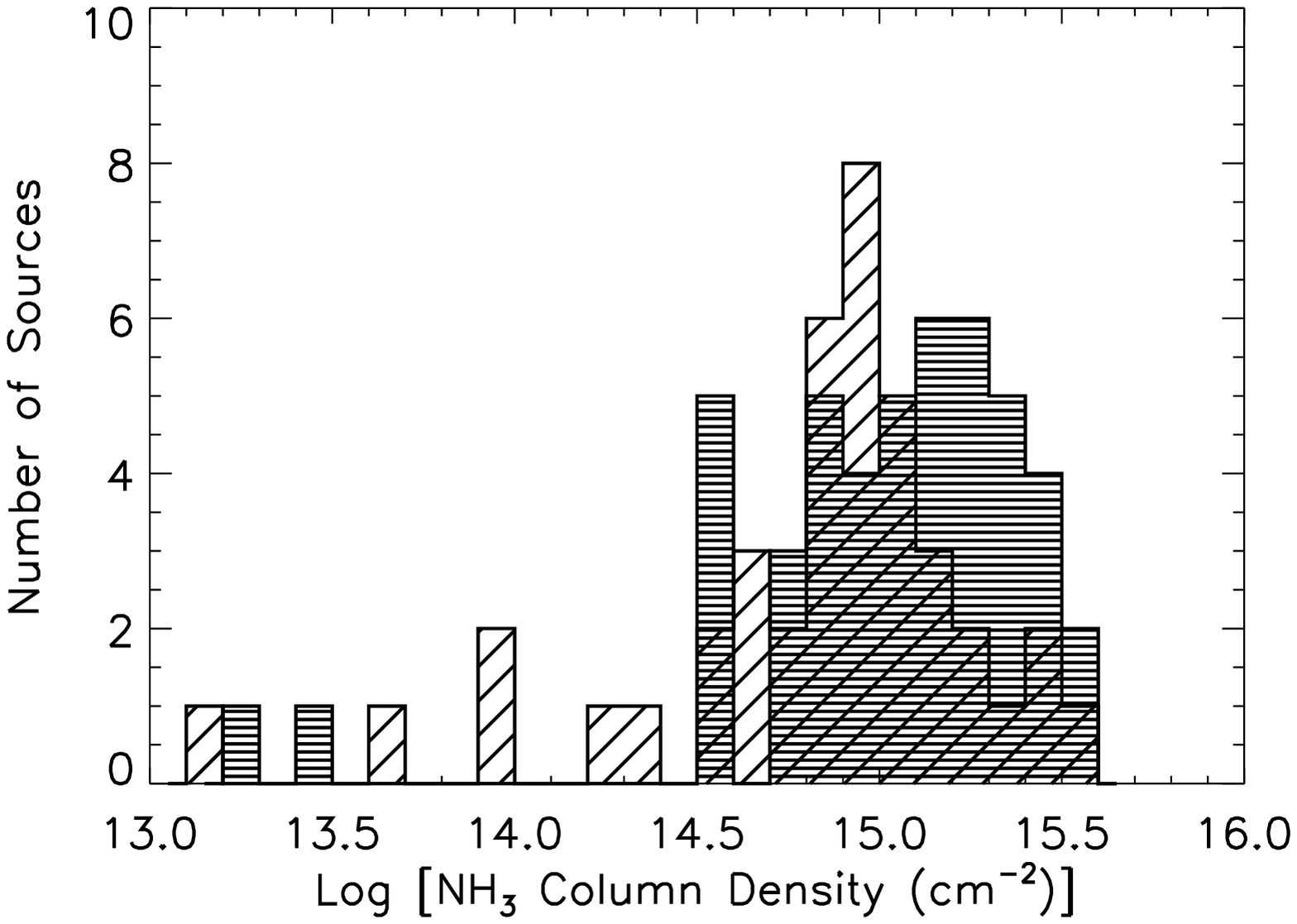}
\includegraphics[scale=0.45]{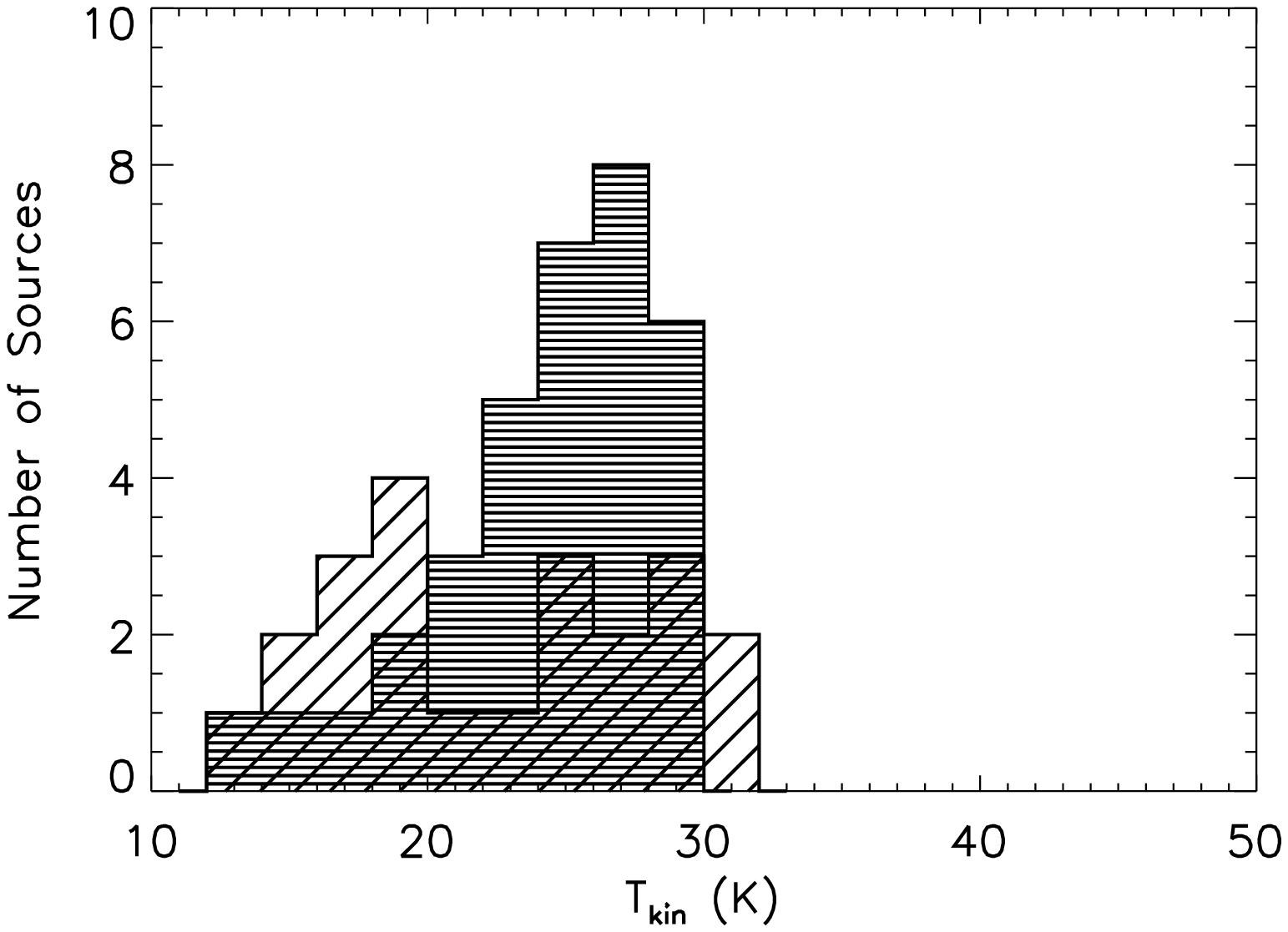}
\caption{\ammonia\/ property distributions for EGOs classified as
``likely'' and ``possible'' MYSO outflow candidates by
\citetalias{egocat}.  ``Likely'' and ``possible'' sources are plotted
as horizontally and diagonally hatched histograms, respectively.  Bin sizes are the same as in Figure~\ref{nh3_prop_fig}.
\label{nh3_prop_likely_poss_fig}}
\end{figure*}

\begin{figure*}
\center{\includegraphics[scale=0.45]{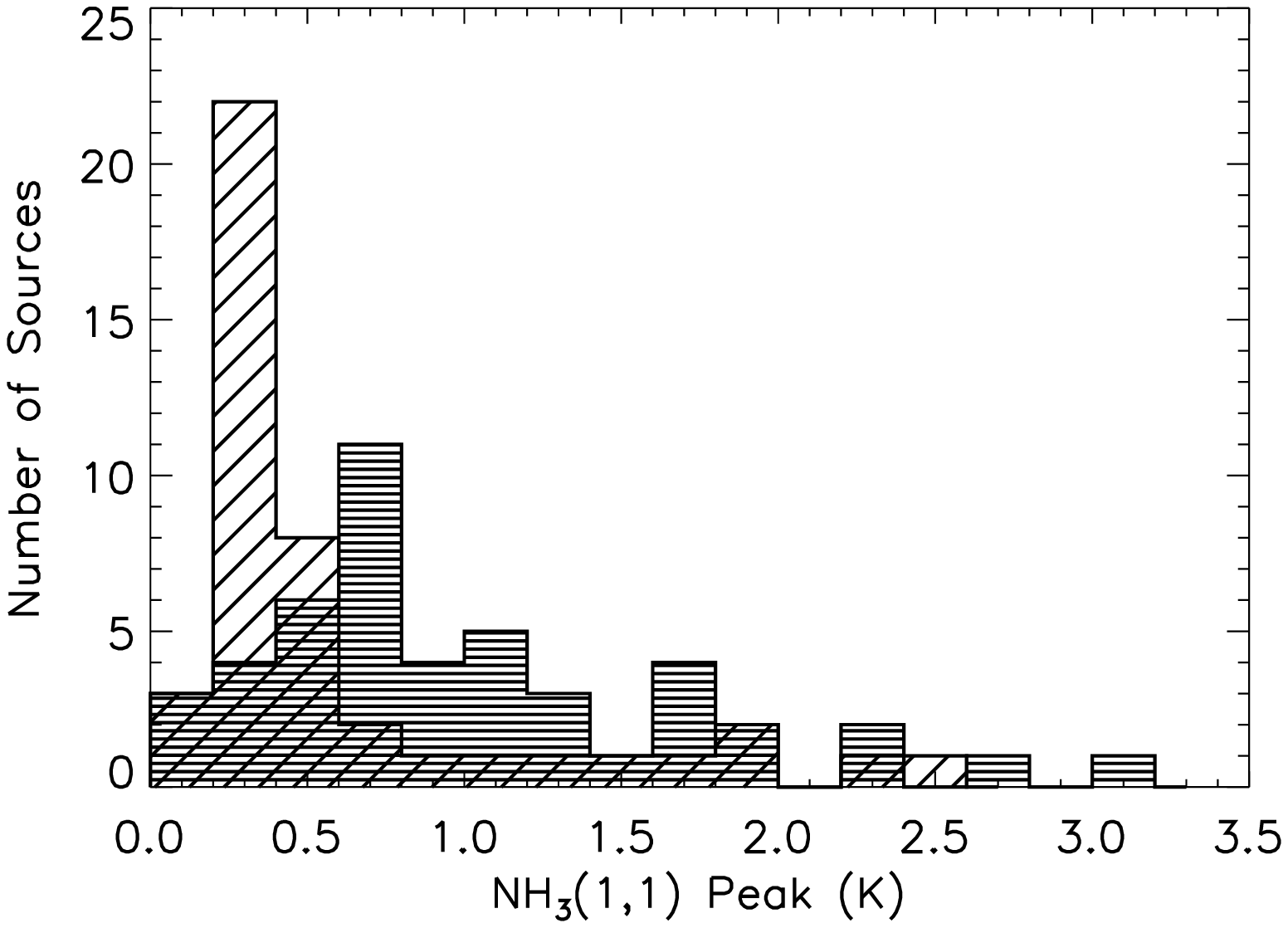}\includegraphics[scale=0.45]{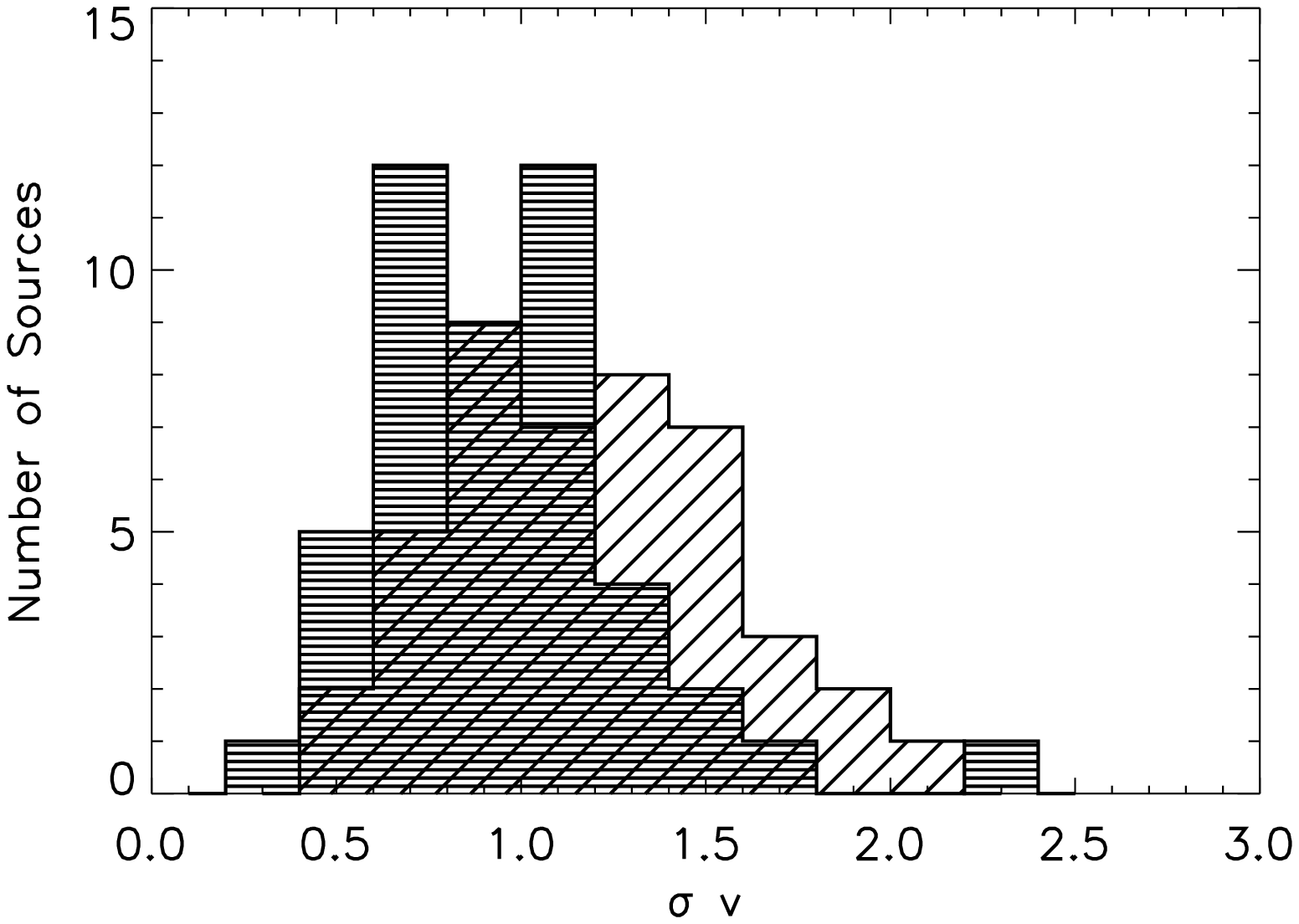}}\\
\includegraphics[scale=0.45]{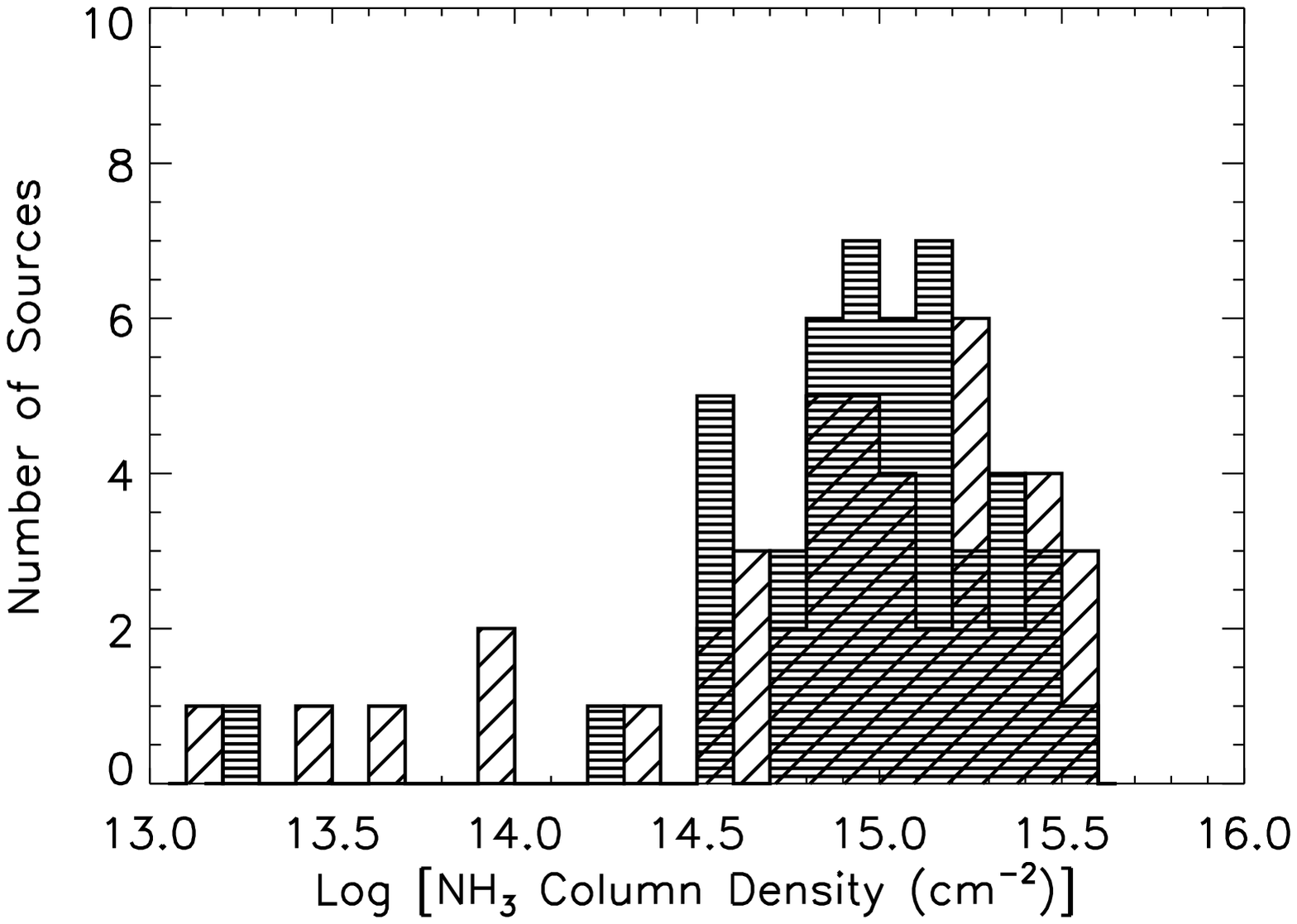}
\includegraphics[scale=0.45]{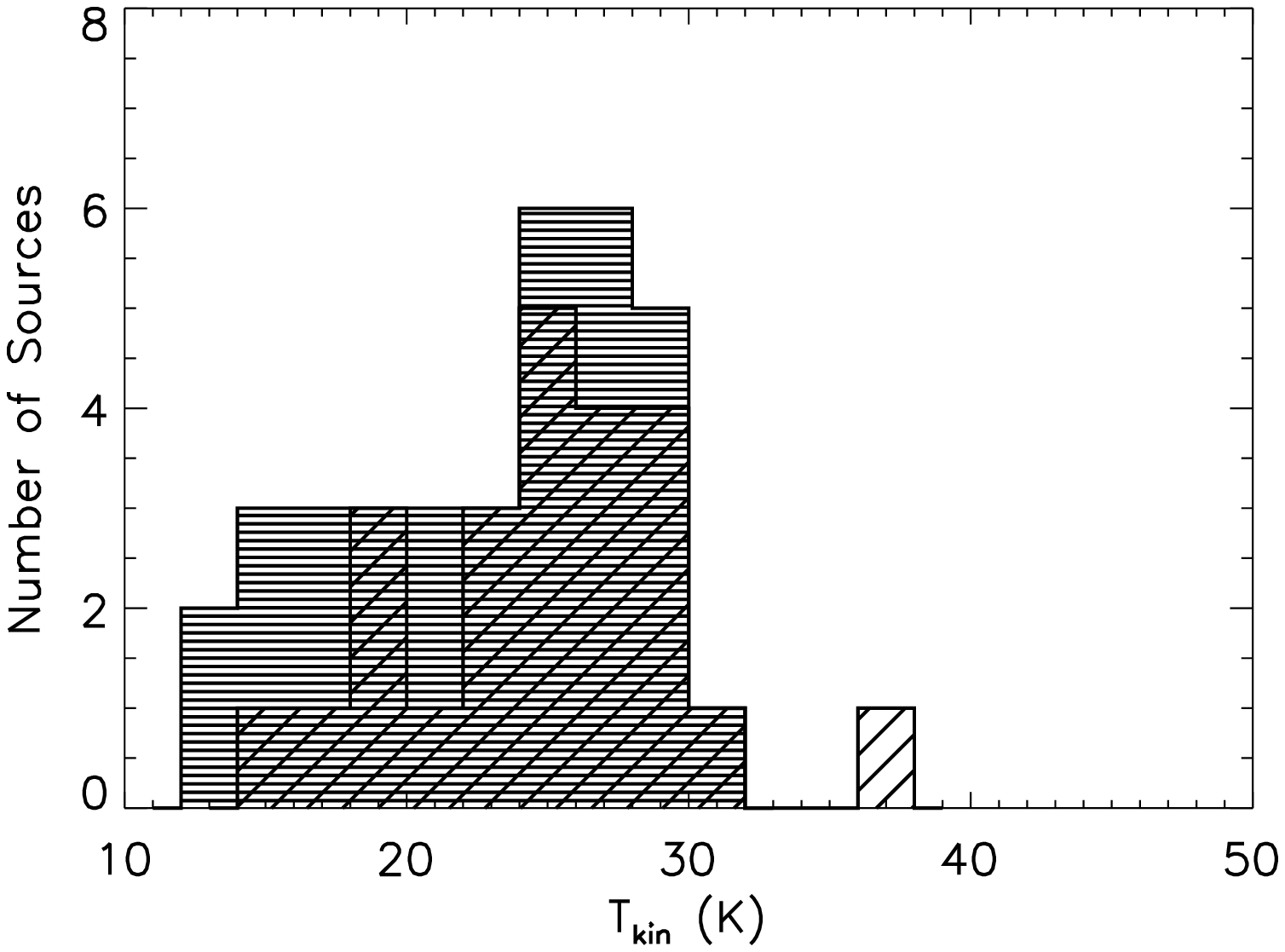}
\caption{\ammonia\/ property distributions for EGOs associated/not associated with IRDCs, plotted
as horizontally and diagonally hatched histograms, respectively. Bin sizes are the same as in Figure~\ref{nh3_prop_fig}.
\label{nh3_prop_irdc_fig}}
\end{figure*}

\begin{figure*}
\center{\includegraphics[scale=0.45]{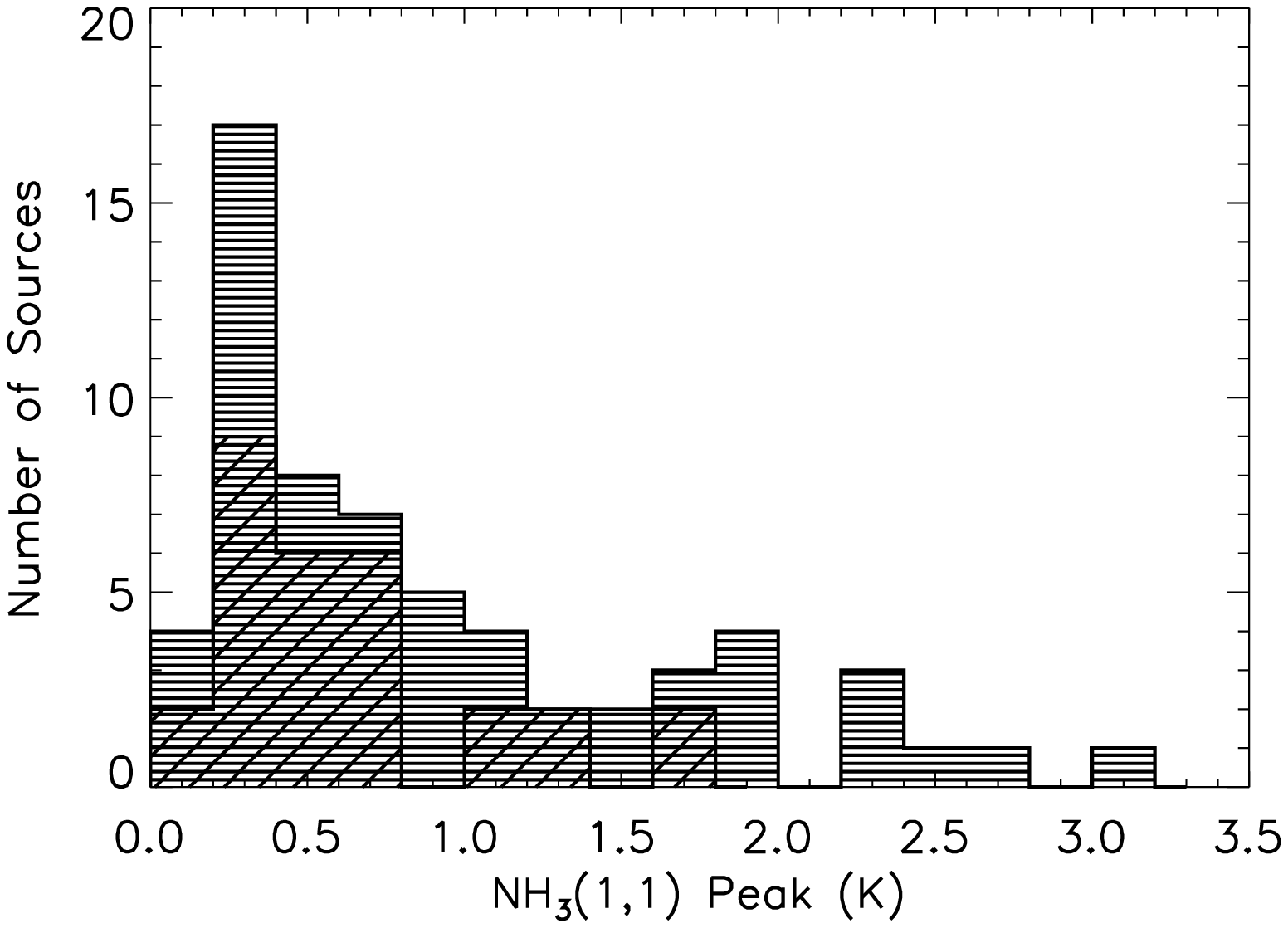}\includegraphics[scale=0.45]{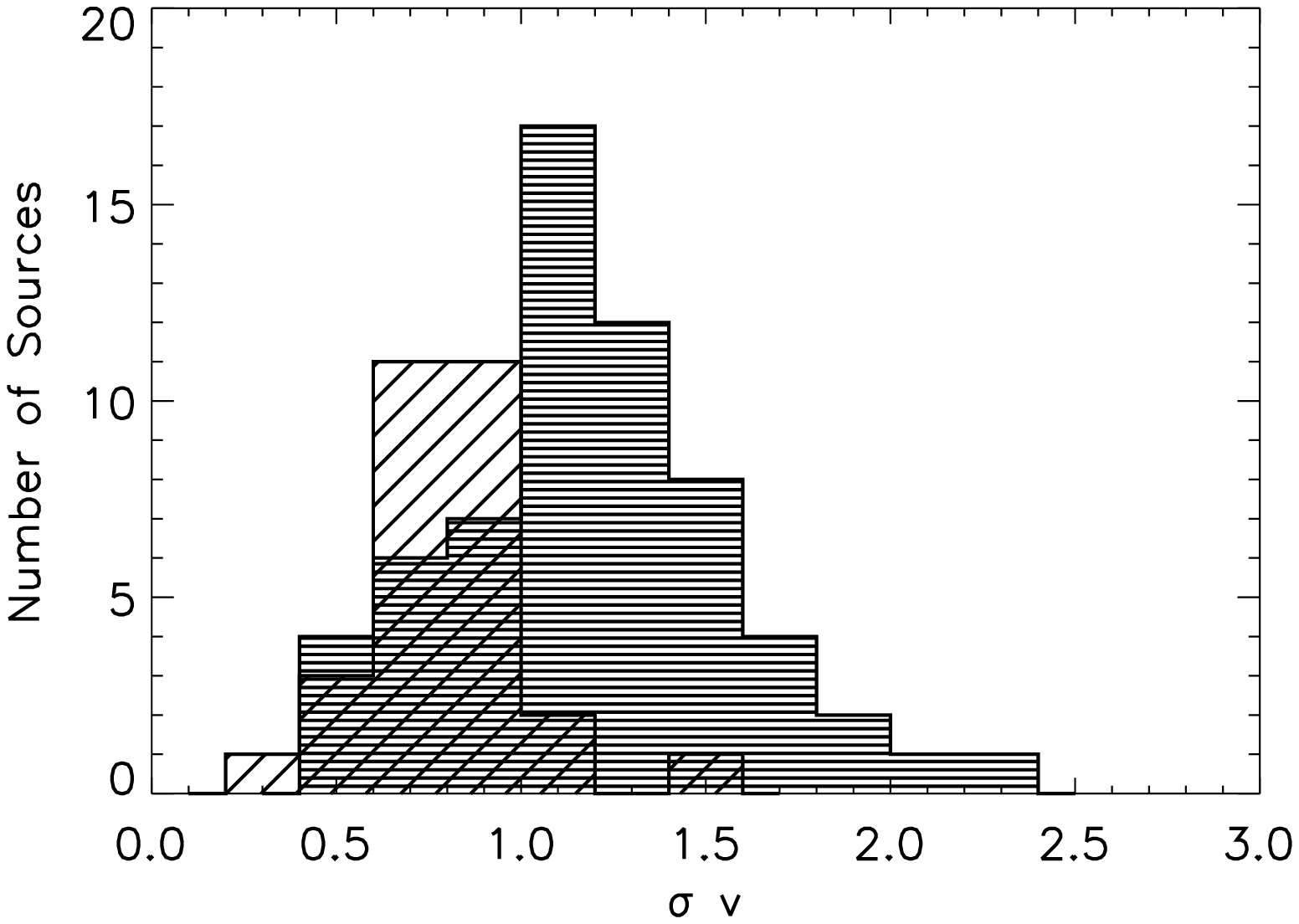}}\\
\includegraphics[scale=0.45]{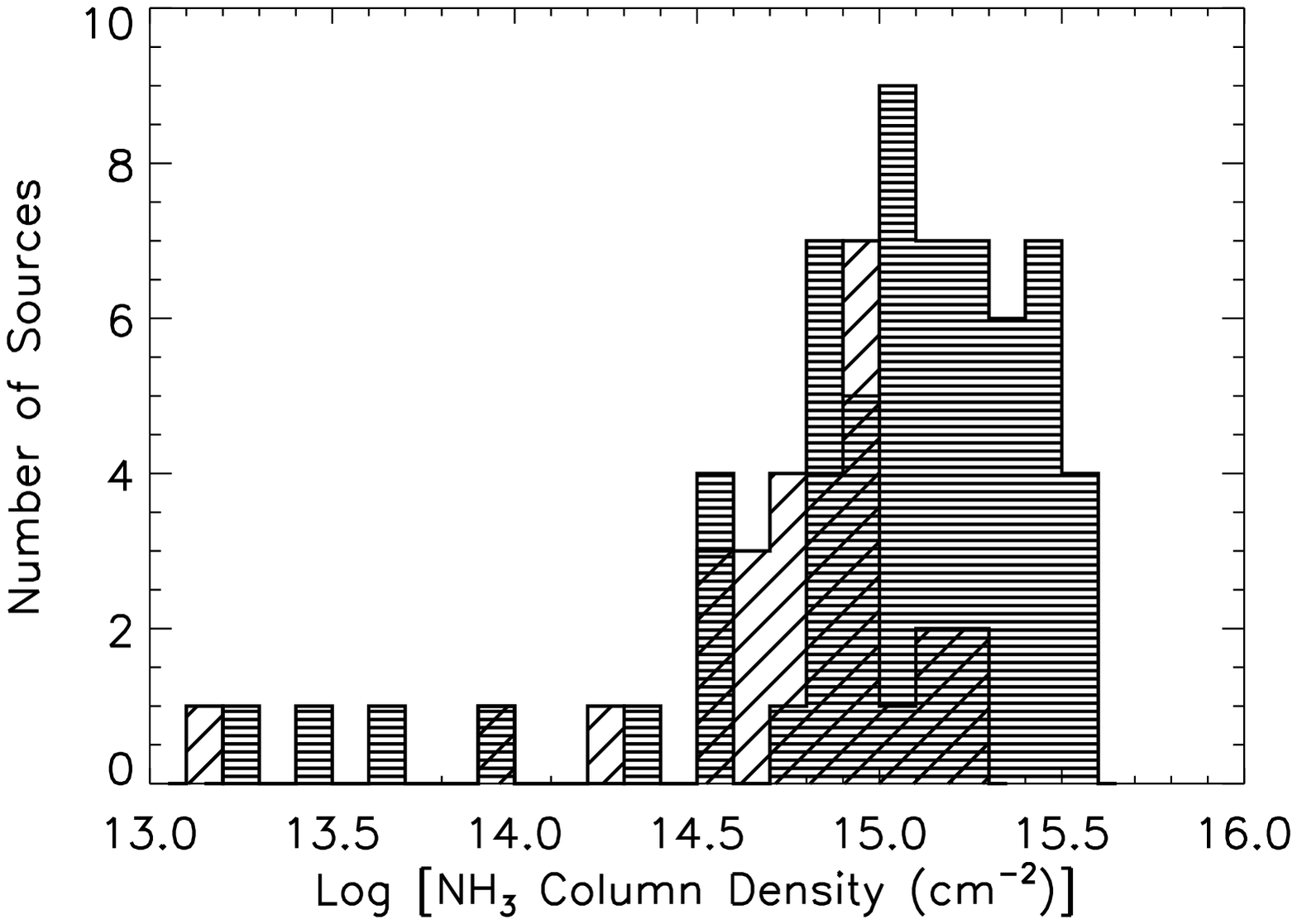}
\includegraphics[scale=0.45]{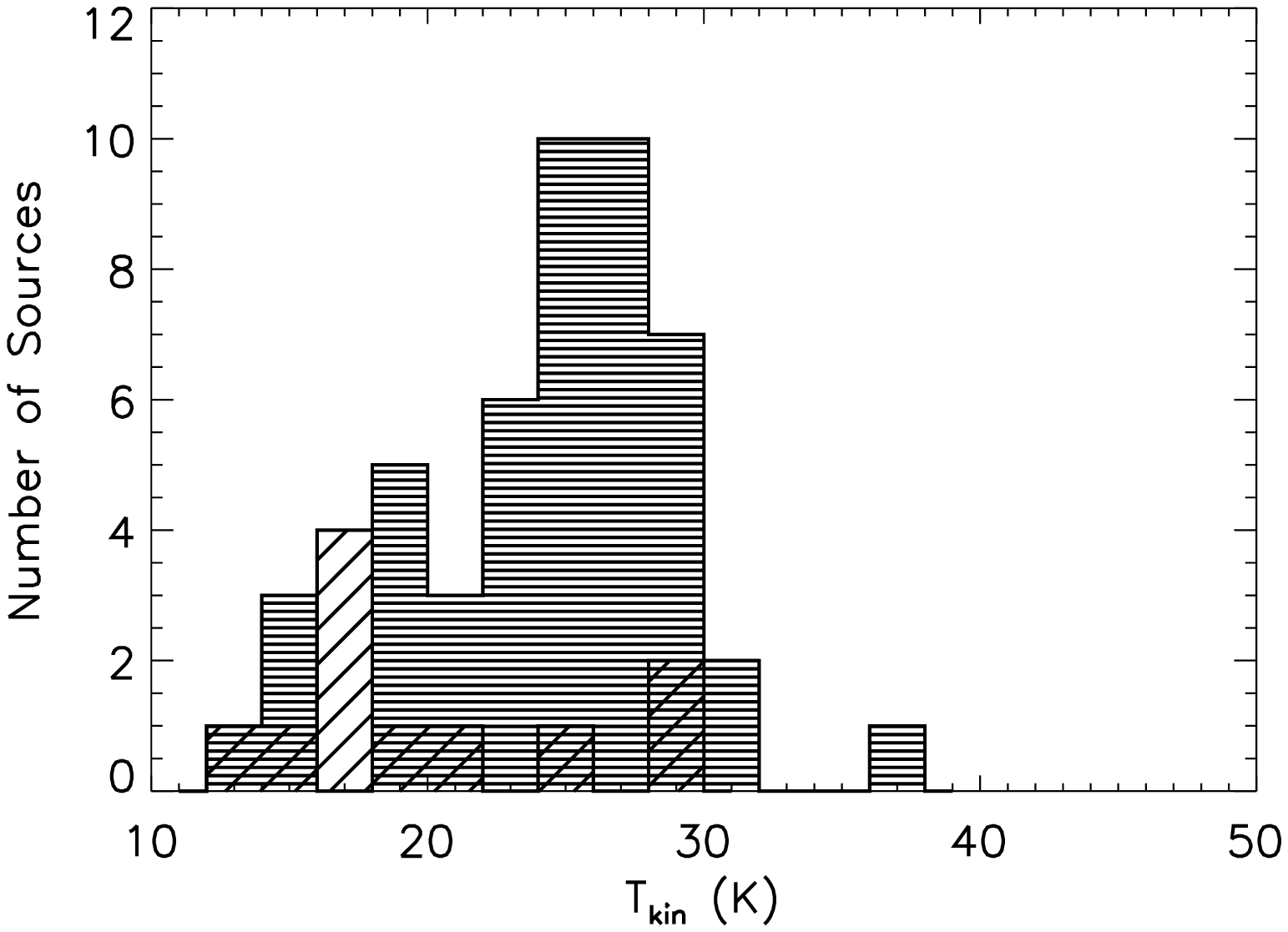}
\caption{\ammonia\/ property distributions for EGOs that are/are not detected in \water\/ maser emission in our survey, plotted
as horizontally and diagonally hatched histograms, respectively. Bin sizes are the same as in Figure~\ref{nh3_prop_fig}.
\label{nh3_prop_water_fig}}
\end{figure*}

\begin{figure*}
\center{\includegraphics[scale=0.45]{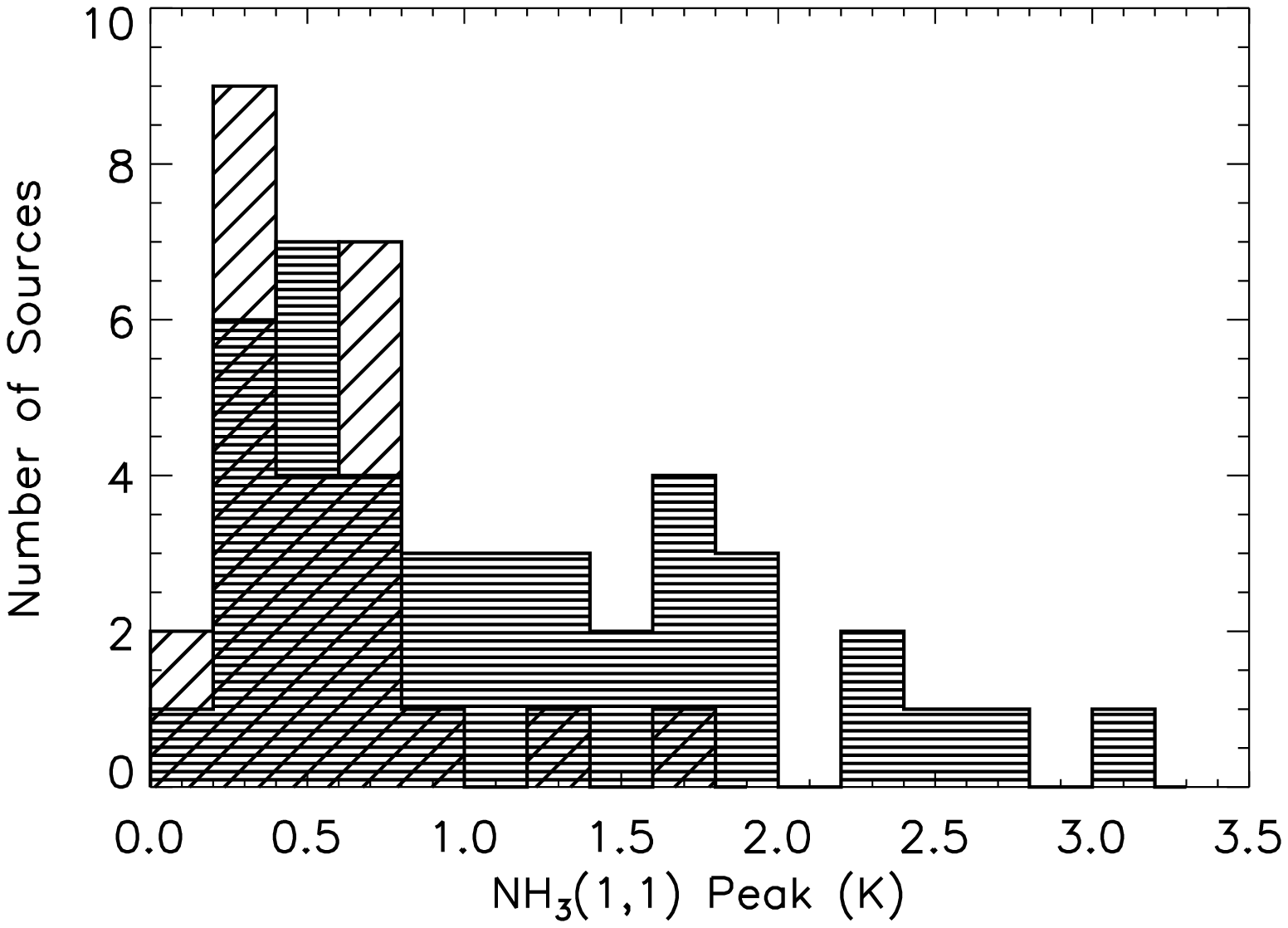}\includegraphics[scale=0.45]{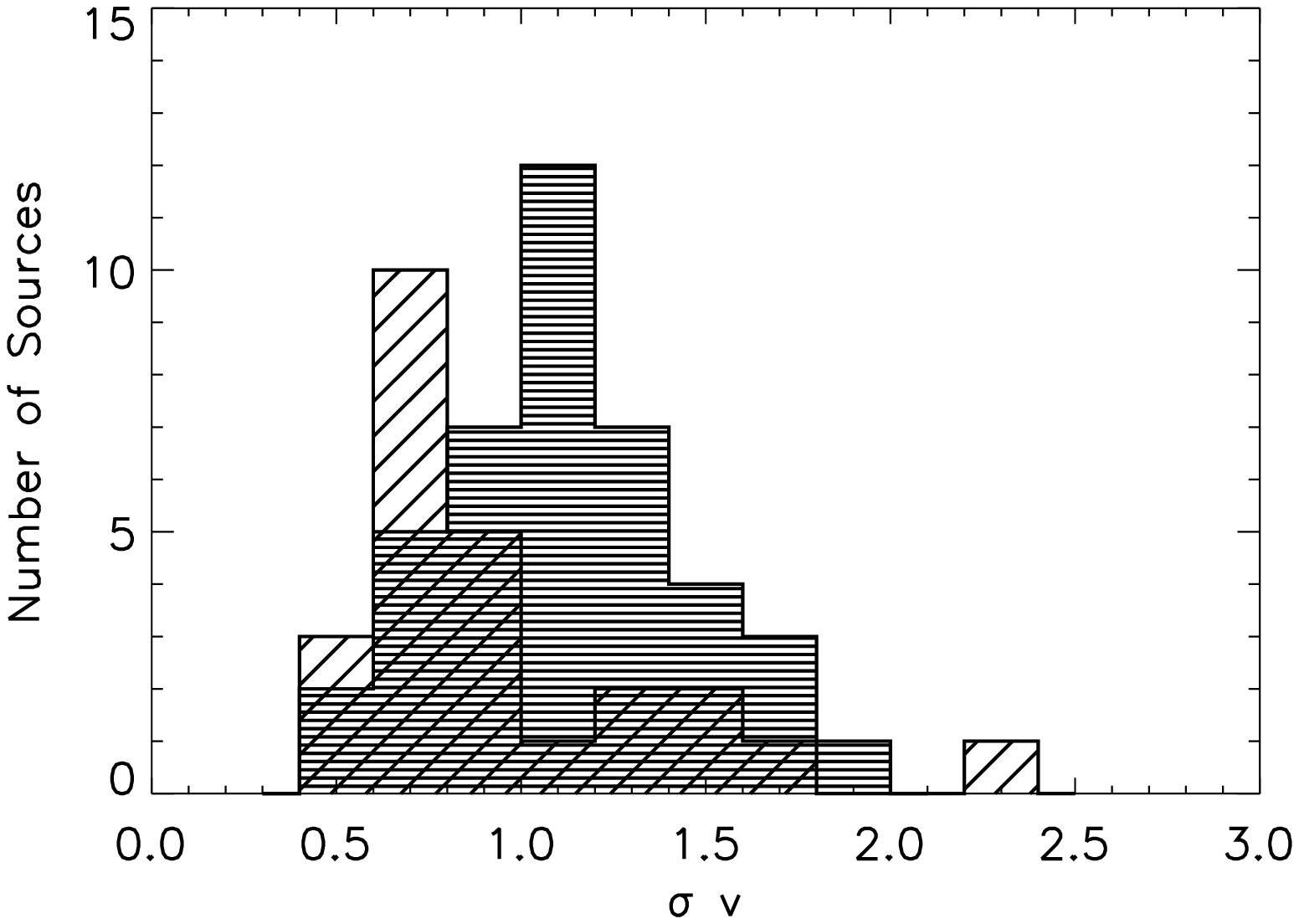}}\\
\includegraphics[scale=0.45]{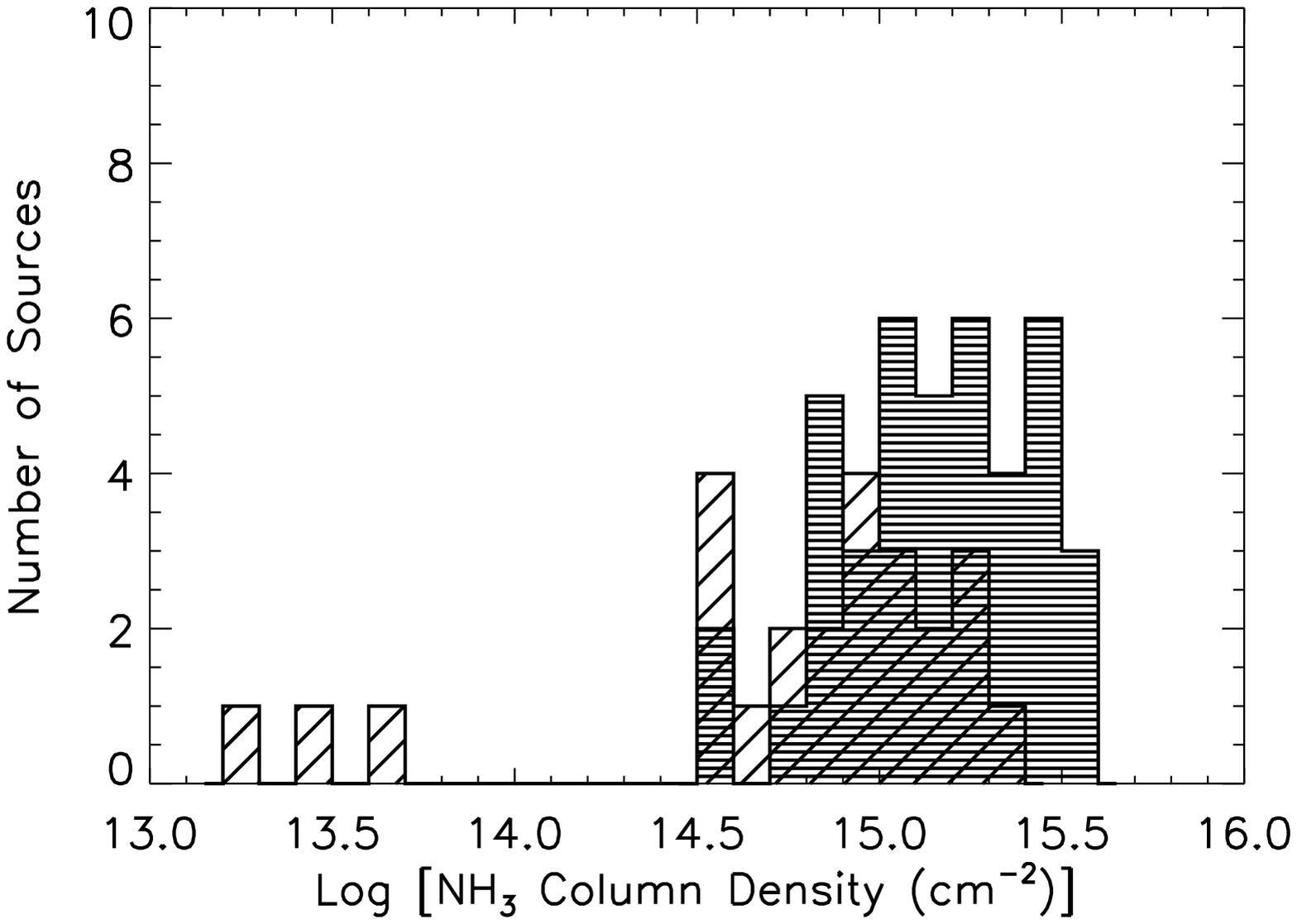}
\includegraphics[scale=0.45]{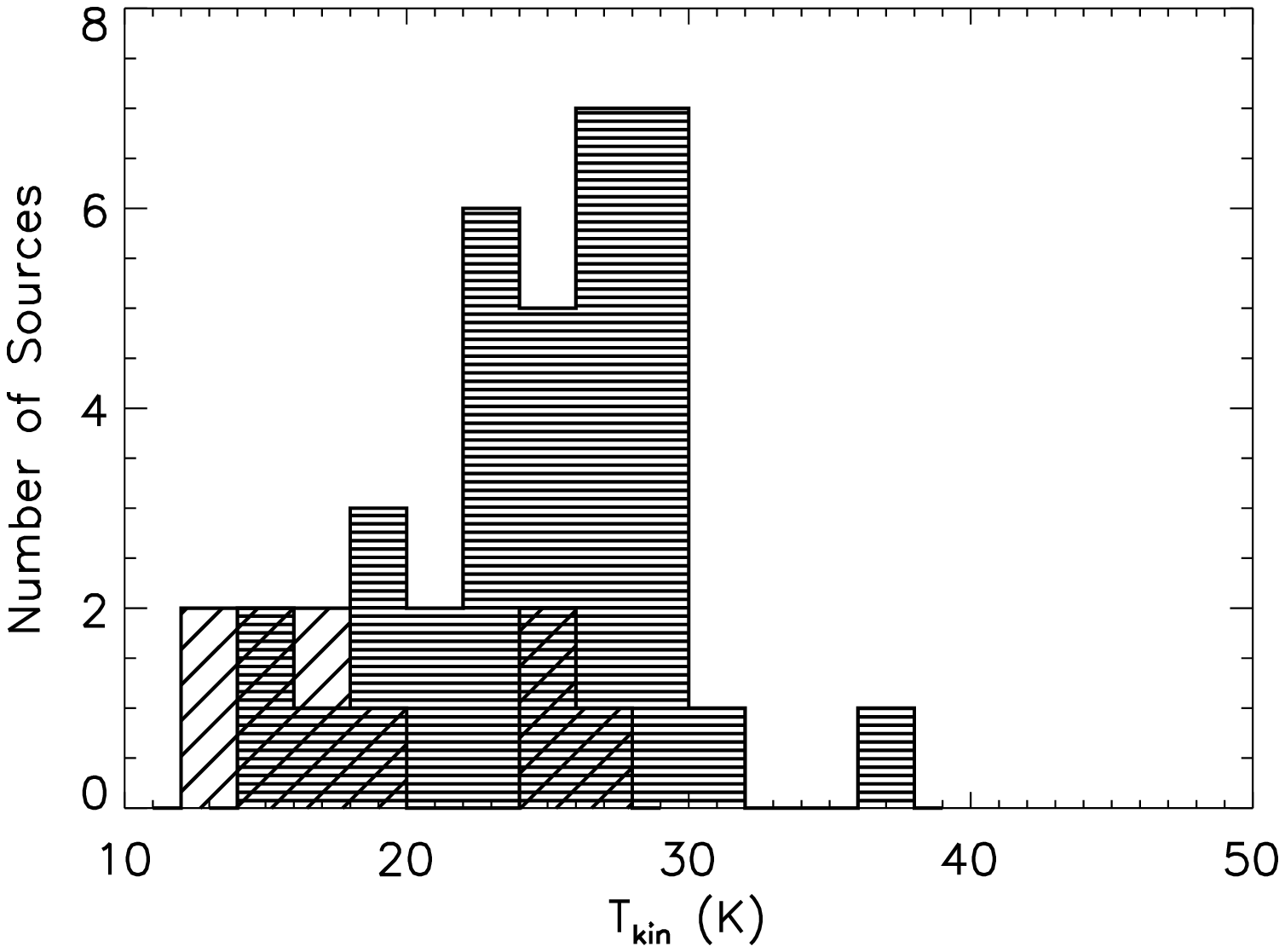}
\caption{\ammonia\/ property distributions for EGOs that are/are not
associated with Class I \meth\/ maser emission \citepalias[in][see
also \S\ref{water_detect}]{Chen11}, plotted as horizontally and
diagonally hatched histograms, respectively. Bin sizes are the same as in Figure~\ref{nh3_prop_fig}.
\label{nh3_prop_class1_fig}}
\end{figure*}

\begin{figure*}
\center{\includegraphics[scale=0.45]{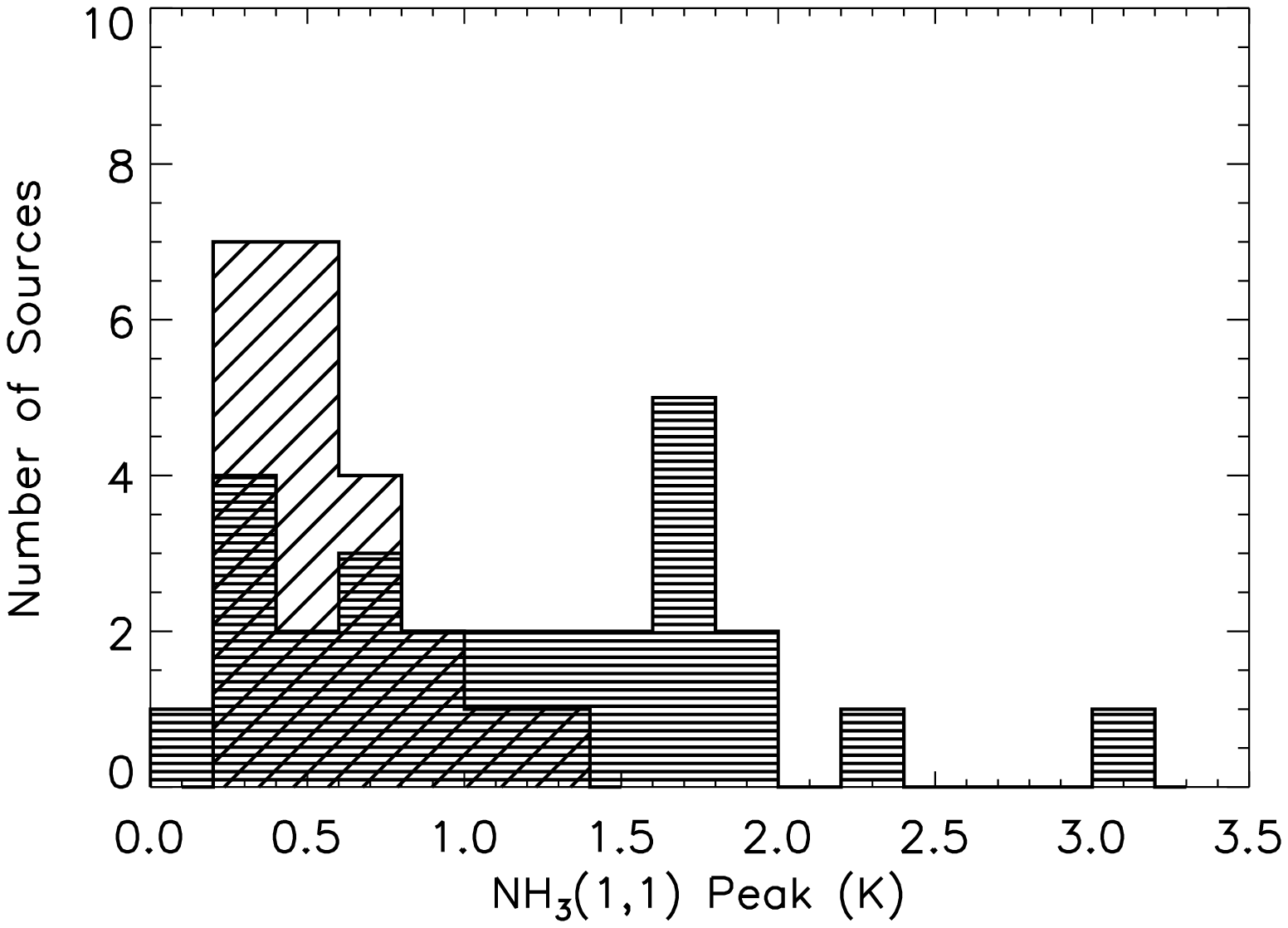}\includegraphics[scale=0.45]{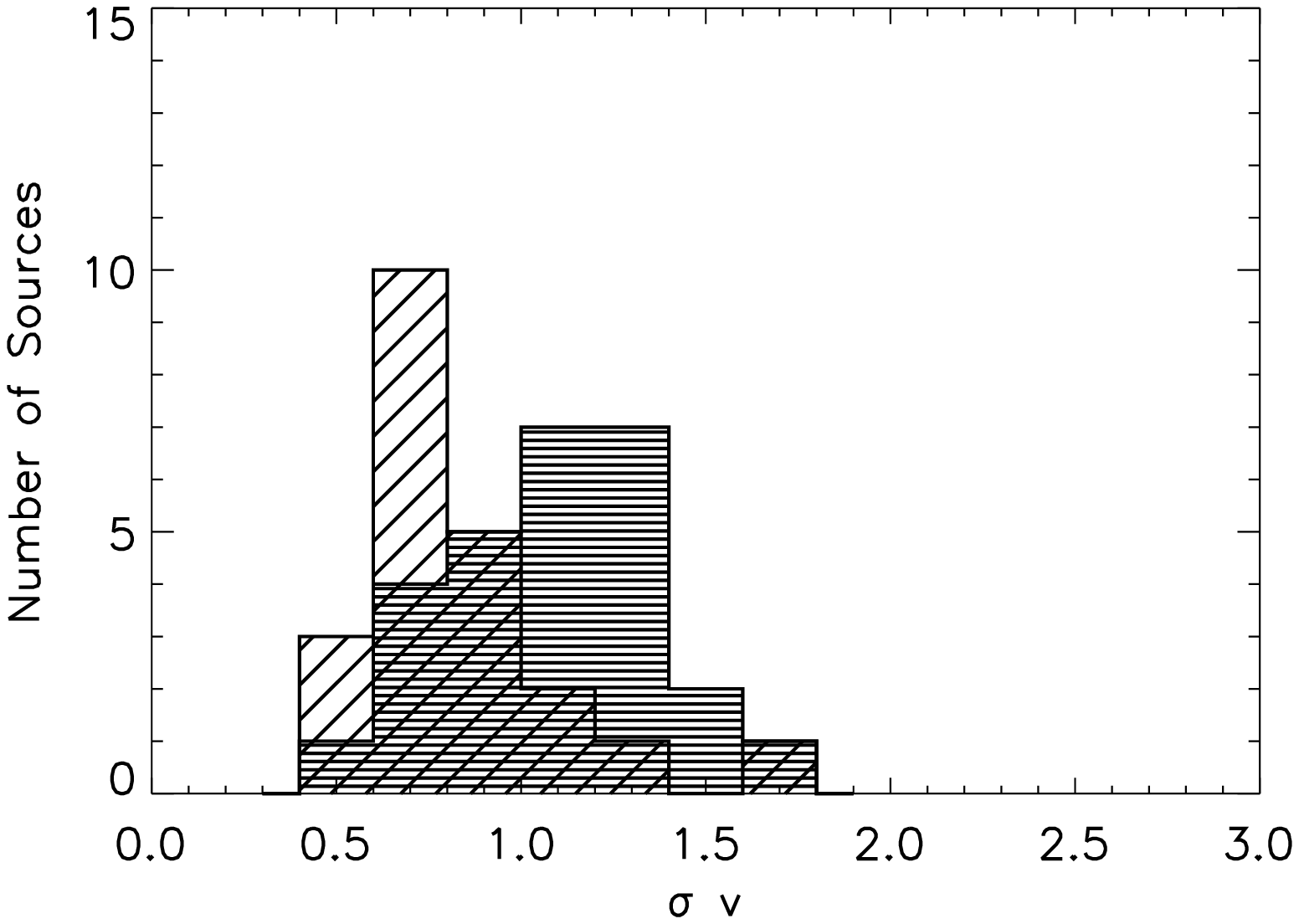}}\\
\includegraphics[scale=0.45]{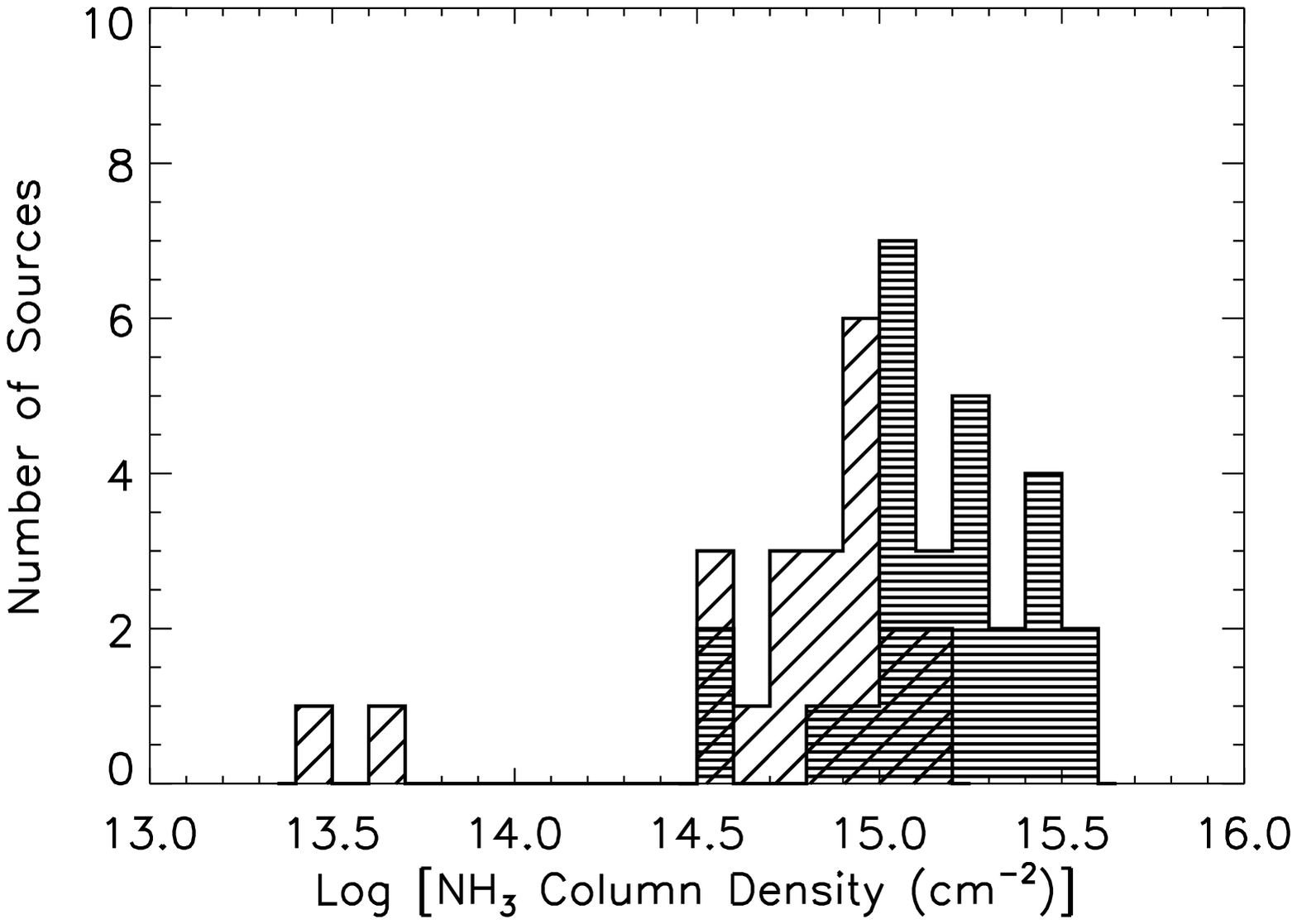}
\includegraphics[scale=0.45]{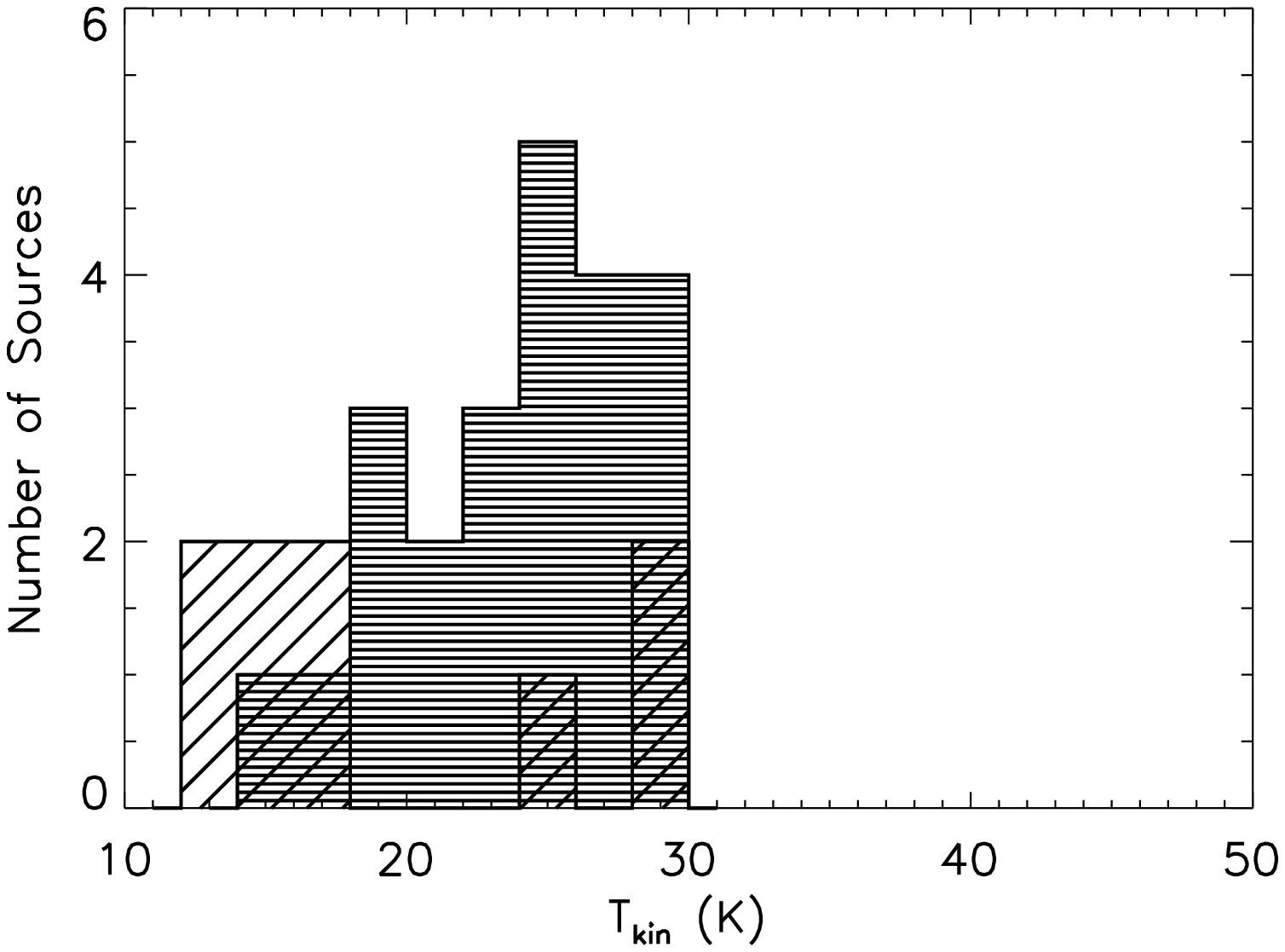}
\caption{\ammonia\/ property distributions for EGOs that are/are not
associated with Class II \meth\/ maser emission \citepalias[in][see
also \S\ref{water_detect}]{Chen11}, plotted as horizontally and
diagonally hatched histograms, respectively. Bin sizes are the same as in Figure~\ref{nh3_prop_fig}.
\label{nh3_prop_class2_fig}}
\end{figure*}

\begin{figure*}
\center{\includegraphics[scale=0.45]{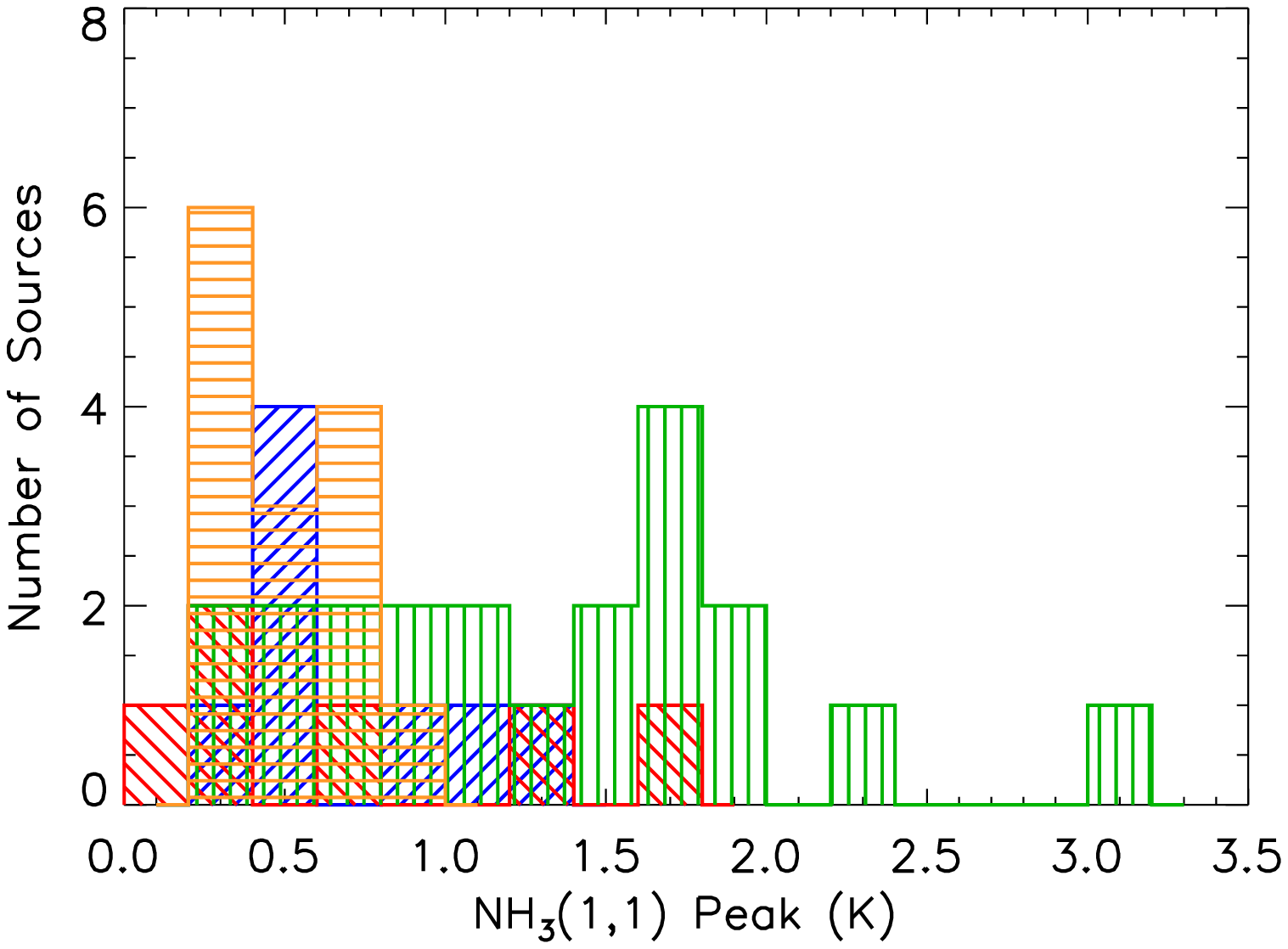}\includegraphics[scale=0.45]{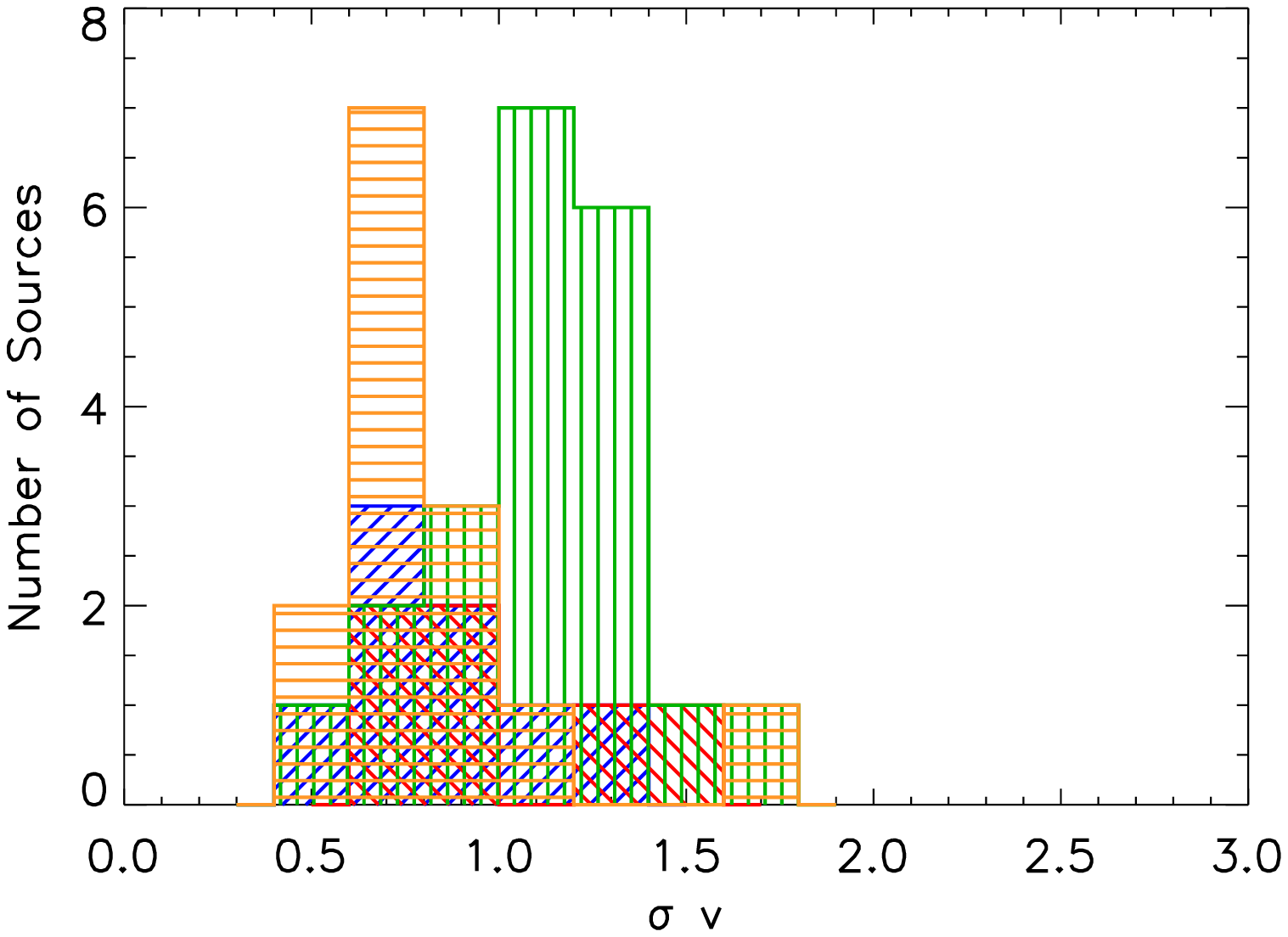}}\\
\includegraphics[scale=0.45]{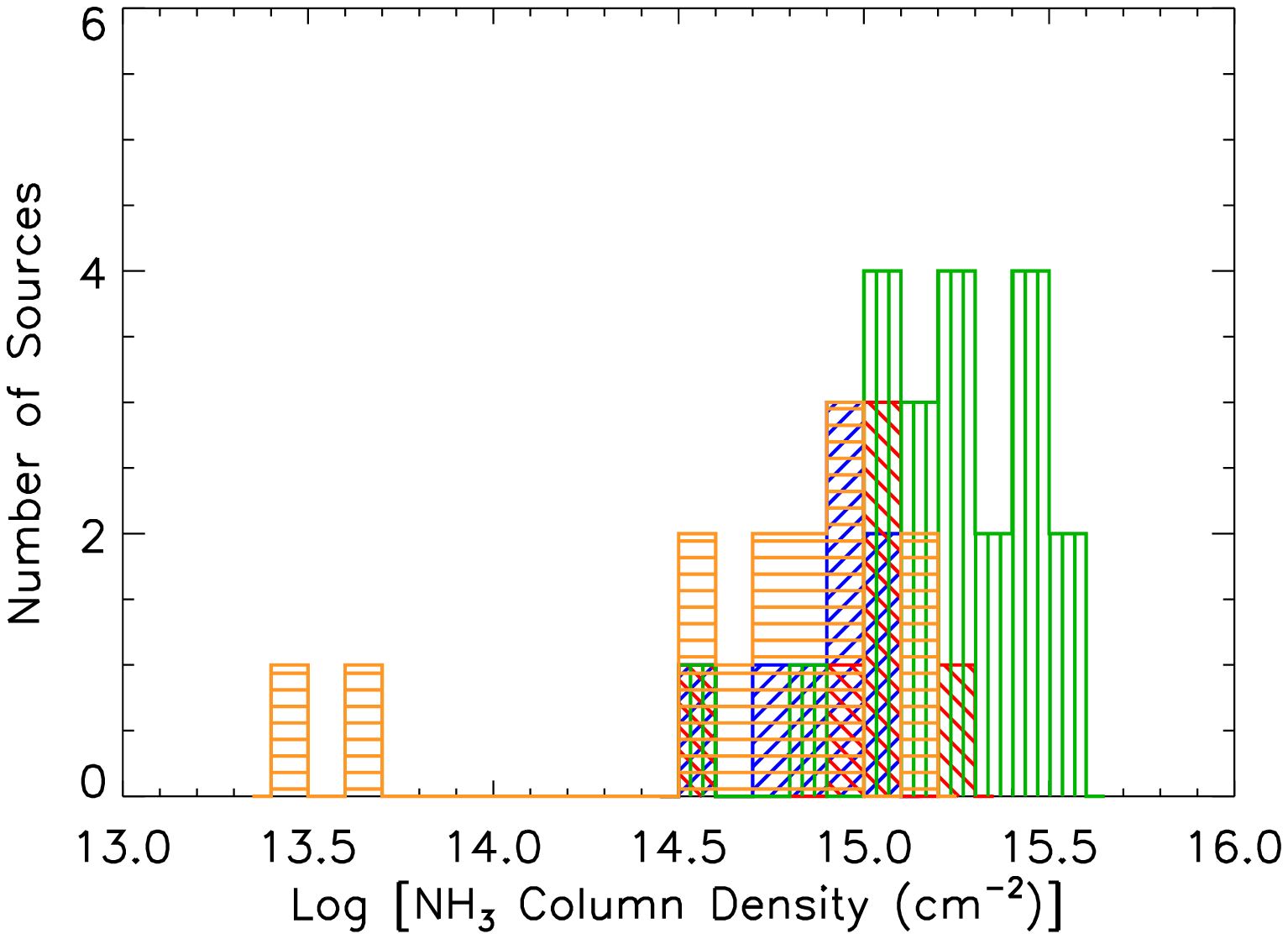}
\includegraphics[scale=0.45]{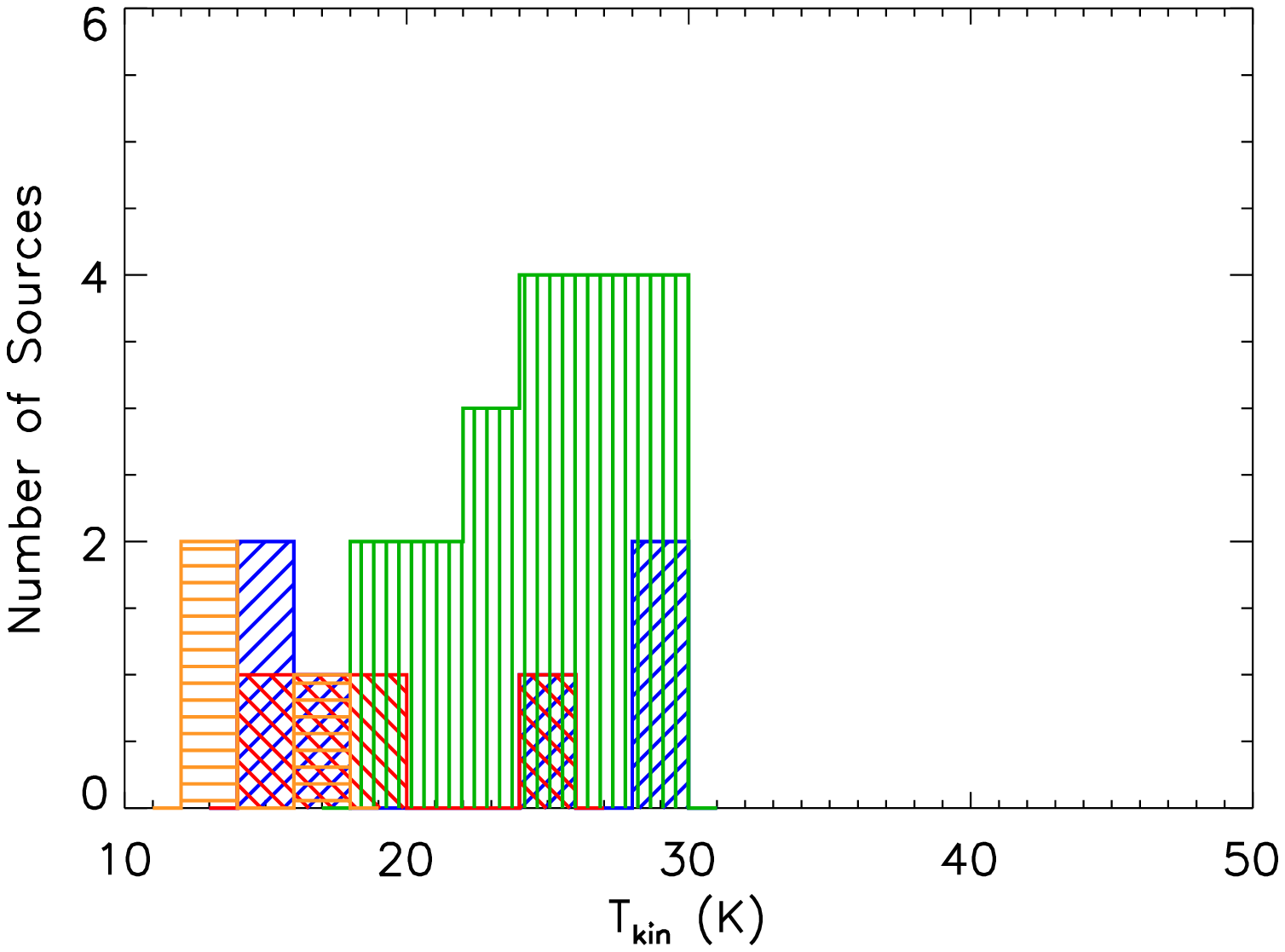}
\caption{\ammonia\/ property distributions for EGOs associated
with both Class I and II \meth\/ masers (green), only Class I \meth\/
masers (blue), only Class II \meth\/ masers (red), and neither type of
\meth\/ maser (orange) \citepalias[\meth\/ maser associations from][see
also \S\ref{water_detect}]{Chen11}.  Bin sizes are the same as in Figure~\ref{nh3_prop_fig}.
\label{nh3_prop_meth_colors_fig}}
\end{figure*}

\begin{figure}
\plotone{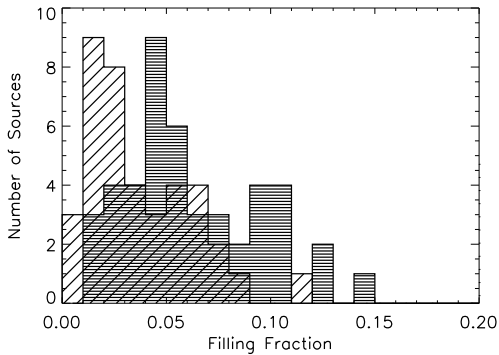}
\caption{Distribution of the beam filling fraction, $\eta_{ff}$, for EGOs associated/not associated with IRDCs, plotted
as horizontally and diagonally hatched histograms, respectively.  The bin size is 0.01.  Sources for which T$_{ex}=$T$_{kin}$ and $\eta_{ff}=$1 are not shown. \label{nh3_beam_ff_irdc_fig}}
\end{figure}

\begin{figure}
\figurenum{14}
\plotone{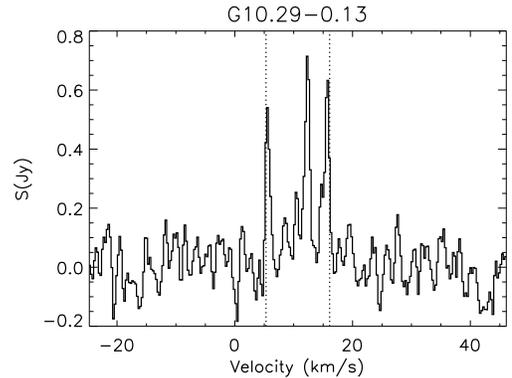}
\caption{H$_{2}$O maser spectrum.  The minimum and maximum velocities of detected maser emission ($>$ 4$\sigma$, Table~\ref{water_detect_tab}) are shown as dashed vertical lines.  The velocity range shown for each EGO extends from V$_{min,water}$-30 \kms\/ to V$_{max,water}$+30 \kms.  A complete figure set including spectra for all EGOs detected in \water\/ maser emission is available in the online journal. \label{water_spectra_fig}}
\end{figure}

\begin{figure*}
\addtocounter{figure}{1}
\center{\includegraphics[scale=0.45]{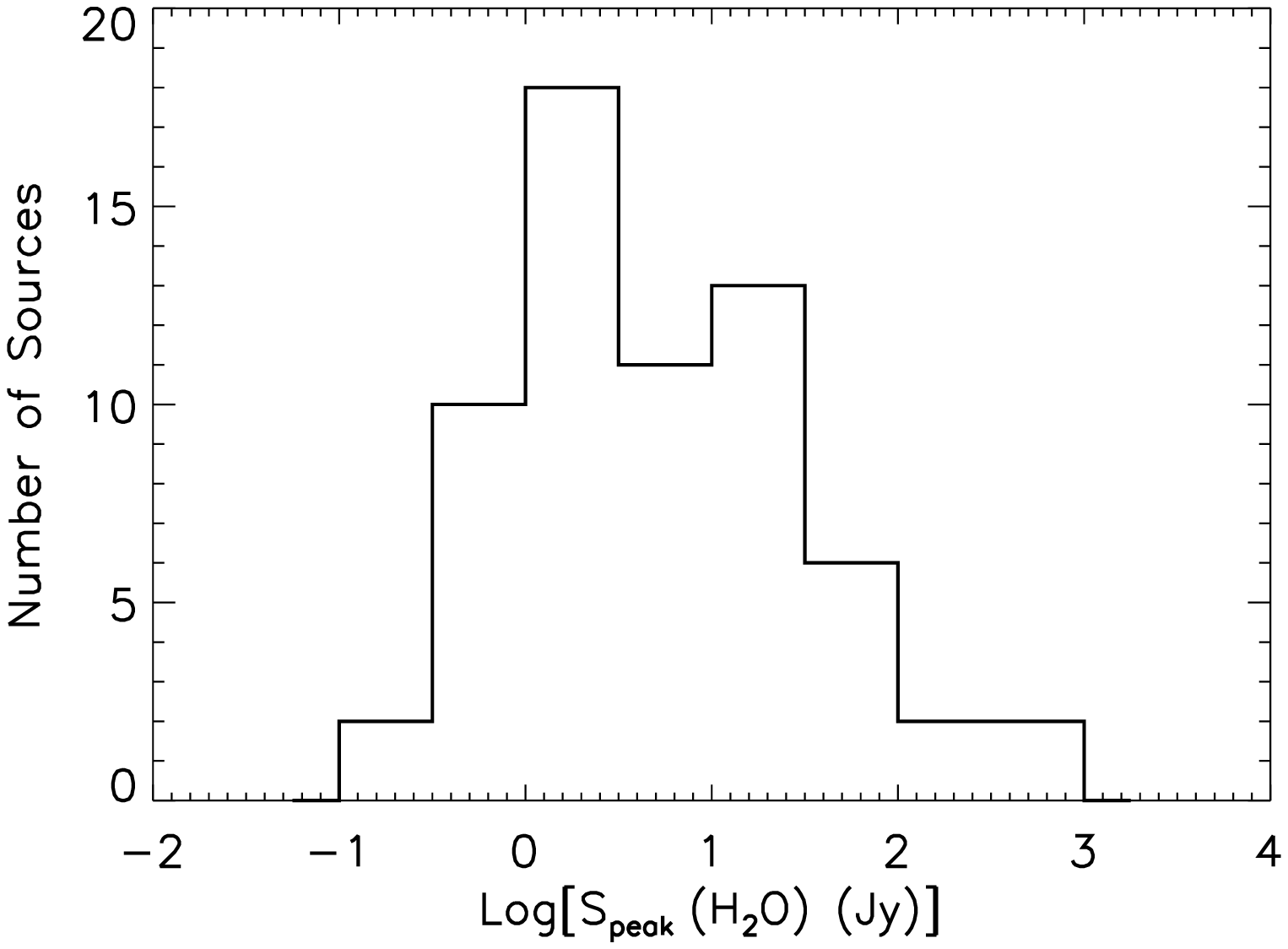}\includegraphics[scale=0.45]{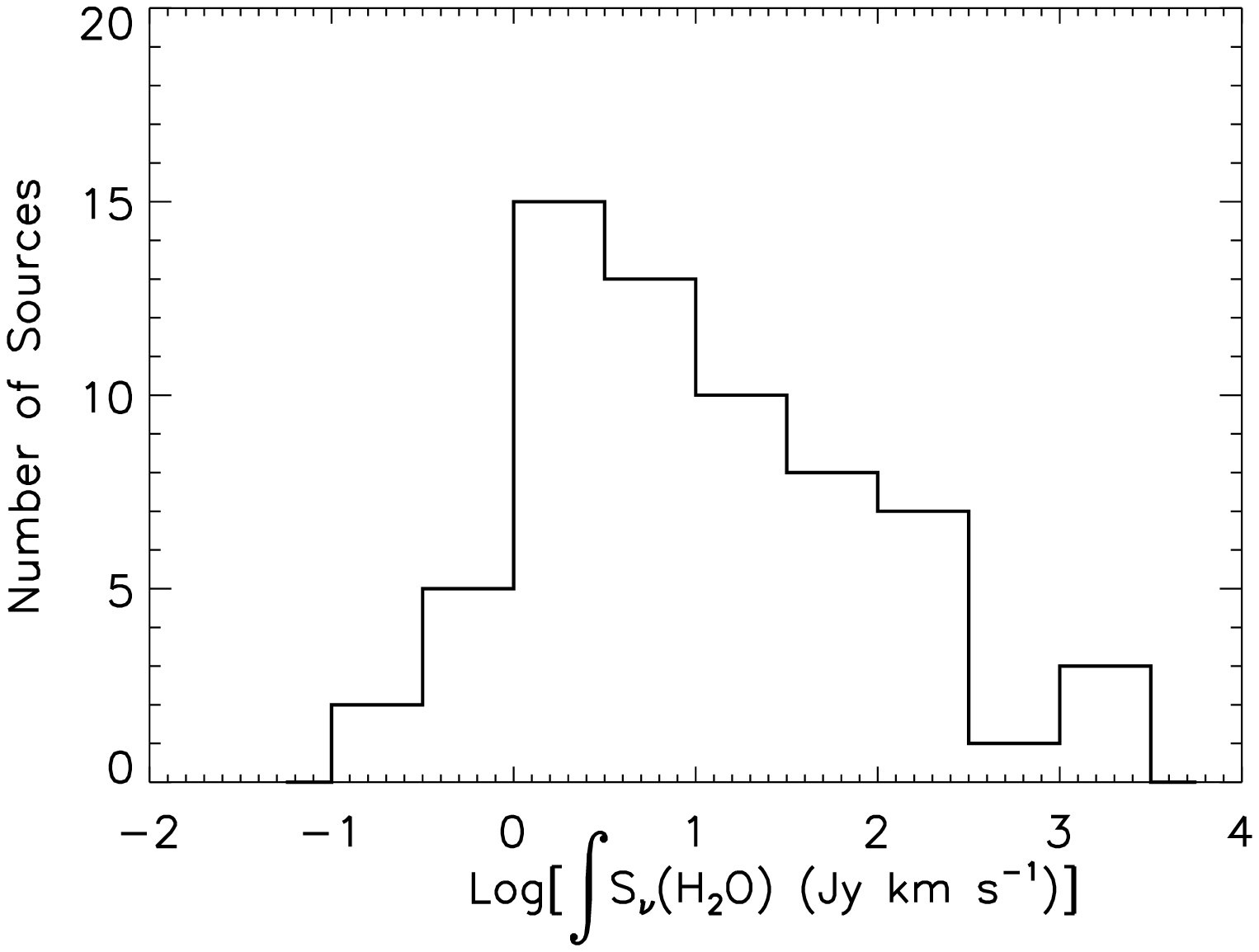}}\\
\includegraphics[scale=0.45]{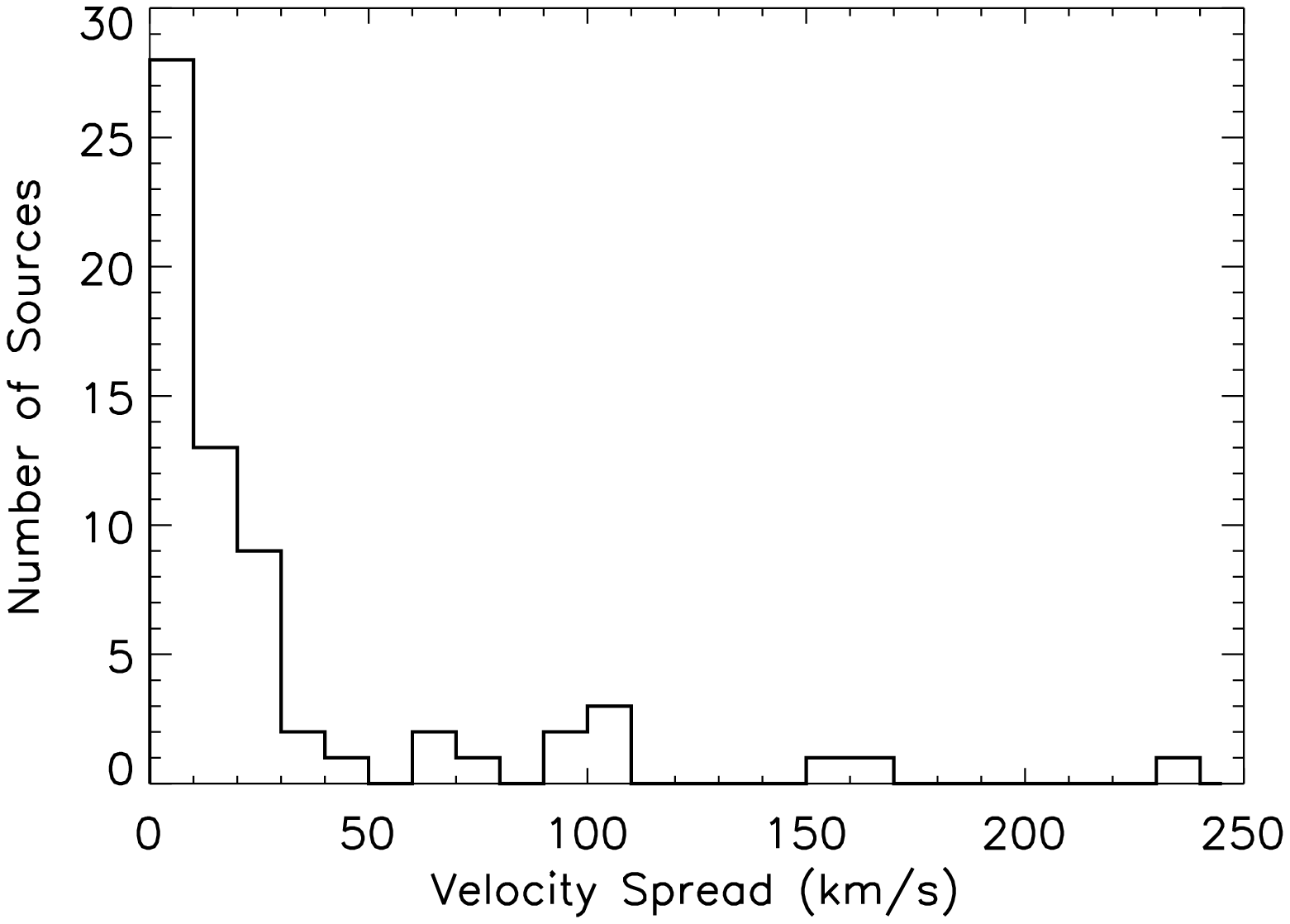}
\includegraphics[scale=0.45]{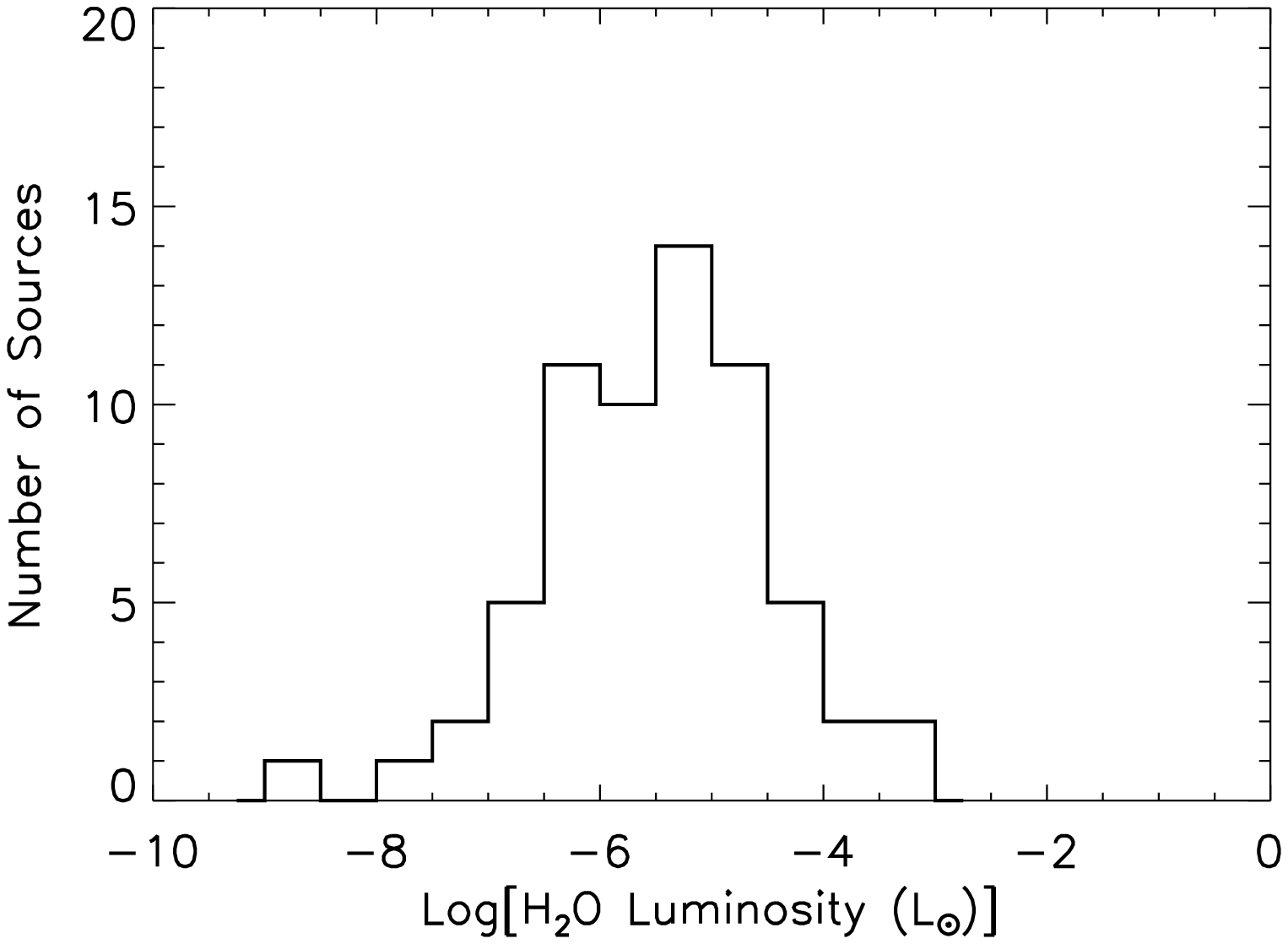}
\caption{Histograms showing the distributions of \water\/ maser properties for all \water\/ maser detections in our sample.  The panels show (clockwise from upper left): peak flux density, integrated flux density, velocity range of maser emission ($>$4$\sigma$), and isotropic \water\/ maser luminosity (\S~\ref{water_dis}).  Bin sizes are 0.5 dex (peak and integrated fluxes and luminosity) and 10 \kms\/ (velocity range). }
\label{water_prop_fig}
\end{figure*}

\subsubsection{Comparison of EGO subsamples}\label{nh3_ego_compare}

As discussed in \S\ref{ammonia_detect}, detection rates for the
higher-excitation \ammonia\/ transitions differ for
various EGO subsamples.  We consider seven pairs of EGO subsamples: (1)
``likely''/``possible'' outflow candidates; (2) sources associated/not
associated with IRDCs; (3) \water\/ maser detections/nondetections in our survey; (4) Class I \meth\/ maser
detections/nondetections (regardless of Class II association); (5)
Class II \meth\/ maser detections/nondetections (regardless of Class I
association); (6) EGOs associated with only Class I/only Class II
\meth\ masers; and (7) EGOs associated with both Class I and II
\meth\/ masers/EGOs associated with neither \meth\/ maser type.  To
assess whether the \ammonia\/ \emph{properties} of these subsamples exhibit
statistically significant differences, we ran two-sided K-S tests of
eight parameters: the \ammonia\/ (1,1), (2,2), and (3,3) peaks
(T$_{\rm MB}$), $\sigma_{\rm v}$, $\tau_{(1,1)}$, $\eta_{ff}$, N(\ammonia), and
T$_{kin}$.  To maximize our sample size, we used the parameters from
the single-component fits.  
To check for biases due to sensitivity limits, we also ran two-sided
K-S tests of distance and the \ammonia(1,1) rms for the same seven EGO
subsamples.  Table~\ref{stat_sig_diff_table} lists the
subsample/parameter combinations that have significantly different
distributions, adopting a moderately conservative threshold of $<$0.01
for the significance of the K-S statistic.  Note that K-S tests
involving the \meth\/ maser subsamples are limited by small sample
sizes, particularly for parameters that require (2,2) or (3,3)
detections.  While we ran K-S tests in all cases where the
subsamples being compared each have $\ge$4 members, we interpret the
small-n results with caution.  Statistically significant differences
are seen most often in the \ammonia(1,1) peak temperature, $\sigma_{\rm v}$,
the \ammonia\/ column density, and the kinetic temperature.  The
distributions of these properties for the various subsamples are shown
in
Figures~\ref{nh3_prop_likely_poss_fig}-\ref{nh3_prop_meth_colors_fig}.

The most dramatic difference is between the $\sigma_{\rm v}$ distributions
for EGOs that are/are not detected in \water\/ maser emission in our
survey (Fig.~\ref{nh3_prop_water_fig}).  The \ammonia\/ lines are
broader towards EGOs associated with \water\/ masers, with median
$\sigma_{\rm v}$ of 1.18 \kms\/ and 0.80 \kms\/ for \water\/ maser detections
and nondetections, respectively (\begin{math} \sigma \rm v=\frac{\rm FWHM}{\sqrt{ \rm 8 ln 2}} \end{math}).  
This is in agreement with previous single-dish
studies of \water\/ masers in star-forming regions.  In their
\ammonia(1,1) survey of 164 \water\/ masers ($\theta_{\rm FWHP} \sim$1.4\arcmin), \citet{Anglada96} found a
correlation between $L_{H_{2}O}$ and the \ammonia\/ line width;
comparing their data with other \ammonia\/ surveys, they found
increased \ammonia\/ linewidths towards star-forming regions with
\water\/ masers.  Both our results and those of \citet{Anglada96} are
consistent with the \water\/ masers being excited in outflows, which
also contribute to gas motions in the surrounding clump, increasing
the \ammonia\/ line width.  Indeed, in high-resolution Karl G. Jansky Very Large Array (VLA)
observations of one of the EGOs in our sample, \citet{rsro} detect a
hot (220 K), blueshifted outflow component in \ammonia\/ emission, coincident with redshifted \water\/ masers.
In our survey, EGOs with \water\/ masers are also generally found in clumps with
higher \ammonia\/ column densities and higher kinetic temperatures
than \water\/ maser nondetections.

The populations of EGOs associated and not associated with IRDCs show
statistically significant differences in three \ammonia\/ properties:
\ammonia(1,1) peak, $\sigma_{\rm v}$, and the beam filling fraction,
$\eta_{ff}$.  EGOs associated with IRDCs have stronger \ammonia(1,1)
emission (higher \ammonia(1,1) peak temperatures) and narrower
\ammonia\/ linewidths (Fig.~\ref{nh3_prop_irdc_fig}).  We note that
the distance distributions for EGOs associated/not associated with
IRDCs are statistically indistinguishable based on our K-S tests
(K-S significance 0.21, median distance 4.0 and 4.3 kpc, respectively; see also
\S\ref{distances}).
\citet{Pillai06} found that
IRDCs had, on average, narrower \ammonia\/ linewidths than
\emph{IRAS}-selected high-mass protostellar objects or UC HII regions.
It is perhaps surprising, however, that we see a difference in the
linewidth distributions for IRDC/non-IRDC EGOs, since we are
specifically targeting active star-forming regions within IRDCs.
The effect may be attributable to emission from more quiescent
regions of IRDCs being included within the Nobeyama beam
(73\pp \q 1.4 pc at a typical distance of 4 kpc).  
As shown in
Figure~\ref{nh3_beam_ff_irdc_fig}, EGOs associated with IRDCs also generally 
have larger (though still small, $<$0.2) beam filling fractions.  This
is consistent with numerous studies that show \ammonia\/ emission
overall follows 8 \um\/ extinction in IRDCs, while exhibiting clumpy
substructure \citep[e.g.][]{Pillai06,Devine11,Ragan11}.

Interestingly, there is little evidence for statistically significant
differences between the \ammonia\/ properties of ``likely'' and
``possible'' outflow candidates.  The only properties for which the
K-S significance meets our criterion are the \ammonia(1,1) and (2,2)
peak temperatures.  However, their significance values are close to
our cutoff (Table~\ref{stat_sig_diff_table}), and no comparable
difference is seen in the distributions of the physical properties
(N(\ammonia), T$_{kin}$, etc.).  
This suggests that the difference in
T$_{\rm MB}$(1,1) and T$_{\rm MB}$(2,2) might not reflect intrinsic source
properties. 
We find no statistically
significant difference in the distance distributions of ``likely'' and ``possible''
EGOs.  Existing data is insufficient to evaluate other possible effects, such as the peak 4.5 \um\/ positions
cataloged by \citetalias{egocat} (and so our pointing positions) being
systematically further from the driving sources in ``possible'' EGOs.

EGO subsamples based on \meth\/ maser associations show notable
differences in \water\/ maser and \ammonia(2,2) and (3,3) detection
rates (\S\ref{detection_rates}).  The K-S test analysis indicates that
these \meth\/ maser subsamples also have statistically significant
differences in their \ammonia\/ properties
(Table~\ref{stat_sig_diff_table}).  EGOs associated with Class I
\meth\/ masers \citepalias[in the study of][
\S\ref{water_detect}]{Chen11} have brighter \ammonia(1,1) emission
(e.g. greater \ammonia(1,1) peak temperatures), broader \ammonia\/
linewidths, and higher \ammonia\/ column densities and kinetic
temperatures than Class I \meth\/ maser nondetections
(Fig.~\ref{nh3_prop_class1_fig}).  Class II \meth\/ maser
detections/nondetections show the same trends in the same properties
(Fig.~\ref{nh3_prop_class2_fig}).  EGOs associated with both Class I
and II \meth\/ masers likewise show stronger \ammonia(1,1) emission
and increased \ammonia\/ linewidths and column densities compared to
EGOs associated with neither type of \meth\/ maser.  Too few EGOs with
neither \meth\/ maser association are detected in \ammonia(2,2) to run a K-S test on T$_{kin}$, but
Figure~\ref{nh3_prop_meth_colors_fig} shows that the kinetic
temperature is indeed also higher towards EGOs with Class I and II \meth\/ masers.  Of the
\meth\/ maser subsamples, the most significant difference (lowest K-S
significance) is between the N(\ammonia) distributions for EGOs
with/without Class II \meth\/ masers.

The majority of our sample of Class II \meth\/ maser detections
(21/28), and about half of our sample of Class I \meth\/ maser
detections (21/41), are comprised of EGOs associated with both Class I
and II \meth\/ masers.  Similarly, the majority of the Class II
nondetections (15/23) and \q 1/2 the Class I nondetections (15/28) are
EGOs with neither type of \meth\/ maser.  Thus, it is not
surprising that the Class I detection/nondetection, Class II
detection/nondetection, and both(Class I and II)/neither EGO
subsamples show similar patterns in their \ammonia\/ properties.  The
sample sizes of EGOs known to be associated with \emph{only} Class I
or \emph{only} Class II \meth\/ masers are small (Table~\ref{detect_rates_table}).  Nonetheless, there
are no indications of systematic differences in the \ammonia\/
properties of Class I-only and Class II-only EGOs, either in the K-S
test results or in the plots shown in
Figure~\ref{nh3_prop_meth_colors_fig}.

\subsection{Water Maser Properties}\label{water_dis}

For each EGO with detected \water\/ maser emission in our survey,
Table~\ref{water_detect_tab} lists the rms, peak flux density,
velocity of peak maser emission, minimum and maximum velocities of
maser emission ($>$4$\sigma$, see also \S~\ref{water_detect}),
integrated flux density, and isotropic maser luminosity.  Spectra are
presented in Figure~\ref{water_spectra_fig} (available online in its
entirety), with the minimum and maximum velocities of detected maser
emission plotted as dotted lines.
In the absence of precise positions, the extreme variability of
\water\/ masers makes it very difficult to establish with confidence
whether or not a newly observed maser is identifiable with one
previously reported \citep[as discussed in][and references therein]{BreenEllingsen11}.  
The present study is, to our knowledge, the first systematic search for \water\/ maser emission towards EGOs.
We note in Table~\ref{water_detect_tab} \water\/ masers detected in
high-resolution studies targeting other samples that fall within the
polygonal EGO apertures from \citetalias{egocat}, but do not attempt
to correlate our Nobeyama spectra with previous single-dish
detections.   As in similar studies \citep[e.g.][]{Anglada96,Urquhart11}, we estimate the
isotropic \water\/ maser luminosity, $\rm L(H_{2}O)$, as
\begin{equation}\label{water_lum_eqn}
\Bigg[ \frac{L(H_{2}O)}{L_{\odot}}\Bigg]= 2.30 \times 10^{-8} \Bigg[ \frac{\int S_{\nu}dV}{\rm Jy ~km~ s^{-1}}\Bigg] \Bigg[\frac{D}{\rm kpc}\Bigg]^2
\end{equation}   
where D is the distance to the source (\S\ref{distances}, Table~\ref{nh3_prop_tab})
and \begin{math} \int S_{\nu}dV \sim \sum_i(S_{i}\Delta v_{i})
\end{math} is calculated over all channels that meet our 4$\sigma$
detection criterion.  For \water\/ maser nondetections, Table~\ref{water_nondetect_tab} lists the rms and upper limit for the isotropic \water\/ maser luminosity (calculated from equation~\ref{water_lum_eqn} for 4$\sigma$ and two channels).
The distributions of \water\/ maser peak and
integrated flux, luminosity, and velocity range for \water\/ maser detections in our sample are
shown in Figure~\ref{water_prop_fig}.

\subsubsection{High Velocity Features}\label{high_vel_dis}

\water\/ masers are known for their wide velocity ranges and
high-velocity features, as compared to other masers found in MSFRs
(e.g. \meth\/ and OH).  The velocity of the strongest \water\/ maser
emission in a given source is nonetheless generally well-correlated
with the \vlsr\/ of the dense gas
\citep[e.g.][]{Churchwell90,Anglada96,Urquhart11}.  Notably, for the
Red \emph{MSX} Source (RMS)\footnote{For additional details on the RMS
sample, see \citet{Hoare05,Urquhart08}.} sample of MIR-bright MYSOs
and UC HIIs, the distribution of $V_{H_{2}O,peak}-V_{NH_{3}}$ is
skewed towards negative velocities.  The offset (from zero) is
statistically significant, and indicates that blueshifted masers are
stronger and more prevalent than redshifted masers \citep{Urquhart11}.
In our sample of 62 sources detected in both \water\/ maser and
\ammonia(1,1) emission, the mean offset $V_{H_{2}O,peak}-V_{NH_{3}}$
is $-$2.43 \kms\/ and the median offset $-$0.54 \kms.  However, in our
(smaller) sample, the offset from zero is not statistically
significant (standard errors 1.37 and 1.72, respectively).  The
distribution of $V_{H_{2}O,peak}-V_{NH_{3}}$ for our EGO sample is
shown in Figure~\ref{voffset_nh3_water_fig}.

The relative frequency of blue- and red-shifted emission can
also be accessed by examining high velocity maser features \citep[generally defined as $V-V_{LSR} \ge 30$ \kms, e.g.][]{CB10,Urquhart11}.
\citet{CB10} recently analyzed high-velocity emission in numerous \water\/ maser subsamples and
proposed that an excess of sources showing only blueshifted high-velocity emission 
is an indicator of youth.  For \water\/ masers associated with 
Class II \meth\/ but not OH masers \citep[from the sample of ][]{Breen10water},
they find a ``blue'' (blueshifted high velocity emission only)
fraction of 16\%, a ``red'' fraction of 8\%, and a ``red+blue'' (both
blue- and redshifted high velocity \water\/ maser features) fraction
of 7\%.  Interestingly, \citet{Urquhart11} find a similar ratio of
``blue'':''red'' sources in their much larger sample of RMS YSOs and
UC HII regions, though a smaller overall fraction (22\%) of their
detected \water\/ masers show some high velocity emission.  Twelve of
our EGO targets (\q 19\%) have high velocity \water\/ maser features
(offset by $\ge$ 30 \kms\/ from the \ammonia\/ \vlsr): 6 ``blue'', 1 ``red'', and 5
``red+blue''.  Of these 12 EGOs, 5 are associated with both Class I
and II \meth\/ masers, 4 are associated with Class I \meth\/ masers
and are classified as Class II ``no information'' in
\citetalias{Chen11}, 1 is associated with Class I but not Class II
\meth\/ masers, and 2 are not included in \citetalias{Chen11}.  Our
sample sizes and those of \citet{CB10} are too small
to warrant detailed comparisons; however, the ``blue'':''red'' excess
we observe is generally consistent with their results for \meth\/
maser sources.

\subsubsection{Comparison of EGO Subsamples}\label{water_ego_compare}

To look for differences in the properties of \water\/ masers
associated with the various EGO subsamples, we ran two-sided K-S tests
for four parameters: velocity range, peak intensity,
integrated intensity, and isotropic luminosity.  The subsample pairs
considered were the same as outlined above (\S\ref{nh3_ego_compare}),
with the exception of \water\/ maser detections/nondetections (since
we are investigating \water\/ maser properties, only detections are
considered).
We find no evidence for statistically significant
differences.  As an example, Figures~\ref{water_liso_shaded_fig}-\ref{liso_meth_color_fig} show histograms of the
\water\/ maser luminosity, shaded by subsample, for the six
subsample pairs.

\subsection{Properties of Associated Dust Clumps}\label{clump_prop}

Of the 94 northern EGOs in our survey, 82 fall with the coverage of
the 1.1 mm Bolocam Galactic Plane Survey
\citep[BGPS, resolution 33\pp;][]{Aguirre11,Rosolowsky10}, and 77 are associated with
BGPS sources.\footnote{The slight difference from the statistics in
\citet{Dunham11} is because we consider G19.01$-$0.03 as a single EGO,
while they treat this EGO and its northern and southern outflow lobes
\citepalias[for which separate photometry is given in][]{egocat} as three
objects.}  The BGPS source extraction algorithm, Bolocat, uses a
seeded watershed approach to identify the boundaries of BGPS sources,
and outputs 'label maps' in which each pixel assigned to a source has
a value of that source's BGPS catalog number \citep[see][for more
details]{Rosolowsky10,Dunham11}.  If the position of an EGO from
\citetalias{egocat} falls within the Bolocat-defined boundary of a
BGPS source, we consider the EGO and BGPS source to be associated.

We calculate clump gas masses from the 1.1 mm dust continuum emission
\begin{equation}
M_{gas}= {\frac{4.79 \times 10^{-14} R S_{\nu}(Jy) D^2(kpc)}{B(\nu,T_{dust})\kappa_{\nu}}},
\label{dust_mass_eqn}
\end{equation}
where S$_{\nu}$ is the integrated flux density from the BGPS catalog corrected by the recommended factor
of 1.5$\pm$0.15 \citep{Aguirre11,gemob1}, D is the distance to the
source (\S\ref{distances}, Table~\ref{nh3_prop_tab}), B($\nu,T_{dust}$) is the Planck function, R is the gas-to-dust
mass ratio (assumed to be 100), and $\kappa_{\nu}$ is the dust mass opacity coefficient in
units of cm$^{2}$ g$^{-1}$.  We follow recent BGPS studies
\citep[e.g.][]{gemob1,Dunham11,nh3-innergal} in adopting $\kappa_{\rm 271 GHz}$/R=0.0114 cm$^{2}$ g$^{-1}$.  Our \ammonia\/ observations provide
a measurement of the clump-scale gas kinetic temperature, T$_{kin}$,
and we assume T$_{dust}$=T$_{kin}$ in calculating the clump masses.  
To estimate the volume-averaged number densities of the clumps, we use
the clump gas mass from equation~\ref{dust_mass_eqn} and the
deconvolved angular source radius from the BGPS catalog
\citep{Rosolowsky10}, assuming spherical geometry.  For consistency
with \citet{Hill05} (see \S\ref{clump_prop_dis}), we adopt a mean mass
per particle $\mu=$2.29 m$_{H}$.  The 1.1 mm flux densities, radii,
gas masses, and volume-averaged number densities for the clumps
associated with our target EGOs are listed in
Table~\ref{bgps_prop_tab}.  For the three BGPS sources in our sample
that could not be stably deconvolved (listed as ``null'' radii in the
BGPS catalog), we adopt half the BGPS beamsize as an upper limit to
the source radius, e.g. R$<$16\farcs5.  The derived number densities
for these sources are thus lower limits, and are indicated as such in
the tables and figures.  We regard this radius upper limit as
conservative because source radii can sometimes be determined for
source diameters smaller than a beam width.  However, given the
substantial uncertainty in relating an emission distribution to a true
radius, particularly at low signal-to-noise, a more aggressive limit
could be incorrect \citep[e.g.][]{Rosolowsky10,RL06}.  If we instead
adopted an upper limit of half the BGPS beamsize for the source
diameter, this would increase the density limits by a factor of 8.

To estimate clump parameters consistently for the largest possible
number of sources in our sample, we first calculate M$_{gas}$ and
n$_{H_{2}}$ as described above using the gas kinetic temperatures derived
from the single-component \ammonia\/ fitting.  
For EGOs undetected in
\ammonia(2,2), we treat the best-fit T$_{kin}$ as an upper limit (see
also \S\ref{nh3_prop}); the derived clump mass and density are thus
lower limits.  
The clump masses estimated using well-determined kinetic temperatures
are in the range of hundreds to thousands of solar masses
(Fig.~\ref{mass_histo_fig}), with a mean (median) of \q 1850 \msun\/
(\q 1010 \msun).  The range of EGO dust clump masses is consistent with
expectations for MYSOs based on bolometer studies of other samples.
For example, \citet{Rathborne06} find a median IRDC mass of \q 940
\msun\/ (range \q 120 to 16,000 \msun), and \citet{Mueller02} report a
similar range and a mean mass of 2020 \msun\/ for a sample of
 \water\/ maser sources with high luminosities (L$_{\rm bol} >$10$^{3}$\lsun).
The star-forming sources (those with Class II \meth\/ masers and/or UC
HII regions) in the \citet{Hill05} dust clump sample similarly span a
mass range of \q10$^{2}$-10$^{4}$ \msun\/ \citep{Hill10}.  Only one
EGO appears to be a potential example of a nearby, low-mass YSO based
on the properties of its associated dust clump: G49.91+0.37, which has
a low ($<$10 \msun) lower-limit mass and a near kinematic distance of
0.53 $^{+0.52}_{-0.53}$ kpc.

The significantly improved fits obtained with two temperature components for \q 1/4 of our \ammonia\/ spectra 
indicate emission from both warmer inner regions and cooler outer
envelopes along our lines of sight.  As noted in \S\ref{nh3_modeling},
the beam filling factor $\eta_{ff}$ and the excitation temperature
T$_{ex}$ are degenerate for the two-component modeling.  To estimate
the relative contributions of the warm and cool components, we assume
T$_{kin}$=T$_{ex}$ and calculate \begin{math}
\eta_{ff}=\frac{T_{ex}-2.73}{T_{kin}-2.73} \end{math} for each
component.  We then assign weights, \begin{math} W_{\rm
warm}=\frac{1}{\frac{\eta_{\rm cool}}{\eta_{\rm warm}}+1} \end{math}
and \begin{math} W_{\rm cool}=\frac{1}{\frac{\eta_{\rm
warm}}{\eta_{\rm cool}}+1} \end{math}, and recalculate the clump mass
as \begin{math} M_{total}=M_{gas,warm}+M_{gas,cool} \end{math} where
\begin{equation}       
M_{gas,warm}= {\frac{4.79 \times 10^{-14} R S_{\nu}(Jy) W_{warm} D^2(kpc)}{B(\nu,T_{dust,warm})\kappa_{\nu}}} 
\end{equation}
and
\begin{equation}       
M_{gas,cool}= {\frac{4.79 \times 10^{-14} R S_{\nu}(Jy) W_{cool} D^2(kpc)}{B(\nu,T_{dust,cool})\kappa_{\nu}}}.
\end{equation}
The volume-averaged number density is then estimated as described
above, using $M_{total}$ in place of the single-temperature isothermal gas mass
calculated from equation~\ref{dust_mass_eqn}.  We can estimate revised
clump masses and number densities in this way for 16 of the 21 sources
with two-component \ammonia\/ fits.  For these sources, the median(mean) mass fraction in the warm component is 5.5\%(10.0\%).     
For five sources with
two-component fits, T$_{ex}$=T$_{kin}$ for one of the modeled
temperature components (the upper limit).  Coincidentally, three of
these five sources fall outside the BGPS survey area.  For the
remaining two sources, we retain the isothermal masses and densities
in our analysis.

\begin{figure}
\plotone{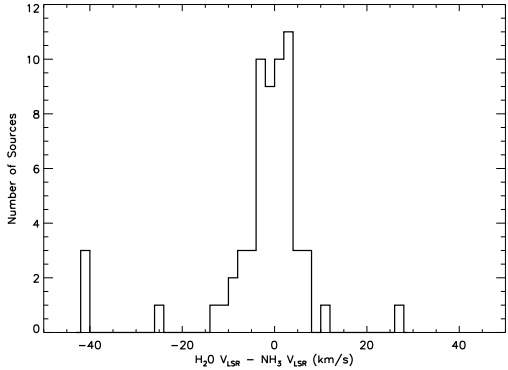}
\caption{Distribution of the difference between the \ammonia\/ \vlsr\/ and the velocity of peak \water\/ maser emission for all sources with both \water\/ maser and \ammonia(1,1) detections.  The bin size is 2 \kms.}
\label{voffset_nh3_water_fig}
\end{figure}

\begin{figure*}
\center{\includegraphics[scale=0.45]{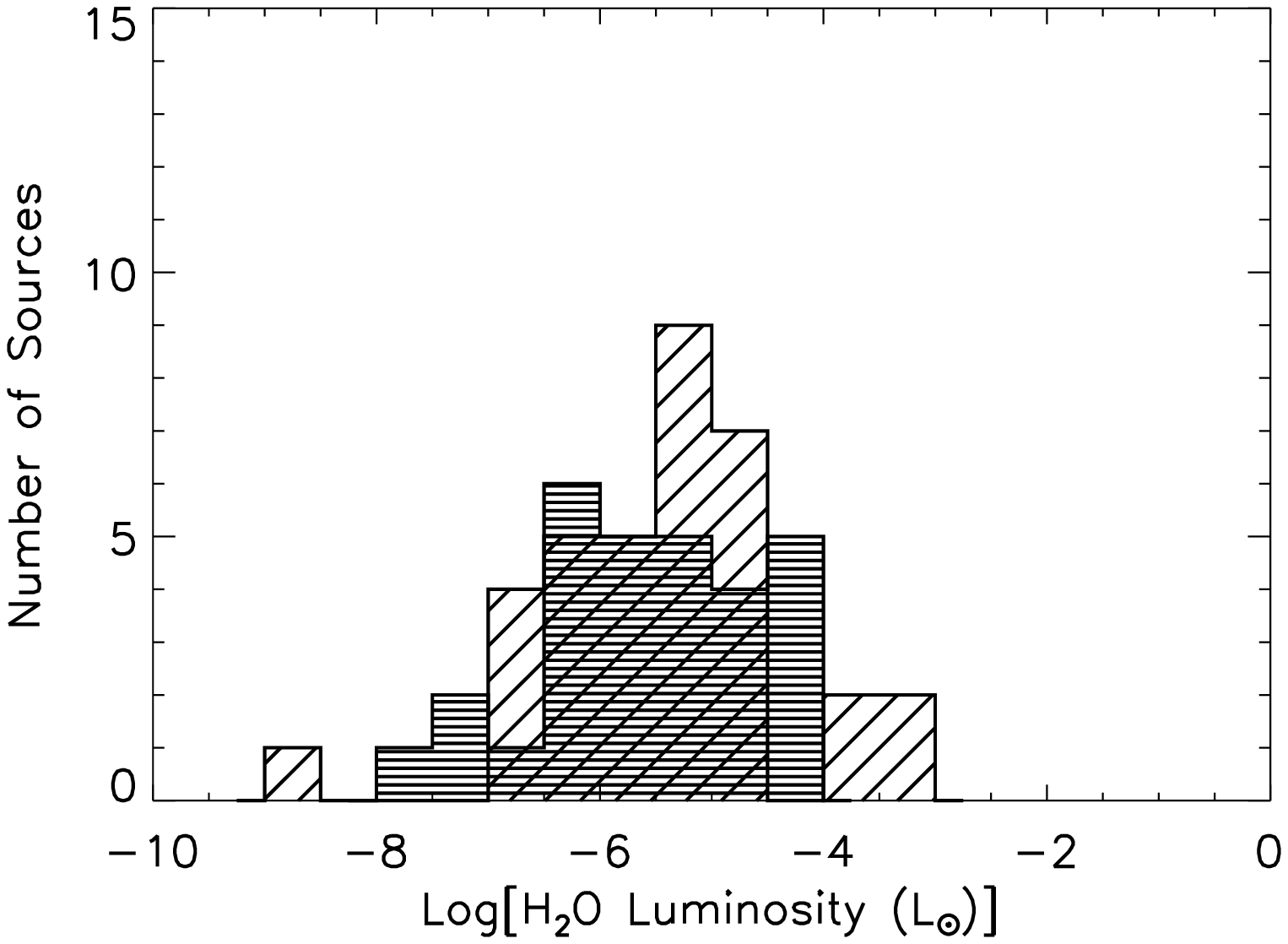}\includegraphics[scale=0.45]{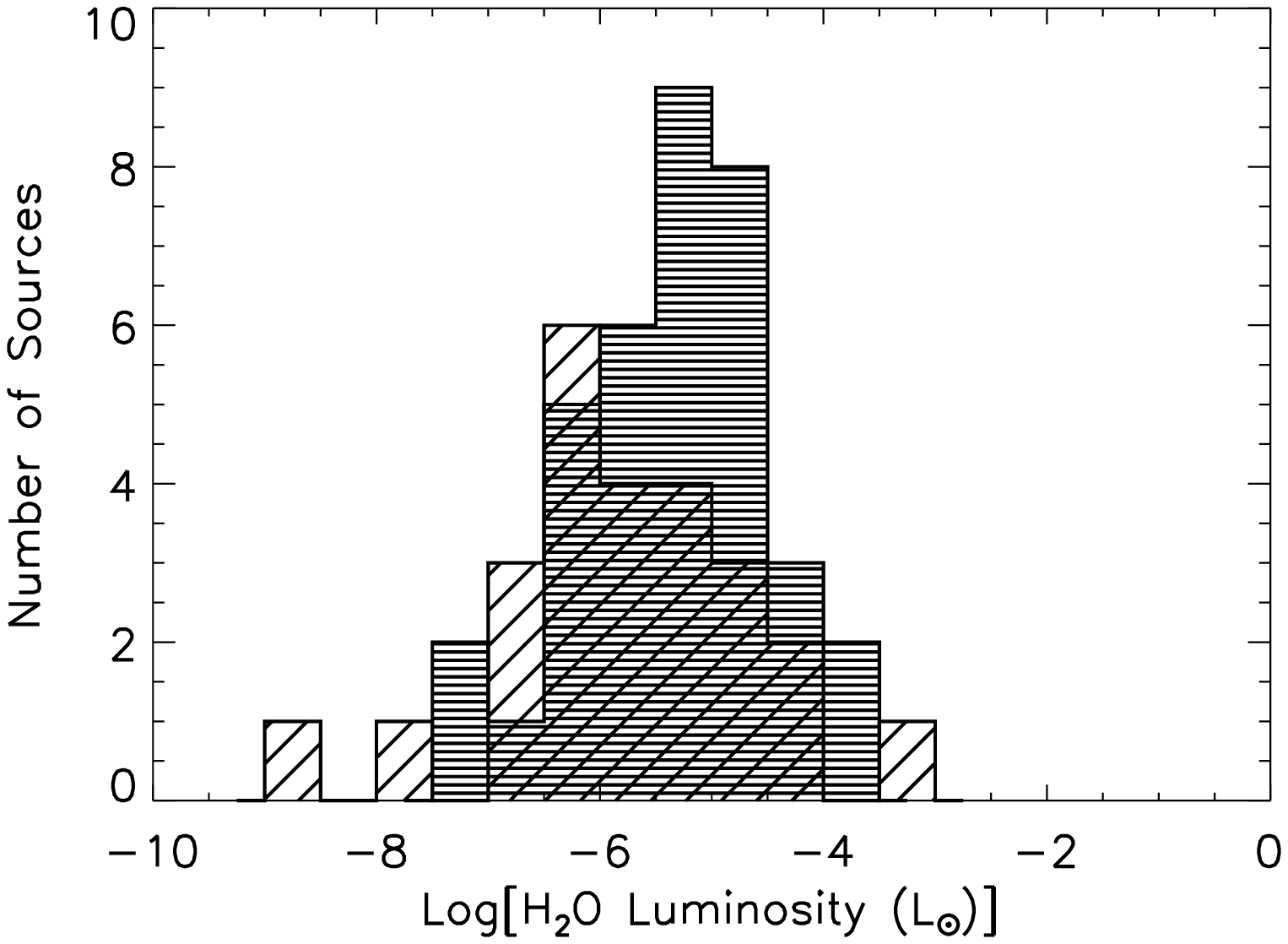}}\\
\includegraphics[scale=0.45]{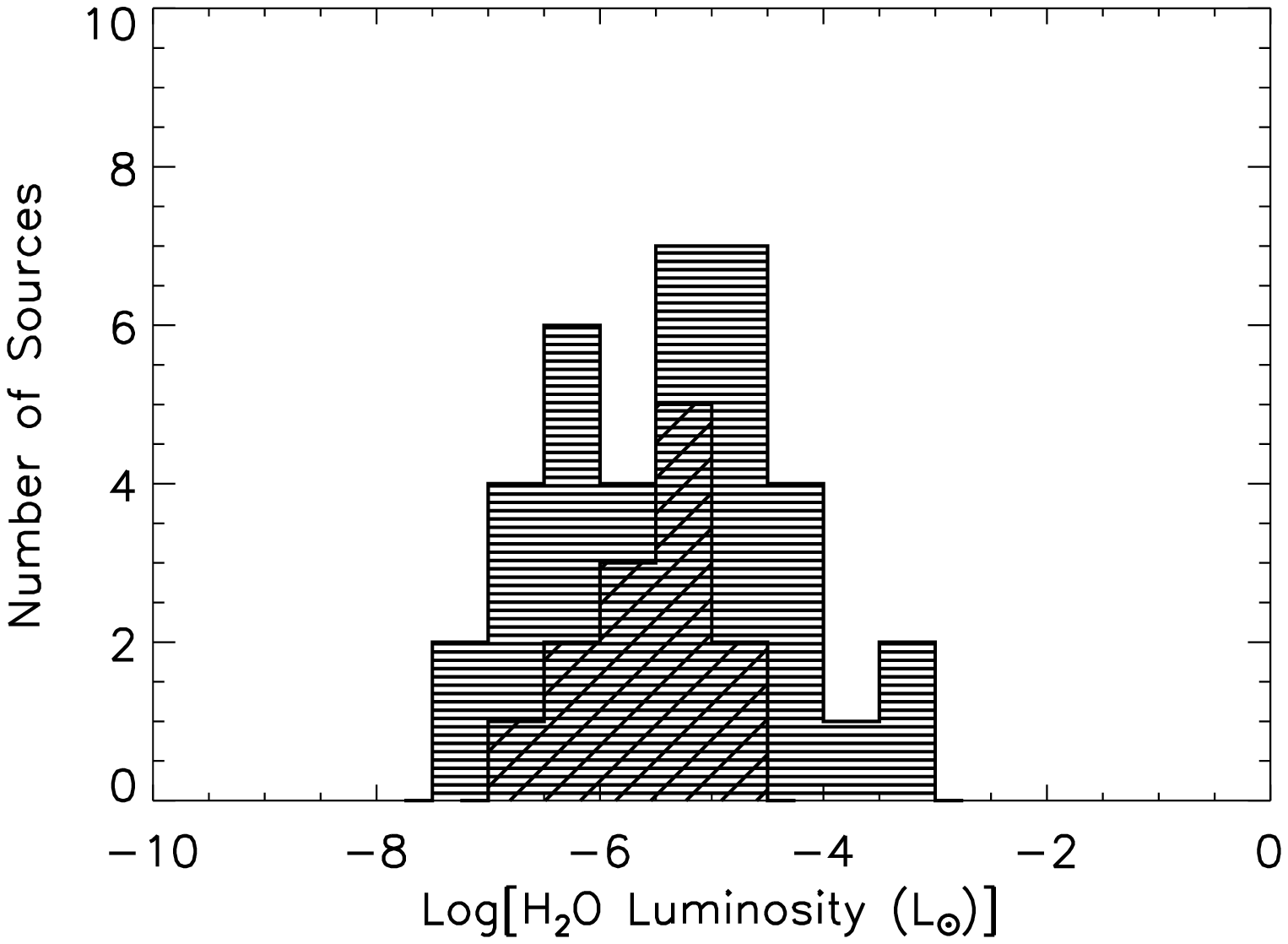}
\includegraphics[scale=0.45]{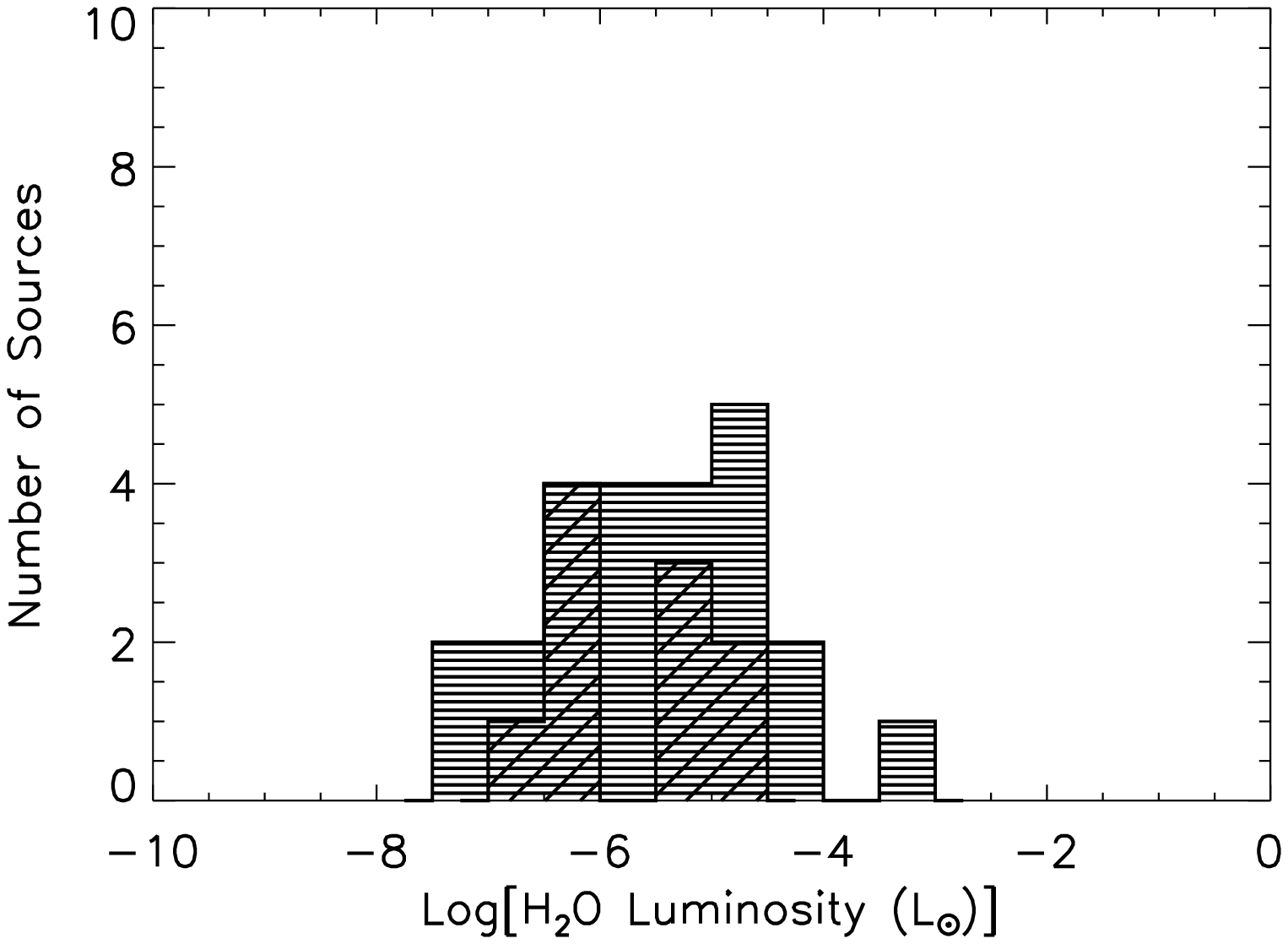}
\caption{Distributions of the \water\/ maser luminosity for different
EGO subsamples.  Upper left: Divided by association with IRDCs.  EGOs
associated and not associated with IRDCs are plotted as horizontally
and diagonally hatched histograms, respectively.  Upper right: Divided
by ``likely''/''possible'' outflow candidates.  EGOs classified as
``likely'' and ``possible'' by \citetalias{egocat} are plotted as
horizontally and diagonally hatched histograms, respectively.  Lower
left: Divided by Class I \meth\/ maser association (regardless of
Class II association/information).  Class I
detections/nondetections are
plotted as horizontally and diagonally hatched histograms,
respectively.  Bottom right: Divided by Class II \meth\/ maser
association (regardless of Class I association/information).  Class II 
detections/nondetections are plotted as horizontally and
diagonally hatched histograms, respectively.  The bin size is 0.5 dex, as in Figure~\ref{water_prop_fig}.  The significance of the
K-S statistics (low values indicate different cumulative distribution
functions) are 0.98 (IRDC/no IRDC), 0.06 (likely/possible), 0.51
(Class I/no Class I), and 0.51 (Class II/no Class II), indicating no
statistically significant differences in the distributions of the
\water\/ maser luminosities.}
\label{water_liso_shaded_fig}
\end{figure*}

%\newpage

\subsubsection{BGPS 1.1 mm Nondetections}

Young, actively accreting MYSOs are expected to be still embedded in
their natal clumps; as discussed above, we find a strong correlation
between EGOs and BGPS 1.1 mm dust sources.  The five EGOs within the
BGPS survey area but not matched to a BGPS source are all detected in
\ammonia(1,1) emission; as a group, they are not particularly distant
(all have D$<$ 5.5 kpc, Table~\ref{nh3_prop_tab}).  The \ammonia(1,1)
detections indicate that dense gas is present; here we briefly
consider the nature of these EGOs and the reasons for their lack of
counterparts in the BGPS catalog.

\begin{figure}
\plotone{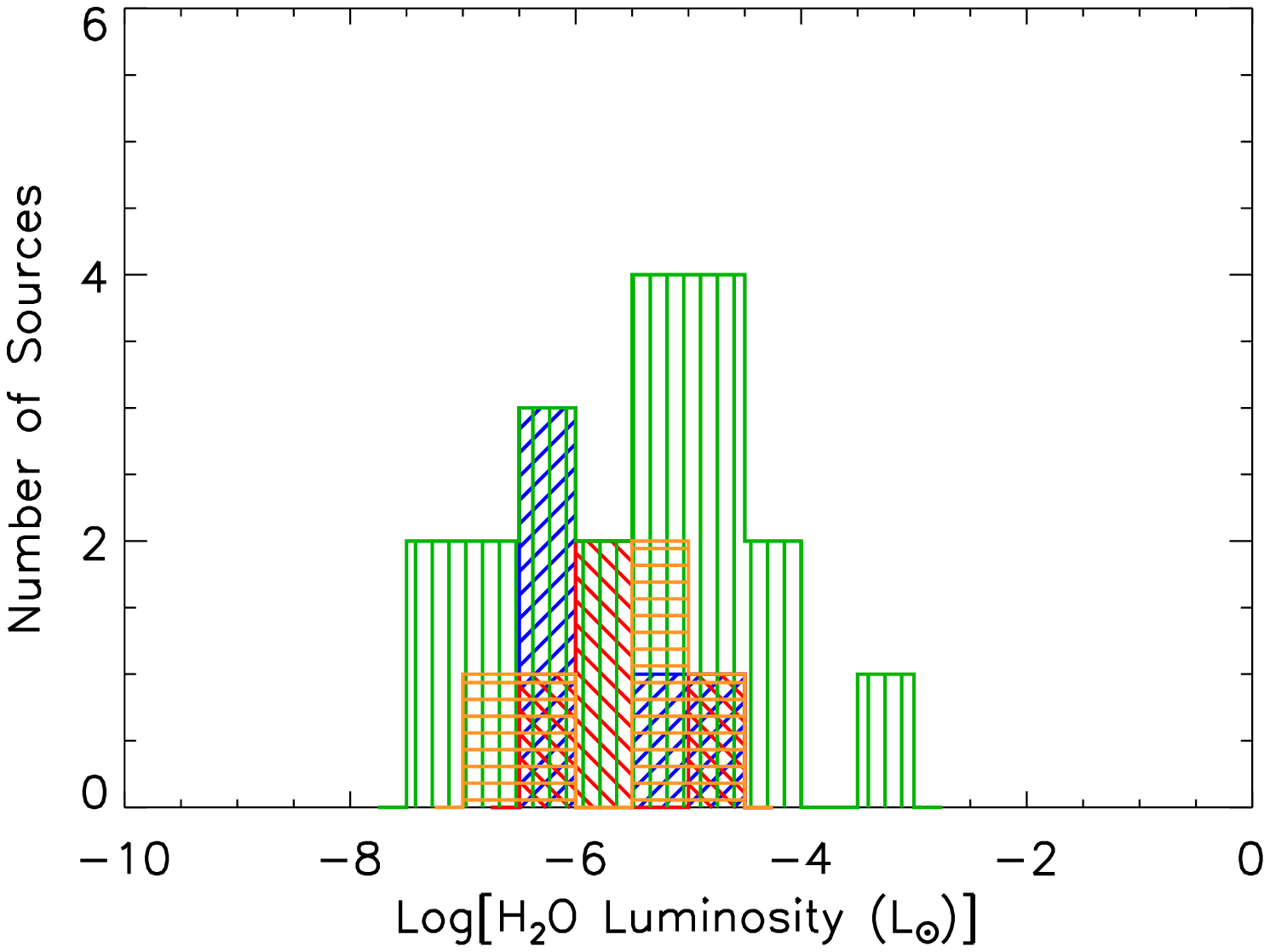}
\caption{Distribution of \water\/ maser luminosity for EGOs associated
with both Class I and II \meth\/ masers (green), only Class I \meth\/
masers (blue), only Class II \meth\/ masers (red), and neither type of
\meth\/ maser (orange).  The bin size is 0.5 dex.}
\label{liso_meth_color_fig}
\end{figure}

\begin{figure}
\plotone{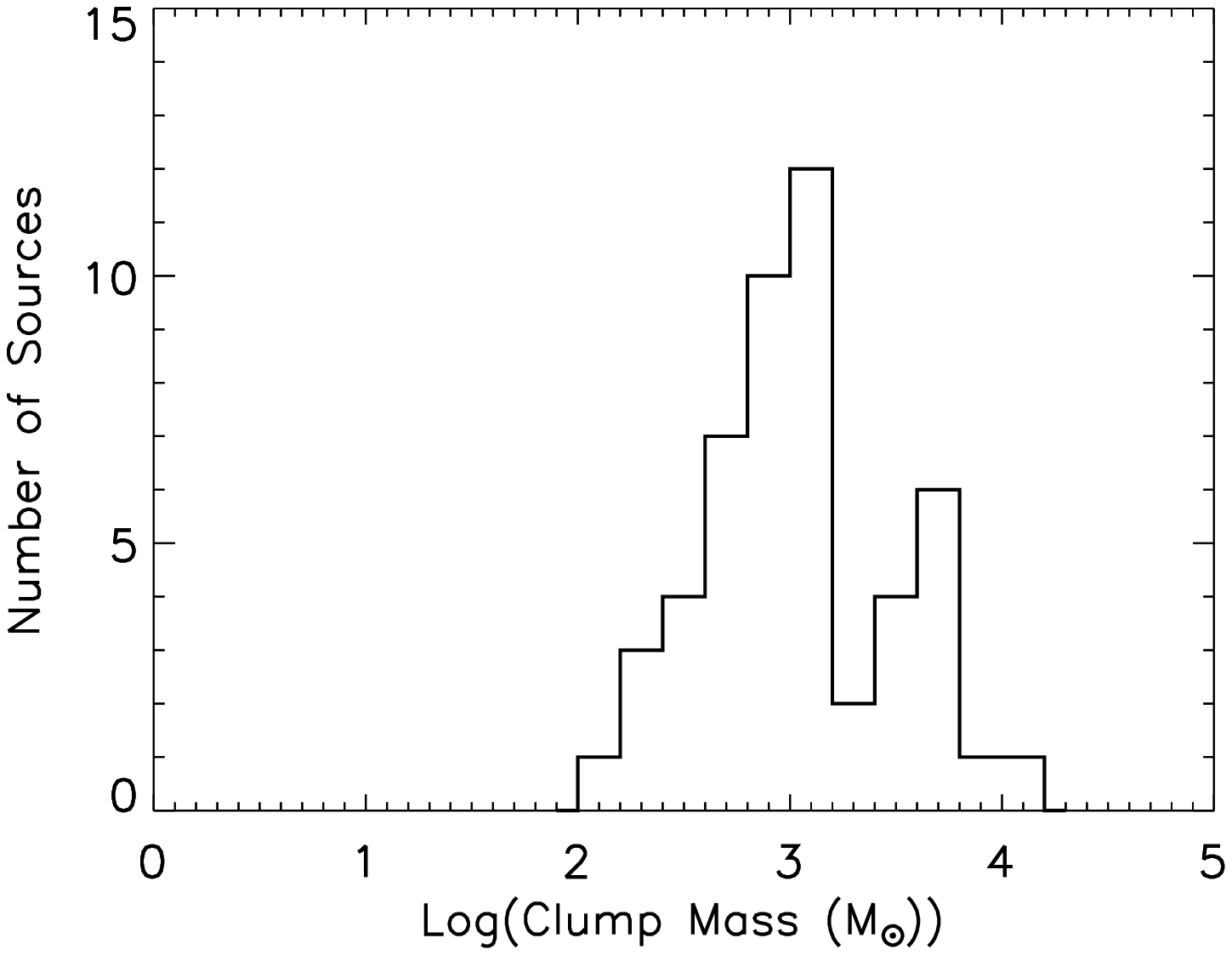}
\caption{Distribution of clump masses estimated from 1.1 mm dust
continuum emission for sources with well-determined kinetic
temperatures (\S\ref{clump_prop}).  For clarity, only nominal mass
values from Table~\ref{bgps_prop_tab} are plotted.  The bin size is
0.2 dex.}
\label{mass_histo_fig}
\end{figure}

The rms noise of the BGPS survey varies with Galactic longitude, and
is locally increased in the vicinity of bright sources
\citep{Aguirre11}.  Two of the unmatched EGOs (G62.70$-$0.51 and
G58.09$-$0.34) are at $l \sim$ 60$^{\circ}$, where the noise is
significantly higher than in most of the inner Galaxy \citep[Fig. 11
of][]{Aguirre11}; G62.70$-$0.51 is also near the edge of the BGPS map.
A third unmatched EGO, G49.27$-$0.32, is in a region of locally high
noise due to its proximity to W51.  Thus, it is possible that these
EGOs are associated with mm dust clumps that would have been detected
elsewhere in the BGPS survey.  The distances of G49.27$-$0.32 and
G62.70$-$0.51 are typical of our sample (D=5.5 and 3.9 kpc
respectively, Table~\ref{nh3_prop_tab}), and their properties are
generally consistent with EGOs detected only in \ammonia(1,1) emission
and matched to BGPS sources.  The increased noise of the BGPS survey
at the locations of these EGOs thus seems to be a likely
explanation for their lack of BGPS counterparts.  G58.09$-$0.34,
however, may be an example of a nearby, low-mass YSO: it has a near
kinematic distance of 0.74$^{+0.65}_{-0.61}$ kpc and exceptionally
narrow \ammonia(1,1) emission ($\sigma$v$\sim$0.23).

Examining the BGPS images suggests that the two other unmatched EGOs
(G50.36$-$0.42 and G29.89$-$0.77) are associated with 1.1 mm emission,
despite not being matched to BGPS catalog sources.  G50.36$-$0.42
appears to be associated with faint 1.1 mm emission that fell below
the threshold for extraction as a BGPS source \citep{Rosolowsky10}.  A
\citetalias{egocat} ``possible'' outflow candidate (D=3.0 kpc),
G50.36$-$0.42 also has detected \water\/ maser emission in our survey.
G29.89$-$0.77 is immediately adjacent to a BGPS source, but the
\citetalias{egocat} position falls outside the BGPS source boundary
defined by the label maps.  Also a \citetalias{egocat} ``possible''
outflow candidate, G29.89$-$0.77 has the strongest \ammonia\/ emission
of the unmatched EGOs; though the (2,2) line is formally undetected by
our 4$\sigma$ criteria, weak \ammonia(2,2) emission is evident in the
spectrum.  Taken together, this evidence suggests G29.89$-$0.77 and
G50.36$-$0.42 are likely similar in nature to EGOs that are matched to
BGPS sources.

\section{Discussion}\label{discussion}

\subsection{EGOs in Context}

\subsubsection{Comparison with Other Samples}\label{rate_dis}

A notable feature of EGOs, compared to other samples of young
massive (proto)stars, is their very strong association with both Class
I and II \meth\/ masers, reflected in notably high detection rates in
\meth\/ maser surveys to date \citepalias[e.g.][]{maserpap,Chen11}.
Since \water\/ maser and \ammonia\/ observations are common tools for
studying massive star formation, our Nobeyama survey allows us to
better place EGOs in their broader context, by comparing their
molecular environments to those of MYSOs selected using other
criteria/tracers.  Table~\ref{lit_detect_comparison_tab} summarizes
\water\/ maser and \ammonia(1,1) detection rates towards a variety of
MYSO samples from the literature, chosen to cover a range of sample
selection criteria, survey parameters, and proposed evolutionary state of the target objects.
The strong correlation of EGOs with 6.7 GHz \meth\/ masers and dust
clumps (\S\ref{clump_prop}) suggests these as natural comparison
samples \citep[indeed, the samples of][include some EGOs, see also
discussion therein]{BreenEllingsen11,Bart11}.
'Active' cores in \citet{Chambers09} are defined by the presence of
``green fuzzy'' and 24 \um\/ emission.  They define ``green fuzzy''
broadly, compared to \citetalias{egocat} EGOs; still, one might
expect these sources to be similar to EGOs associated with IRDCs.
In contrast, MYSO and UC HII samples
compiled using the \emph{IRAS} or \emph{MSX} point source catalogs
comprise sources that are more MIR-bright than EGOs, and so likely more luminous and/or more evolved \citepalias[see also][]{egocat}.

As illustrated by Table~\ref{lit_detect_comparison_tab}, \water\/
maser detection rates towards massive (proto)star samples span a broad
range, from $<$20\% to $>$80\%: our overall detection rate of 68\% is
towards the upper end of this range.  Notably, our \water\/ maser
detection rate towards EGOs associated with both Class I and II
\meth\/ masers (95\%) exceeds, to our knowledge, any reported in the literature.
Our
much lower detection rate towards EGOs with neither \meth\/ maser type
(33\%) is nonetheless higher than towards quiescent dust clumps or IRDC
cores.
In general, the \water\/ maser associations of EGO subsamples are
similar to those of the most comparable subsamples in
Table~\ref{lit_detect_comparison_tab}.  For example, our detection
rate for EGOs associated with Class II masers (regardless of Class I
association) is roughly comparable to those for Class II \meth\/ maser
and dust clump/Class II \meth\/ maser samples.  
Sensitivity is of course an important consideration, particularly in
light of recent evidence that \water\/ maser flux density increases as
sources evolve, then turns over at a late (UC HII region) stage
\citep{BreenEllingsen11}.  While \water\/ masers are variable, the
fact that we fail to detect \water\/ maser emission towards EGO
G11.11$-$0.11, where a weak (\q0.3 Jy) \water\/ maser was reported by
\citet{Pillai06G11}, indicates that some EGOs are associated with
\water\/ masers below the detection limit of our survey.  Most of the
surveys in Table~\ref{lit_detect_comparison_tab} have sensitivity 
comparable to or better than our Nobeyama data.

The properties of the \water\/ masers detected towards EGOs are
typical of \water\/ masers detected towards MYSOs.  For example, the
distributions of the velocity range of detected masers and of the
velocity offset between dense gas and peak maser emission
(Fig.~\ref{water_prop_fig}-\ref{voffset_nh3_water_fig}, see also
\S\ref{high_vel_dis}) are generally similar to those reported in the
literature, including for more evolved UC HII region samples
\citep[e.g.][]{Churchwell90,Anglada96,Urquhart11}.  Based on their
study of MIR-bright MYSOs and UC HII regions from the RMS sample,
\citet{Urquhart11} argue that \water\/ maser properties (in
particular, L$_{\rm iso}$) are driven by the bolometric luminosity of the
central MYSO (see also \S\ref{clump_prop_dis}).  The distributions of
\water\/ maser peak and integrated flux density and isotropic
luminosity for the MIR-bright RMS sample have high-end tails
\citep[e.g. Fig. 8 of][]{Urquhart11}; the strongest RMS water masers
are several orders of magnitude brighter and more luminous than the
strongest water masers we detect towards EGOs.  However, two-sided K-S
tests on these parameters indicate that the differences are not
statistically significant (K-S significance 0.055, 0.249, and 0.027
for S$_{\rm peak}$, S$_{\rm int}$, and L$_{\rm iso}$, respectively).  The
K-S tests are consistent with the RMS and EGO water masers being drawn
from the same parent distribution.  

As discussed in \S\ref{nh3_ego_compare} and \S\ref{water_ego_compare},
we find evidence for statistically significant differences among EGO
subsamples in \ammonia\/ but not in \water\/ maser properties.  Other
\ammonia\/ studies of large MYSO samples similarly find significant
internal variations. 
The mean kinetic temperature, \ammonia\/ linewidth, and \ammonia\/
column density of BGPS sources increase with the number of associated
MIR sources \citep[albeit with considerable scatter, particularly in
T$_{kin}$, e.g. Fig. 23 of][]{nh3-innergal}.  In the RMS sample,
\citet{Urquhart11} find that the mean kinetic temperature, \ammonia\/
column density, and \ammonia\/ linewidth are higher for UC HII regions
than for MYSOs.  Overall, the clump-scale \ammonia\/ properties of
EGOs are roughly comparable to those of other MYSO samples.  Comparing
Figure~\ref{nh3_prop_fig} to Figure 4 of \citet{Urquhart11}, for
example, the linewidth, T$_{kin}$, and N(\ammonia) distributions are
broadly similar (accounting for the conversion between $\sigma_{\rm v}$ and
FWHM linewidth), though our sample is considerably smaller.  The
distribution of \ammonia\/ column density extends to lower values for
EGOs than for the RMS sample; however, this is a beam-averaged
quantity, and the Nobeyama beam (\q73\pp) is considerably larger than
that of the GBT.  For BGPS sources, the low end of the \ammonia\/
column density range (also based on GBT observations) extends to
\q1.7$\times$10$^{13}$, more comparable to our EGO results.  The EGO
T$_{kin}$ distribution (from the single-component fitting, for
consistency with other studies) lacks the high temperature ($>$40 K)
tail seen in RMS, UC HII region, and even BGPS samples
\citep{Urquhart11,nh3-innergal,Churchwell90}.  The mean T$_{kin}$ for
the EGO sample as a whole (23.6 K) is higher than that of the
\citet{nh3-innergal} sample (17.4 K, for their 'T$_K$ subsample'
consisting of (2,2) detections) and similar to that of the RMS
sample as a whole (\q 22 K).  

These general comparisons illustrate that the \water\/ maser and clump-scale \ammonia\/
properties of EGOs are consistent with their being a population of
young MYSOs.  However, we emphasize that the differences within
samples (EGOs, RMS sources, BGPS sources) are often as
great or greater than the differences between them.
These \emph{intra}-sample differences emphasize the importance of
studying multiple star formation tracers across wavelength regimes.

\begin{figure}
\plotone{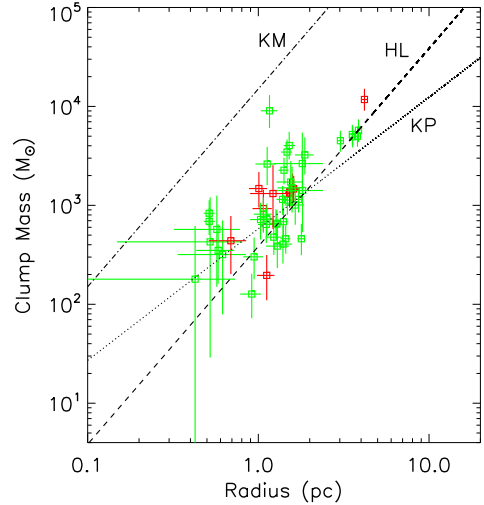}
\caption{Clump mass vs. radius for BGPS sources
associated with EGOs.  Clump masses are calculated assuming T$_{dust}$=T$_{kin}$ from single-component \ammonia\/ fitting.  Open squares indicate nominal values from
Table~\ref{bgps_prop_tab}.  The error bars indicate the range in
radius associated with the uncertainty in distance
(Table~\ref{bgps_prop_tab}) and the range in mass associated with
the combined uncertainties in the BGPS integrated flux density, the
BGPS flux correction factor, and the distance.  The star formation
thresholds of \citet{KM08}, \citet{Heiderman10} and \citet{Lada10},
and \citet{KP10} are indicated as dot-dashed, dashed, and dotted lines
respectively (see \ref{clump_prop}).  Only sources for which the
T$_{\rm kin}$ and radius are well-determined (non-limit) are plotted.
\water\/ maser detections
are plotted in green, and \water\/ maser nondetections in red.}
\label{mass_radius_fig}
\end{figure}

\subsubsection{Comparison with Star Formation Criteria}\label{sf_thres_dis}

By combining our Nobeyama \ammonia\/ data with the BGPS, we can also
consider the dust clumps associated with EGOs in the context of
proposed star formation thresholds.  Unlike purely mm-selected samples
\citep[e.g.][]{nh3-innergal}, all of the clumps we consider are
associated with EGOs, and thus demonstrably star-forming (many are
also associated with other MIR sources).
Figure~\ref{mass_radius_fig} shows a mass-radius plot for clumps with
well-determined (non-limit) T$_{\rm kin}$ and radius, with the clump
mass estimated assuming T$_{\rm dust}$=T$_{\rm kin}$ from the
single-component \ammonia\/ fits.  The errors bars shown in
Figure~\ref{mass_radius_fig} indicate the range in radius associated
with the distance uncertainty from Table~\ref{bgps_prop_tab}, and the
range in mass associated with the combined uncertainties in the BGPS
integrated flux density, the BGPS flux correction factor, and the
distance (see also \S\ref{clump_prop}).  The error bars do not include systematic
uncertainty in the radius estimate due to different geometries
\citep[see also][]{Rosolowsky10}.  Three
proposed star formation thresholds are indicated on
Figure~\ref{mass_radius_fig}: (1) the \citet{KM08} threshold for
massive star formation of 1 g cm$^{-2}$ (4788 \msun\/ pc$^{-2}$); (2)
the average of the \citet{Lada10} and \citet{Heiderman10} thresholds
for ``efficient'' star formation (122.5 \msun\/ pc$^{-2}$); and (3)
the \citet{KP10} and \citet{Kauffmann10} threshold for massive star
formation.  We refer to these as the KM, HL, and KP thresholds, respectively.  As in the recent BGPS study of \citet{nh3-innergal}, we
scale the KP criterion of M(r)$>$ 870 \msun\/ (r/pc)$^{1.33}$ to
M(r)$>$ 580 \msun\/ (r/pc)$^{1.33}$ to account for the difference between
our assumed dust opacity and that adopted by \citet{KP10}.
Adopting the nominal clump mass and radius values from
Table~\ref{bgps_prop_tab}, 70\% (35/50) of the sources shown in
Figure~\ref{mass_radius_fig} exceed the KP threshold, and 76\%
(38/50) exceed the HL threshold; as in the \citet{nh3-innergal} study
of BGPS sources, none of our EGO clumps meet the KM criterion.
We emphasize that the points in
Figure~\ref{mass_radius_fig} represent \emph{average} surface
densities over entire BGPS sources, and that the BGPS and Nobeyama
observations probe large scales.
At a typical distance of 4 kpc, the 33\pp\/ BGPS beam is \q 0.64 pc,
and the 73\pp\/ Nobeyama beam \q 1.4 pc.  Interferometric
observations of EGOs, and of other MYSOs, provide ample evidence for
substructure (e.g. cores and (proto)clusters) and variations in gas
temperature on much smaller scales \citep[e.g.][]{C11,rsro}.
 
Having placed clumps on the mass-radius plot using (primarily) the
BGPS data, we use our Nobeyama survey data to look for differences in
the properties of clumps above/below the HL and KP thresholds.  As in
our comparison of EGO subsamples (\S\ref{nh3_ego_compare}), we ran
two-sided K-S tests on eight \ammonia\/ parameters (the \ammonia\/
(1,1), (2,2), and (3,3) peaks (T$_{\rm MB}$), $\sigma_{\rm v}$,
$\tau_{(1,1)}$, $\eta_{ff}$, N(\ammonia), and T$_{kin}$).  We find
statistically significant differences only for the \ammonia(1,1) and
(2,2) peak temperatures and the filling fraction $\eta_{ff}$\footnote{We note $\eta_{ff}$ is mildly degenerate with T$_{\rm MB}$(1,1).}, with
clumps below the HL and KP thresholds having lower values of these
parameters.  The K-S tests indicate no statistically significant
differences in the distributions of the physical properties
$\sigma_{\rm v}$, T$_{kin}$, and N(\ammonia) for clumps above/below
the thresholds.  Interestingly, and perhaps counterintuitively, the \water\/ maser detection rates are
\emph{higher} for EGOs associated with clumps \emph{below} the HL and
KP thresholds (Fig.~\ref{mass_radius_fig}).  The
\water\/ maser detection rate is 0.74($\pm$0.07) for sources that meet
the KP criterion, and 0.93($\pm$0.06) for sources that do not
(uncertainties in detection rates calculated using binomial
statistics).  Similarly, the \water\/ maser detection rates are
0.76($\pm$0.07) and 0.92($\pm$0.08) for sources that do/do not meet the HL
criterion, respectively.     

\begin{figure}
\plotone{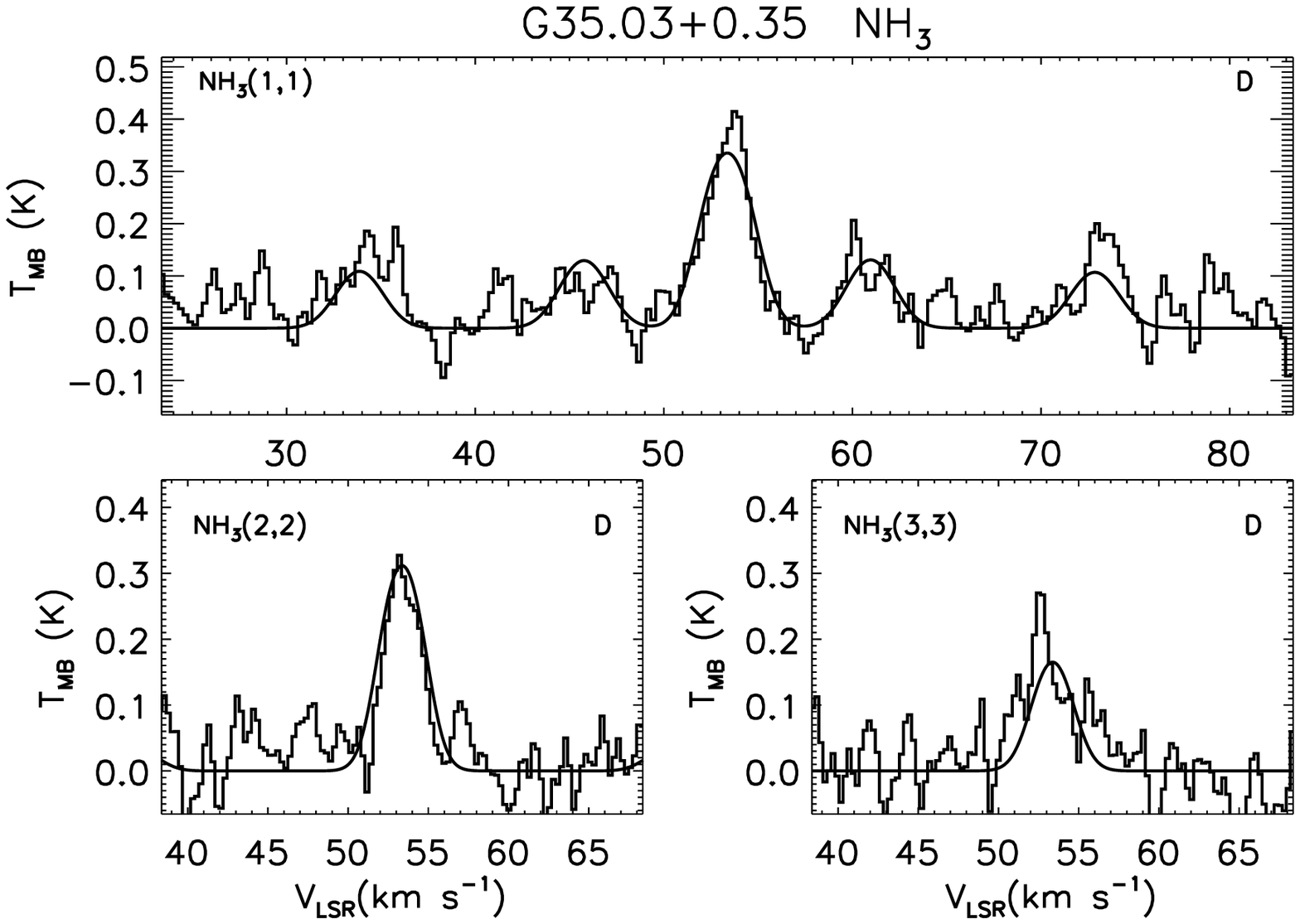}
\caption{\ammonia\/ spectra of the EGO G35.03+0.35, in which \citet{rsro} detect an \ammonia(3,3) maser.}
\label{g35_nh3_spec}
\end{figure}

The nature of the EGOs associated with clumps that fall below the KP
threshold requires further investigation.  The higher \water\/ maser
detection rate towards clumps below the KP threshold is surprising,
and the lack of difference in \ammonia\/ properties suggests a
continuum, rather than a sharp distinction.  Additionally, one source
that falls below the KP threshold, G24.94+0.07, is associated with 6.7
GHz Class II \meth\/ maser and cm continuum emission
\citepalias{maserpap,C11vla}, both indicative of the presence of an
MYSO.  We note that the placement of clumps on a mass-radius plot is
sensitive to assumptions about clump temperature structure (or lack
thereof).  For EGOs in our study fit with warm and cool components,
the warm component constitutes a small fraction of the clump mass; the
bulk of the material generally has temperature
T$_{cool}<$T$_{single~comp.}$, and so the isothermal assumption
underestimates the clump mass (\S\ref{clump_prop},
Table~\ref{bgps_prop_tab}).  Interferometric \ammonia\/ observations
show significant temperature structure on scales within the Nobeyama
beam for G35.03+0.35 \citep[Fig. 3 of][]{rsro}, a source that did
\emph{not} require two temperature components to fit its Nobeyama
\ammonia\/ spectrum (Fig.~\ref{g35_nh3_spec}).  On larger scales,
many of the BGPS sources associated with EGOs (and plotted in
Fig.~\ref{mass_radius_fig}) extend beyond the Nobeyama beam.  If
isothermal clump masses for EGOs tended to be underestimates--due to
temperature structure on small or large scales--this would move points
up in Fig.~\ref{mass_radius_fig}, and increase the proportion of
sources above the KP threshold.  Additional data--such as \ammonia\/
maps with sufficient resolution to probe the temperature structure of
the BGPS clumps--are needed to address this issue.  Interferometric
(sub)mm observations, to resolve the dust continuum emission and
detect individual cores, and improved constraints on bolometric
luminosity (e.g. from HiGal) will also help to clarify the nature of
the driving sources.

\begin{figure}
%\begin{center}
%\includegraphics[scale=0.6]{f22_a.eps}\\
%\includegraphics[scale=0.6]{f22_b.eps}
\plotone{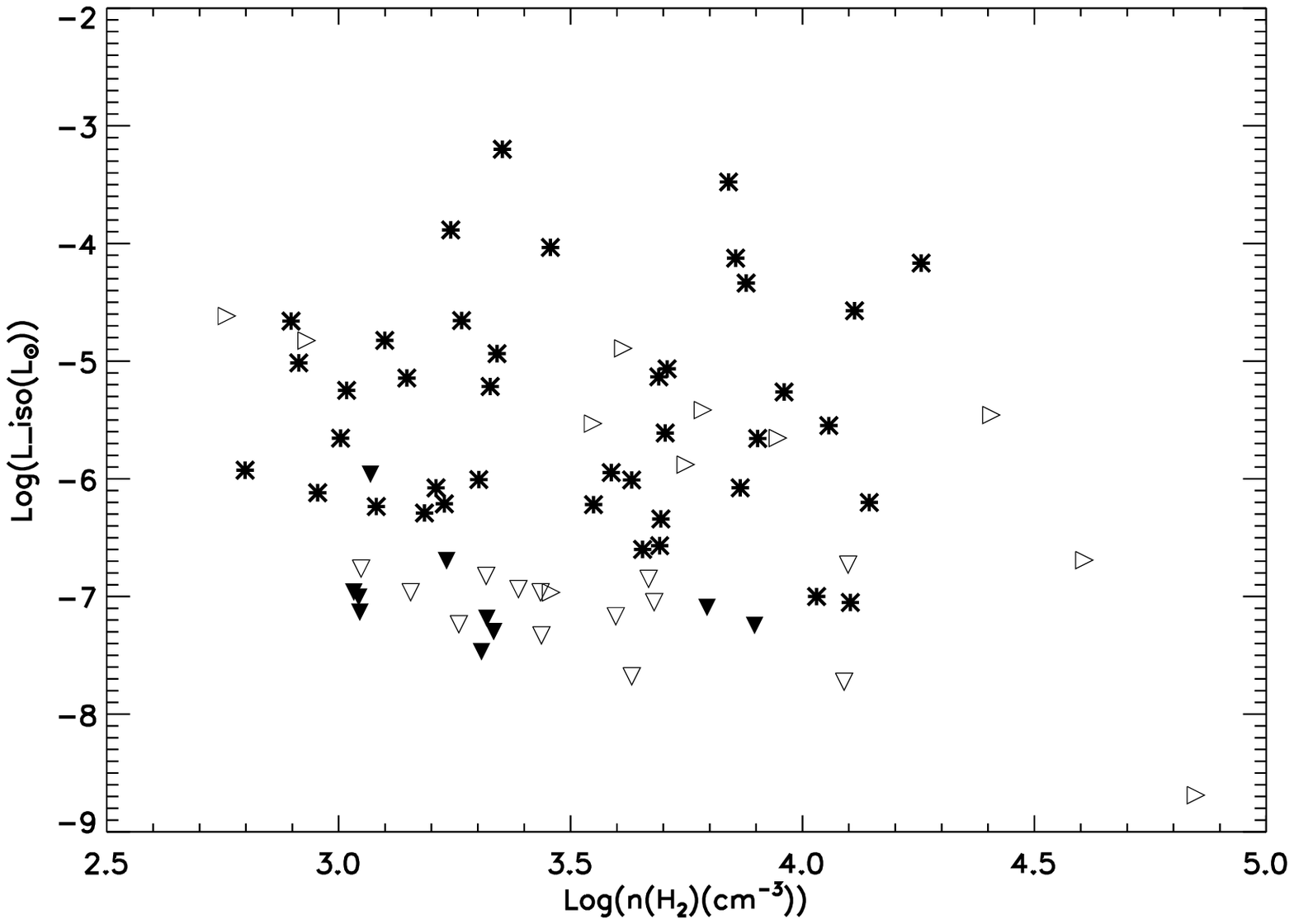}\\
\plotone{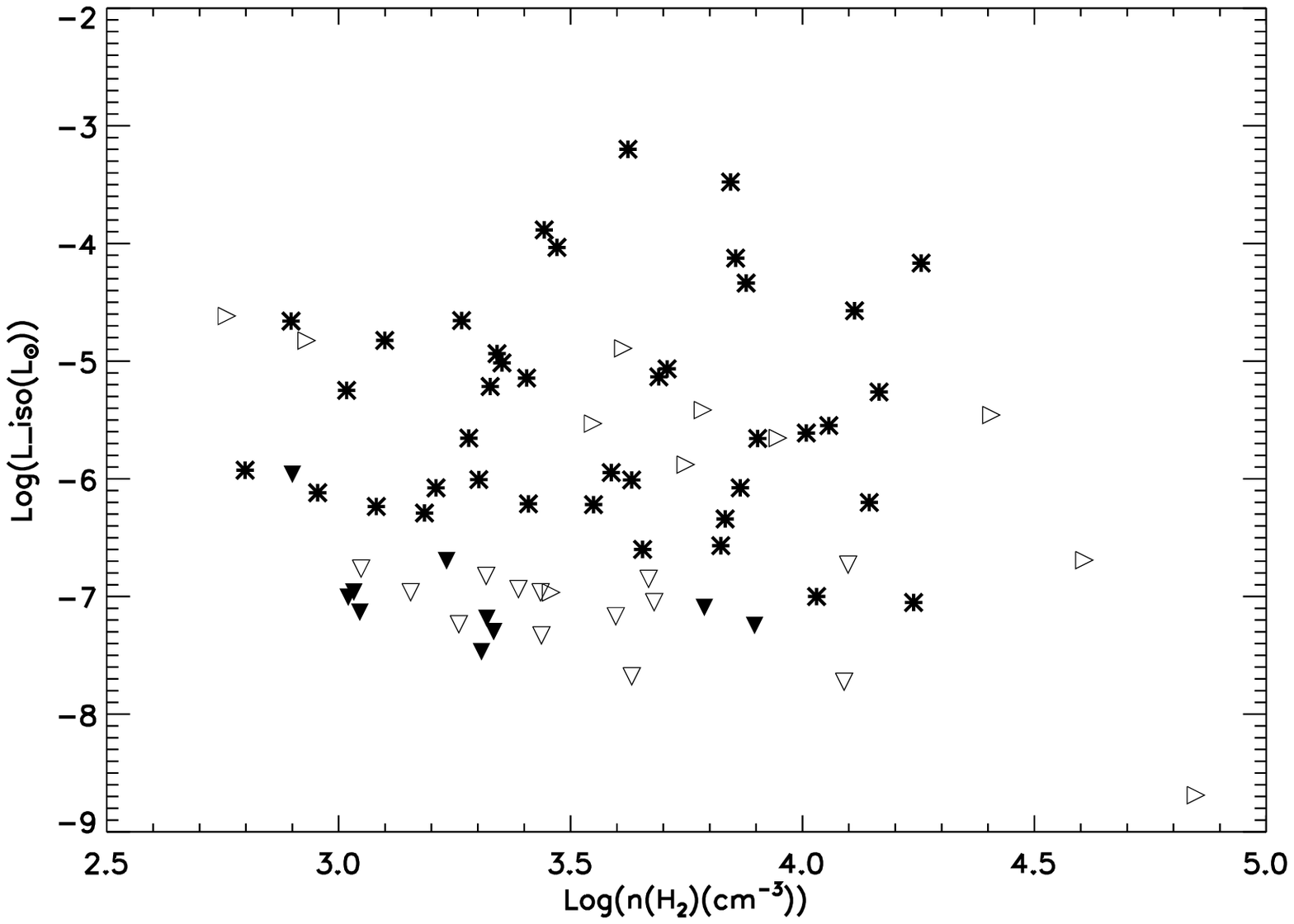}
\caption{Top: isotropic \water\/ maser luminosity vs. volume-averaged
number density estimated using T$_{kin}$ from single-component
\ammonia\/ fitting.  * indicates EGOs with \water\/ maser detections in
our survey and well-determined density estimates (e.g. neither
T$_{kin}$ nor R is a limit).  Filled downward-pointing triangles
indicate 4$\sigma$ L(\water) upper limits for EGOs with
well-determined density estimates that are \water\/ maser
nondetections in our survey.  EGOs for which the estimated density is
a lower limit are represented as open triangles: open right-facing
triangles indicate \water\/ maser detections, and open
downward-pointing triangles 4$\sigma$ L(\water) upper limits for
\water\/ maser nondetections.  Bottom: Same as top, except the density
estimate accounts for warm and cool components when the \ammonia\/
spectrum is fit with two-components (see \S\ref{clump_prop}).}
\label{liso_v_density_fig}
%\end{center}
\end{figure}

\begin{figure}
%\begin{center}
\plotone{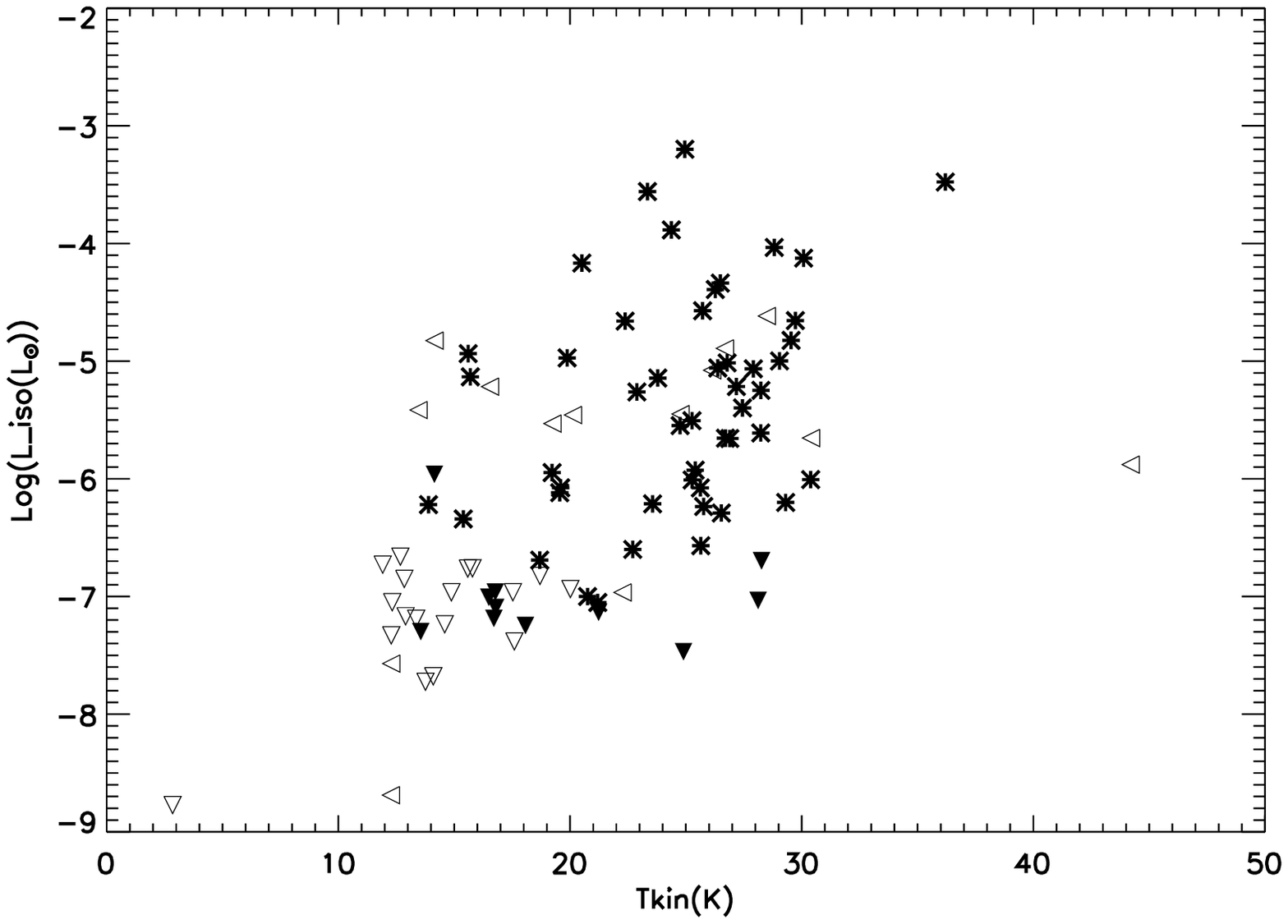}\\
\plotone{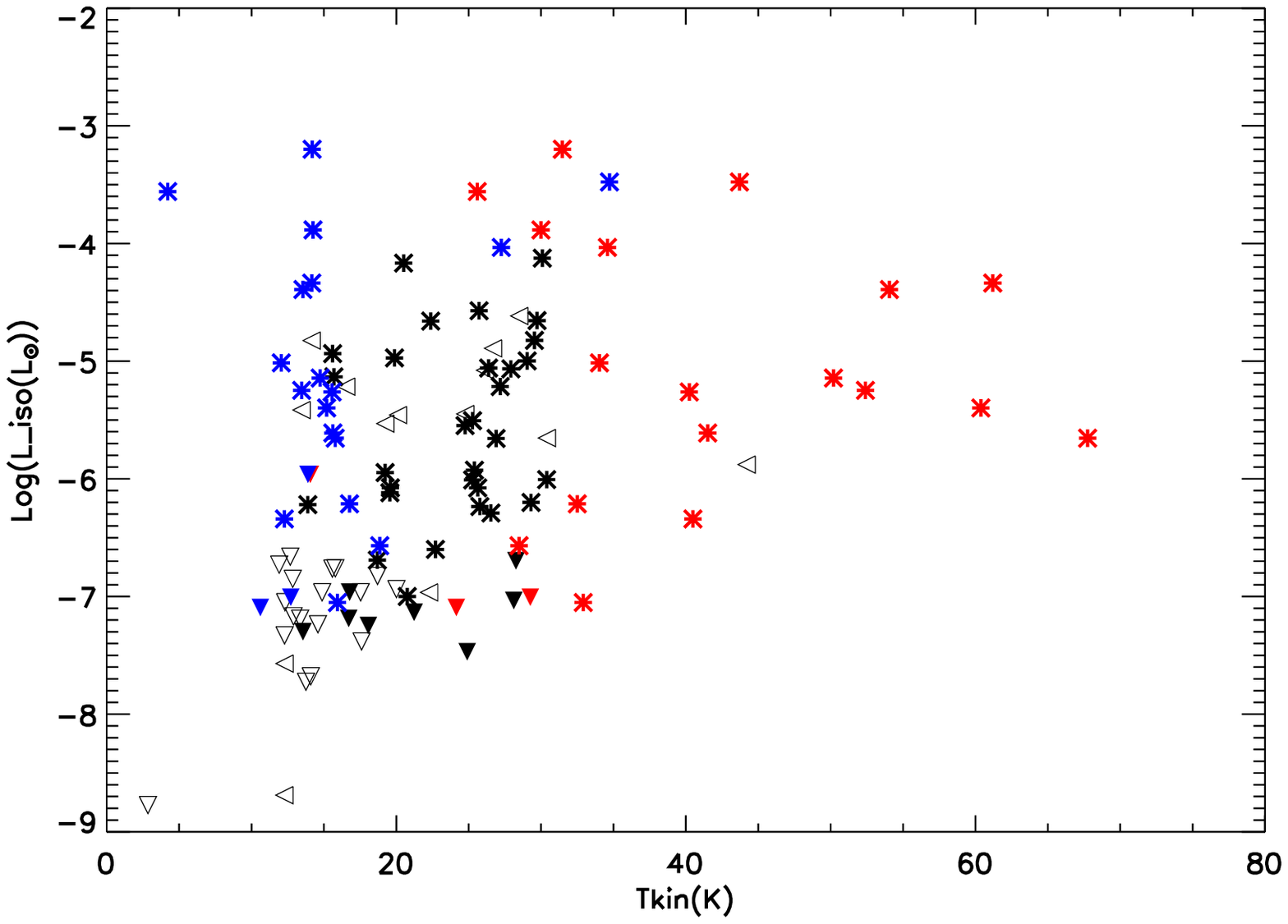}
\caption{Top: isotropic \water\/ maser luminosity vs. T$_{kin}$ from single-component \ammonia\/ fitting.  * indicates EGOs with \water\/ maser and \ammonia(2,2) detections in our survey (e.g. T$_{kin}$ well-determined).  Filled downward-pointing triangles indicate 4$\sigma$ L(\water) upper limits for EGOs undetected in \water\/ maser emission but detected in \ammonia(2,2).  EGOs undetected in \ammonia(2,2)--for which the best-fit T$_{kin}$  is treated as an upper limit--are represented as open triangles: open left-facing triangles indicate \water\/ maser detections, and open downward-pointing triangles 4$\sigma$ L(\water) upper limits for \water\/ maser nondetections.  Bottom: Same as top, except for sources fit with two \ammonia\/ components, T$_{kin}$(cool) is plotted in blue and  T$_{kin}$(warm) in red.}
\label{liso_v_tkin_fig}
%\end{center}
\end{figure}

\begin{figure}
%\begin{center}
\plotone{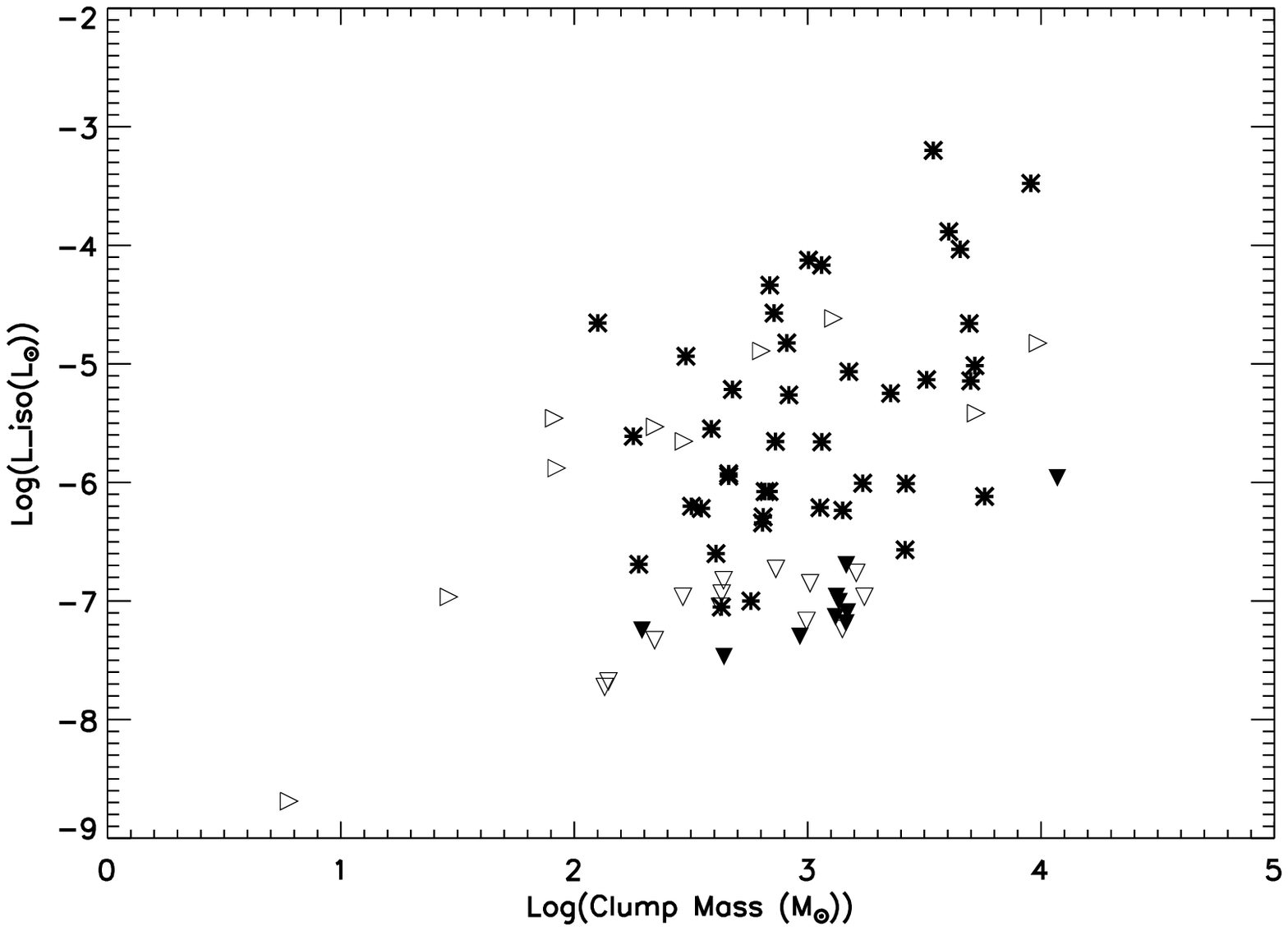}\\
\plotone{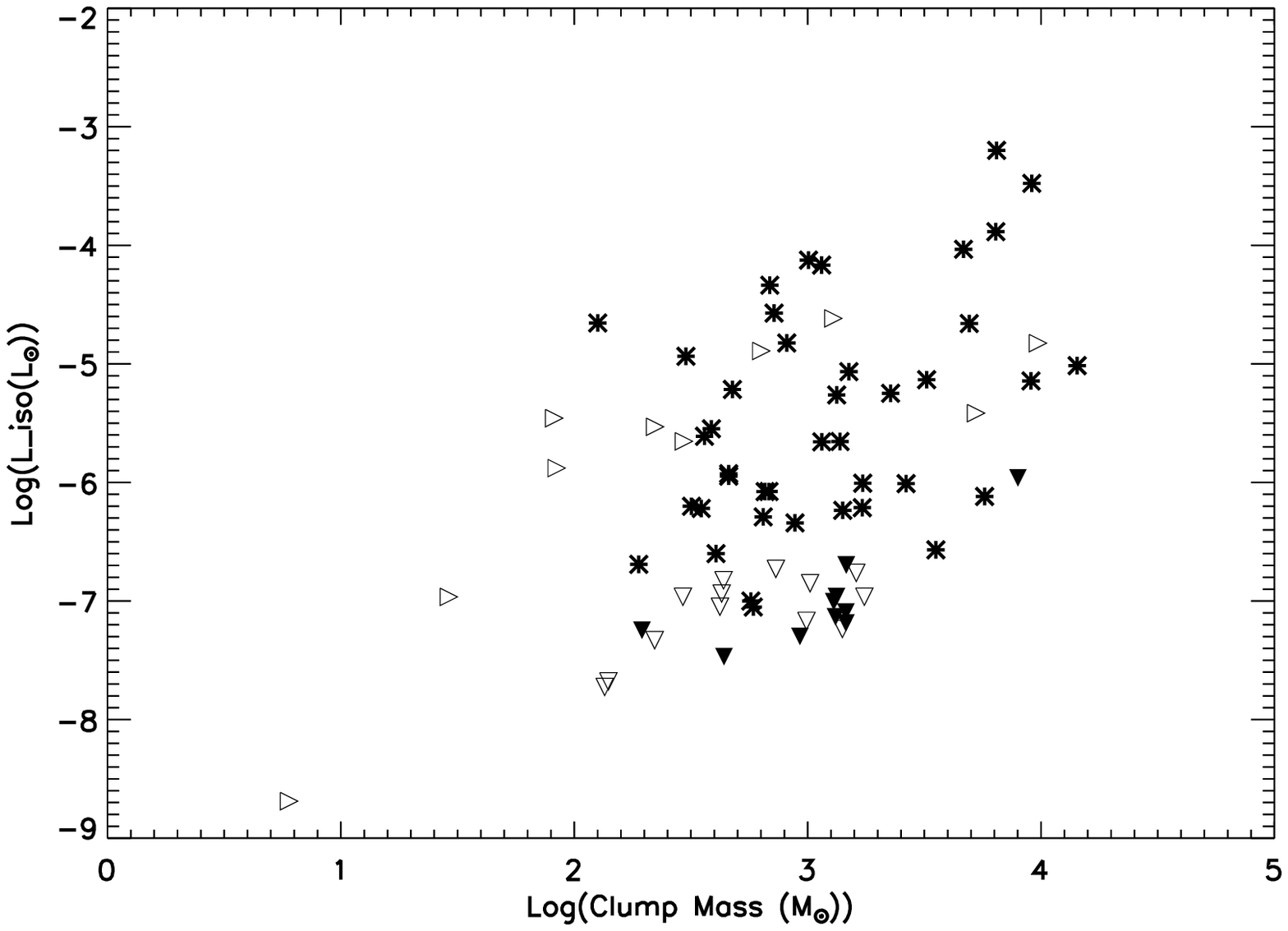}
\caption{Top: isotropic \water\/ maser luminosity vs. clump mass, assuming T$_{dust}$=T$_{kin}$ from single-component \ammonia\/ fitting.  * indicates EGOs with \water\/ maser and \ammonia(2,2) detections in our survey (e.g. T$_{kin}$ well-determined).  Filled downward-pointing triangles indicate 4$\sigma$ L(\water) upper limits for EGOs undetected in \water\/ maser emission but detected in \ammonia(2,2).  EGOs undetected in \ammonia(2,2)--for which the best-fit T$_{kin}$  is an upper limit and the clump mass thus a lower limit (\S\ref{clump_prop})--are represented as open triangles: open right-facing triangles indicate \water\/ maser detections, and open downward-pointing triangles 4$\sigma$ L(\water) upper limits for \water\/ maser nondetections.  Bottom: Same as top, except the mass estimate accounts for warm and cool components for sources fit with two \ammonia\/ components (\S\ref{clump_prop}) .}
\label{liso_v_mass_fig}
%\end{center}
\end{figure}

\subsection{Correlations between \water\/ Maser and Clump Properties?}\label{clump_prop_dis}

Over the past decades, numerous authors have investigated possible
correlations amongst clump, \water\/ maser, and driving sources properties in
MYSO samples \citep[e.g.][]{Churchwell90,Anglada96,BreenEllingsen11,Urquhart11}.  Recently, two studies have
reported correlations between \water\/ maser luminosity and the
properties of the driving source or surrounding clump.  For their sample
of \q 300 RMS sources with \water\/ maser detections,
\citet{Urquhart11} find that \water\/ maser luminosity is positively
correlated with bolometric luminosity for both MYSOs and HII regions.
In contrast, \citet{BreenEllingsen11} report an anticorrelation
between clump \h\/ number density and \water\/ maser luminosity,
which they attribute to an evolutionary effect: more evolved sources
have more luminous water masers and are associated with lower-density
clumps.  All of these studies have combined \water\/ maser and \emph{either}
\ammonia\/ or (sub)mm dust continuum data.  \citet{BreenEllingsen11},
in particular, caution that the clump densities used in their study
\citep[from][]{Hill05} were calculated assuming a single temperature
for all clumps, and that temperature differences could create the
apparent density trend.  Our \ammonia\/ and \water\/ maser survey, in
combination with the BGPS, provides the necessary data to fully
explore correlations between maser and clump properties, and test
evolutionary interpretations.

Figure~\ref{liso_v_density_fig} shows that when clump densities are
calculated for our sample using measured clump temperatures, there is no correlation
between \water\/ maser luminosity and clump density: the log-log plot
of L(\water) vs. number density is a scatter plot.  This remains the
case even when accounting for the contributions of warm and cool
gas for sources that require two-component \ammonia\/ fits
(Fig.~\ref{liso_v_density_fig}, bottom; \S\ref{clump_prop}).  The
partial correlation coefficients, computed with the distance squared
as an independent parameter, are 0.04 and 0.06 for the one- and
two-temperature component density estimates, respectively (only
sources with \water\/ maser detections and non-limit densities are
included in the calculation).  These low values confirm that \water\/
maser luminosity and clump number density are uncorrelated in our
data.

In contrast, \water\/ maser luminosity is weakly correlated with clump
temperature, as shown in Figure~\ref{liso_v_tkin_fig}.  For EGOs
detected in both \water\/ maser and \ammonia(2,2) emission, the
partial correlation coefficient is 0.36 for T$_{kin}$ derived from the
single-component fits (again computed with the distance squared as an
independent parameter).  Interestingly, if we recompute the partial
correlation coefficient using the T$_{kin}$ of the warm component for
sources that require two-component fits (and the single-component
T$_{kin}$ for all other sources), the value is reduced to 0.22.  This
is somewhat surprising, since the warm component traces gas nearer to,
and heated by, the central MYSO.

We also find a weak positive correlation between \water\/ maser
luminosity and clump mass (Fig.~\ref{liso_v_mass_fig}).  Calculating
clump masses assuming T$_{dust}$=T$_{kin}$ from the single-component
\ammonia\/ fits, the partial correlation coefficient is 0.44 (for EGOs
detected in both \water\/ maser and \ammonia(2,2) emission, so that
T$_{kin}$ is well-determined).  The calculated partial correlation
coefficient is very similar (0.43) if the presence of two temperature
components is accounted for when estimating the clump mass
(\S\ref{clump_prop}).  A K-S test indicates no statistically
significant difference between the mass distributions of clumps
with/without \water\/ masers, in contrast to earlier studies
\citep{Chambers09,BreenEllingsen11}.  The K-S statistic is 0.26 using
the isothermal clump masses (for EGOs with (2,2) detections and so
well-determined T$_{kin}$, as above), and increases to 0.45 if clump
masses are estimated accounting for the two temperature components.
Both previous studies assumed dust temperatures, and
\citet{Chambers09} found that the probability their cores with/without
\water\/ masers were drawn from the same distribution increased
dramatically (by a factor of $>$50, to 0.11) if they assumed a higher
temperature for active cores (compared to assuming a single
temperature for all cores).  The $\theta_{\rm FWHM}$ of the BGPS data
(\q33\pp) is larger than that of the SIMBA data used by
\citet{BreenEllingsen11} \citep[\q24\pp;][]{Hill05} or the IRAM 30-m
data used by \citet{Chambers09} \citep[\q11\pp;][]{Rathborne06}.
Additional data (such as temperature measurements for the
\citet{Chambers09} and \citet{BreenEllingsen11} sources) would be
required to assess whether this difference in scale contributes to the
difference in findings.

Our results are consistent with the positive correlation between
\water\/ maser and bolometric luminosity reported by
\citet{Urquhart11} for RMS sources.  In this picture, the key factor
is the bolometric luminosity of the driving MYSO, with more luminous
MYSOs exciting more luminous \water\/ masers.  The observed
correlations of \water\/ maser and clump properties (temperature and
mass) are then understood in terms of the relationship between a clump
and the massive star(s) it forms.  The final mass of an actively
accreting MYSO is limited by the available mass reservoir, and studies
of more evolved sources (UC HIIs) indicate that higher-mass clumps
form higher-mass (and thus more luminous) stars
\citep[e.g.][]{Katharine09}.  The more luminous an MYSO, the more
energy it will impart to its environs, and the more it will heat the
gas and dust of the surrounding clump.

\subsection{\ammonia(3,3) Masers}\label{33_masers}

While \ammonia(3,3) maser emission in a MSFR was first reported
several decades ago \citep[DR21(OH);][]{Mangum94}, the number of known
examples--all detected with the VLA--has remained small
\citep[e.g. W51, NGC6334I, IRAS
20126+4106, G5.89$-$0.39:][]{ZhangHo95,Kraemer95,Zhang99,Hunter08}.  Two
recent, large-scale single dish surveys each report a single
\ammonia(3,3) maser candidate: a blind survey of 100 deg$^{2}$ of the
Galactic plane \citep[HOPS:][]{Walsh_hops}, and a targeted survey of
\q 600 RMS sources \citep{Urquhart11}.  This paucity of candidates led
\citet{Urquhart11} to suggest that bright \ammonia(3,3) masers are
rare.

One of our targets, G35.03+0.35, was recently observed in
\ammonia(1,1)-(6,6) with the VLA \citep{rsro}.  In addition to complex
thermal \ammonia\/ emission from a (proto)cluster, nonthermal
\ammonia(3,3) and (6,6) emission are clearly detected \citep[][Fig. 2; peak (3,3) intensity $<$70 \mjb]{rsro}.
Figure~\ref{g35_nh3_spec} shows our Nobeyama \ammonia\/ spectra of G35.03+0.35: while there is a narrow
\ammonia(3,3) emission feature that is not
well fit by the model, the signal-to-noise 
is insufficient to identify it as a
candidate maser from the single-dish data.  This comparison
demonstrates that single-dish surveys readily miss weak \ammonia(3,3)
masers detected with interferometers; sensitive interferometric
observations are required to assess the prevalence of \ammonia\/
masers in MSFRs, and their association with other maser types
\citep[see also][]{rsro,brogan_iau}.

\subsection{Future Work} \label{future}

Our analysis of our Nobeyama EGO survey shows that the presence of
\ammonia(2,2) and (3,3) emission, \water\/ masers, and Class I and II
\meth\/ masers are strongly correlated.  These star formation
indicators tend to occur in concert (at least on the scales probed by
single-dish surveys), and identify a (sub)population of EGOs in which
central MYSO(s) are substantially affecting their environments,
heating the surrounding gas and exciting maser emission.  Notably,
maser emission and warm dense gas appear to pinpoint such sources more
effectively than MIR indicators such as the ``likely''/''possible''
classification of \citetalias{egocat} or the presence/absence of
IRDCs.  These sources are excellent targets for high-resolution followup
observations aimed at understanding the importance of different
(proto)stellar feedback mechanisms in massive star forming regions, as
demonstrated by the SMA, CARMA, and VLA studies of \citet{C11, C11vla}
and \citet{rsro}.  These EGOs are also important testbeds for proposed
maser evolutionary sequences, as discussed in more detail below.

Less clear is the nature of those EGOs detected only in \ammonia(1,1)
emission in our survey.  An examination of their GLIMPSE images
suggests they are a heterogeneous group,
including both EGOs in IRDCs (e.g. G12.02$-$0.21) and
EGOs adjacent to 8 and 24 \um-bright nebulae (e.g. G29.91$-$0.81).
Some examples of each of these MIR source types are detected in
\water\/ maser emission, while others are not.  The MIR morphologies
of EGOs without detected \water\/ masers in our survey are similarly
heterogeneous, and some \water\/ maser nondetections are associated
with \ammonia(2,2) and (3,3) emission.  Higher-resolution observations
are required to localize the \ammonia\/ and \water\/ maser emission
detected in our Nobeyama data with respect to the MIR emission

We emphasize that high-resolution observations are crucial for
building an evolutionary sequence for MYSOs, and placing EGOs within
it.  In general, multiple MIR sources are present within the Nobeyama
beam, and detailed studies of EGOs to date reveal mm and cm-$\lambda$
multiplicity on \q0.1 pc scales.  Furthermore, the members of
(proto)clusters associated with EGOs exhibit a range of star formation
indicators, suggestive of a range of evolutionary states
\citep[e.g.][]{C11,rsro}.

EGOs are notably rich in maser emission, and maser studies have and continue to provide key insights into the
nature of EGOs; their copious maser emission likewise
provides opportunities to use EGOs to advance our understanding of masers in
massive star-forming regions.  \water, Class I and II \meth, and OH
masers are ubiquitous in regions of massive star formation, and much
effort has been devoted to placing these different maser types into an
evolutionary sequence.  Of particular interest is which maser type
appears first--and thus pinpoints the earliest stages of massive star
formation.  In most proposed sequences, Class I \meth\/ masers are
identified with the earliest stages of MYSO evolution, with the
youngest sources being those associated only with Class I \meth\/
masers \citep[e.g.][]{Ellingsen06,Ellingsen07,Breen10evol}.  However,
recent work suggests that Class I \meth\/ masers may be excited by
shocks driven by expanding HII regions as well as by outflows
\citep[e.g.][]{Voronkov10}, such that Class I \meth\/ may outlast the
Class II maser stage and/or arise more than once during MYSO evolution
\citep[e.g.][]{Chen11,Voronkov12}.  \citet{BreenEllingsen11} and
\citet{CB10} have also recently proposed that \water\/
masers--particularly those with blueshifted high-velocity
features--may be the earliest signposts of MYSO formation, preceding
the Class II \meth\/ maser stage.  

Statistical comparisons of ``Class I only'' and ``Class II only'' EGOs
based on our data are limited by the small sample sizes.  It is
notable, however, that the \ammonia(2,2) and \water\/ maser detection
rates towards these subsamples are comparable, particularly
considering the small number statistics.  Likewise,
Figures~\ref{nh3_prop_meth_colors_fig} and \ref{liso_meth_color_fig} show
no clear patterns in their \ammonia\/ or \water\/ maser properties
that would suggest a trend in evolutionary state.  The parameter space
occupied by Class I-only and Class II-only sources in these plots also
largely overlaps with that occupied by EGOs associated with both
\meth\/ maser types.  
Though the comparison is
again limited by small-number statistics, the difference in the \ammonia(3,3) detection
rates (63\%/14\% for Class I/II-only sources) is intriguing,
particularly given the association of Class I \meth\/ and
\ammonia(3,3) masers \citep[e.g.][]{rsro}.

Progress in our understanding of masers as evolutionary
indicators for MSF requires identifying candidate youngest sources,
and studying them in detail.  The (small) samples of EGOs with \water+Class I
\meth\/ and \water+Class II \meth\/ masers identified in our
survey will be promising targets for such studies, as will the samples of \water-only, Class I \meth-only, and Class II
\meth-only sources.  Sensitive, high-resolution maser
observations are needed: (1) to localize the maser emission, and
determine whether or not all maser species are associated with the
same MYSO and (2) to search for weak masers and establish whether
maser types undetected in single-dish surveys are truly absent.  The
expanded capabilities of the Karl G. Jansky Very Large Array (VLA) are well-suited to such studies.
High-resolution cm-(sub)mm wavelength line and continuum observations
will also constrain the properties of compact cores (temperature,
density, mass, chemistry) and outflows, allowing maser activity to be
correlated with other signposts of star formation at the scale of
individual active sources.

\section{Conclusions}\label{conclusions}

We have surveyed all 94 GLIMPSE EGOs visible from the northern
hemisphere ($\delta \gtrsim -$20$^{\circ}$) in \water\/ maser and
\ammonia(1,1), (2,2), and (3,3) emission with the Nobeyama 45-m
telescope.
Our results provide strong evidence that EGOs, as a population, are
associated with dense gas and active star formation, and also reveal statistically significant variation amongst EGO subsamples:

\begin{itemize}
\item{\water\/ masers, which are associated with outflows and require high densities (n(\h)\q10$^{8}$-10$^{10}$ cm$^{-3}$), are detected towards \q68\% of EGOs surveyed. }

\item{The \ammonia(1,1) detection rate is \q97\%, confirming that EGOs are associated with dense molecular gas.}

\item{Two-component models provide a significantly improved fit for \q
23\% of our \ammonia\/ spectra, indicating contributions from both
warm inner regions and cooler envelopes along the line of sight.}

\item{\water\/ maser emission is strongly correlated with the presence
of warm, dense gas, as indicated by emission in the higher-excitation
\ammonia\/ transitions.  The \water\/ maser detection rate is 81\%
towards EGOs detected in \ammonia(2,2) and/or (3,3) emission, and only
44\% towards EGOs detected only in \ammonia(1,1).  We find
statistically significant differences in the distributions of
\ammonia\/ column density, kinetic temperature, and \ammonia\/
linewidth for EGOs with/without \water\/ maser detections: EGOs with
detected \water\/ masers have greater median N(\ammonia), T$_{kin}$,
and $\sigma_{\rm v}$.
}

\item{\water\/ maser and \ammonia(2,2) and (3,3) detection rates are
higher towards EGOs classified as ``likely'' outflow candidates based
on their MIR morphology than towards EGOs classified as ``possible''
outflow candidates.  However, statistical tests show significant
differences only in the distributions of the \ammonia(1,1) and (2,2)
peaks (T$_{\rm MB}$), not in physical properties.
}

\item{EGOs associated with IRDCs have higher \ammonia(2,2) and (3,3)
detection rates, but a lower \water\/ maser detection rate, than EGOs
not associated with IRDCs.  We find statistically significant
differences in the distributions of \ammonia(1,1) peak (T$_{\rm MB}$),
\ammonia\/ linewidth, and \ammonia\/ beam filling fraction for EGOs
associated/not associated with IRDCs: the median \ammonia(1,1) T$_{\rm
MB}$ is higher, and the median $\sigma_{\rm v}$ lower, for EGOs
associated with IRDCs.
}

\item{The \water\/ maser, \ammonia(2,2), and \ammonia(3,3) detection
rates towards EGOs with both Class I and II \meth\/ masers are the
highest of any EGO subsample we consider: 95\%, 90\% and 81\%,
respectively.  In contrast, we detect \water\/ masers and the
higher-excitation \ammonia\/ lines towards only 33\% (\water), 20\%
(2,2) and 7\% (3,3) of EGOs with neither type of \meth\/ maser.  We
find statistically significant differences in the distributions of
\ammonia(1,1) peak temperature (T$_{\rm MB}$), \ammonia\/ column density,
and \ammonia\/ linewidth for EGOs associated with both types/neither
type of \meth\/ masers: EGOs associated with both Class I and II
\meth\/ masers have higher median \ammonia(1,1) T$_{\rm MB}$, N(\ammonia),
$\sigma_{\rm v}$, and T$_{kin}$.

}

\item{While \water\/ maser detection rates vary across EGO subsamples, we
find no evidence for statistically significant differences in the
properties of detected \water\/ masers.}

\end{itemize}   

Our \water\/ maser and \ammonia\/ survey, in combination with the 1.1
mm continuum BGPS, provides the necessary data to explore connections
between \water\/ maser and clump properties: \water\/ maser spectra,
clump-scale T$_{kin}$ measurements from \ammonia, and clump masses and
densities from the 1.1 mm dust continuum emission and T$_{kin}$
measurements.  These combined data show no correlation between
isotropic \water\/ maser luminosity and volume-averaged clump density.
\water\/ maser luminosity is weakly positively correlated with clump
temperature and with clump mass, consistent with reported correlations
between \water\/ maser luminosity and the bolometric luminosity of the
driving source.

We interpret the observed correlations of \water\/ maser and clump
properties in terms of the relationship between a clump and the
massive star(s) it forms.  For more evolved sources (UC HIIs), studies
indicate that higher-mass clumps form higher-mass (and thus more
luminous) stars \citep[e.g.][]{Katharine09}.  For an actively
accreting MYSO, the available mass reservoir sets the limit on its
final, stellar mass.  The more luminous (and massive) an MYSO, the
more energy it will impart to its environs, and the more it will heat
the gas and dust of the surrounding clump.

We find that \ammonia(2,2) and (3,3) emission, \water\/ masers, and
Class I and II \meth\/ masers are strongly correlated, at least on the
scales probed by single-dish surveys.  These star formation indicators
pinpoint EGOs in which the central MYSO(s) are substantially affecting
their environments, more effectively than MIR indicators
(such as the ``likely''/''possible'' classification of
\citetalias{egocat} or the presence/absence of IRDCs).  We also
identify small samples of EGOs associated with only one maser type;
the \water-only and Class I \meth-only sources are candidates for
extremely young MYSOs.  Constructing an evolutionary scheme for MYSOs
requires localizing maser and dense gas emission at the scale of
individual (proto)stars.  The expanded capabilities of the Karl
G. Jansky VLA will enable such studies for statistically meaningful
samples.

\acknowledgments

We thank the staff at the Nobeyama Radio Observatory for their support
during our observing runs.  This research has made use of NASA's
Astrophysics Data System Bibliographic Services and the SIMBAD
database operated at CDS, Strasbourg, France.  Support for this work
was provided by NSF grant AST-0808119.  C.J.C. was partially supported
during this work by a National Science Foundation Graduate Research
Fellowship, and is currently supported by an NSF Astronomy and
Astrophysics Postdoctoral Fellowship under award AST-1003134.
C.J.C. thanks M. Reid for helpful discussions about kinematic
distances, and J. Brown, L. Chomiuk, and H. Kirk for IDL insight.
E.R. is supported by a Discovery Grant from NSERC of Canada.

\clearpage
\LongTables

\begin{deluxetable*}{lcccccc}
\tablewidth{0pt}
\tablecaption{EGO Sample: Properties from the Literature\label{sample_table}}
\tablehead{ 
\colhead{Source Name} &
\multicolumn{2}{c}{J2000 Coordinates\tablenotemark{a}} &
\colhead{EGO Catalog\tablenotemark{a}} &
\colhead{IRDC?\tablenotemark{a}} &
\multicolumn{2}{c}{CH$_{3}$OH maser?\tablenotemark{b}} \\
\colhead{} & 
\colhead{$\alpha$ ($^{\rm h}~~^{\rm m}~~^{\rm s}$)} &
\colhead{$\delta$ ($^{\circ}~~{\arcmin}~~{\arcsec}$)} &
\colhead{} &
\colhead{} &
\colhead{Class II} &
\colhead{Class I} 
}
\tablecolumns{7}
%\tabletypesize{\scriptsize}
\startdata
G10.29$-$0.13  &  18 08 49.3  &  -20 05 57  &  2  &  Y  &   Y  &   Y  \\
    G10.34$-$0.14  &  18 09 00.0  &  -20 03 35  &  2  &  Y  &   Y  &   Y  \\
    G11.11$-$0.11  &  18 10 28.3  &  -19 22 31  &  3  &  Y  &   Y  &   N  \\
    G11.92$-$0.61  &  18 13 58.1  &  -18 54 17  &  1  &  Y  &   Y  &   Y  \\
    G12.02$-$0.21  &  18 12 40.4  &  -18 37 11  &  1  &  Y  &   N  &   N  \\
    G12.20$-$0.03  &  18 12 23.6  &  -18 22 54  &  4  &  N  &   Y  &   Y  \\
    G12.42+0.50  &  18 10 51.1  &  -17 55 50  &  4  &  N  &   N  &   Y  \\
    G12.68$-$0.18  &  18 13 54.7  &  -18 01 47  &  4  &  N  &   Y  &   Y  \\
\enddata
\tablenotetext{a}{From \citetalias{egocat}.  The table number from \citetalias{egocat} is given in the ``EGO Catalog'' column.  Tables 1 and 2 of \citetalias{egocat} listed ``likely'' outflow candidates; tables 3 and 4 ``possible'' outflow candidates.    Table 5 sources are those for which only ``outflow-only'' photometry was presented; we do not include them in our analysis of ``likely'' and ''possible'' subsamples.}  
\tablenotetext{b}{From \citetalias{Chen11} (only).  EGOs in our sample that are not included in \citetalias{Chen11} are indicated by \nodata.  - indicates a source with a single-dish 6.7 GHz \meth\/ maser detection but no positional information, considered as having ``no information'' at 6.7 GHz by \citetalias{Chen11} (see \S\ref{water_detect}).}
\end{deluxetable*}

\begin{deluxetable*}{lccccccccc}
\tablewidth{0pt}
\tablecaption{Detection Statistics\tablenotemark{a}\label{detect_rates_table}}
\tablehead{ 
\colhead{} & 
\colhead{} & 
\multicolumn{2}{c}{\ammonia\/ (1,1)} &
\multicolumn{2}{c}{\ammonia\/ (2,2)} &
\multicolumn{2}{c}{\ammonia\/ (3,3)\tablenotemark{b}} &
\multicolumn{2}{c}{Water Masers} \\
\colhead{Category\tablenotemark{c}} & 
\colhead{N$_{obs}$} & 
\colhead{N$_{detect}$} & 
\colhead{Rate} & 
\colhead{N$_{detect}$} & 
\colhead{Rate} & 
\colhead{N$_{detect}$} & 
\colhead{Rate} & 
\colhead{N$_{detect}$} & 
\colhead{Rate} 
}
\tablecolumns{10}
%\tabletypesize{\scriptsize}
\startdata
Overall & 94 & 91 & 0.97(0.02) & 59 & 0.63(0.05) & 43 & 0.46(0.05) & 64 & 0.68(0.05) \\
\\
IRDC Assoc. & 47 & 47 & 1.00 & 35 & 0.74(0.06) & 29 & 0.62(0.07) & 29 & 0.62(0.07) \\
No IRDC Assoc. & 47 & 44 & 0.94(0.04) & 24 & 0.51(0.07) & 14 & 0.30(0.07) & 35 & 0.74(0.06) \\
\\
'Likely' & 48 & 47 & 0.98(0.02) & 34 & 0.71(0.07) & 29 & 0.60(0.07) & 36 & 0.75(0.06) \\
'Possible' & 43 & 41 & 0.95(0.03) & 22 & 0.51(0.08) & 12 & 0.28(0.07) & 25 & 0.58(0.08) \\
'Outflow-only' & 3 & 3 & 1.00 & 3 & 1.00 & 2 &  0.67(0.27) & 3 & 1.00 \\
\enddata
\tablenotetext{a}{Uncertainties in detection rates calculated using binomial statistics.}
\tablenotetext{b}{Includes only sources detected in \ammonia(1,1) and (2,2) as well as (3,3), see \S\ref{ammonia_detect}.}
\tablenotetext{c}{IRDC associations and 'likely/possible' designations from \citepalias{maserpap}, see also \S\ref{intro}.}
\tablenotetext{d}{For consistency, all data on Class I and II \meth\/ maser associations are taken from Table 1 of \citetalias{Chen11}, see \S\ref{water_detect}.  'Class I': all EGOs with a Class I maser detection in \citetalias{Chen11}, regardless of Class II association (or lack of Class II information).  'Class I ND': all EGOs listed as Class I nondetections in \citetalias{Chen11}, regardless of Class II association (or lack of Class II information).  'Class II': all EGOs listed as Class II maser detections in \citetalias{Chen11}, regardless of Class I association.  'Class II ND': all EGOs listed as Class II nondetections in \citetalias{Chen11}, regardless of Class I association.  'Class I Only': EGOs listed as Class I detections and Class II nondetections in \citetalias{Chen11}. 'Class II only': EGOs listed as Class I nondetections and Class II detections in \citetalias{Chen11}.  'Class I and II': EGOs listed as both Class I and Class II detections in \citetalias{Chen11}.  'Neither': EGOs listed as both Class I nondetections and Class II nondetections in \citetalias{Chen11}.
} 
\end{deluxetable*}

\clearpage
\LongTables
\begin{landscape}
\tabletypesize{\scriptsize}
\begin{deluxetable}{cccccccccccccc}
\tablewidth{0pt}
\tablecaption{\ammonia\/ Properties: Single Component Fits\tablenotemark{a} \label{nh3_prop_tab}}
\tablehead{ 
\colhead{Source Name} &
\colhead{\vlsr} &
\colhead{$\sigma_{\rm v}$} &
\colhead{Distance\tablenotemark{b}} & 
\colhead{T$_{\rm MB}$(1,1)\tablenotemark{c}} & 
\colhead{$\tau_{(1,1)}$} &
\colhead{N(\ammonia)} &
\colhead{$\eta_{ff}$} & 
\colhead{T$_{ex}$} & 
\colhead{T$_{kin}$\tablenotemark{d}} &
\colhead{T$_{\rm MB}$(2,2)\tablenotemark{c}} & 
\colhead{T$_{\rm MB}$(3,3)\tablenotemark{c}} & 
\colhead{2 comp?\tablenotemark{e}} & 
\colhead{\water} \\
\colhead{} & 
\colhead{(\kms)} &
\colhead{(\kms)} & 
\colhead{(kpc)} & 
\colhead{(K)} & 
\colhead{} &
\colhead{(cm$^{-2}$)} &
\colhead{} &
\colhead{(K)} & 
\colhead{(K)} & 
\colhead{(K)} & 
\colhead{(K)} & 
\colhead{} & 
\colhead{maser?} \\
\colhead{} & 
\colhead{} &
\colhead{} & 
\colhead{} & 
\colhead{} & 
\colhead{} &
\colhead{$\times$10$^{14}$} &
\colhead{} &
\colhead{} & 
\colhead{} & 
\colhead{} & 
\colhead{} & 
\colhead{} & 
\colhead{} }
\tablecolumns{14}
\setlength{\tabcolsep}{0.06in}
%\tabletypesize{\scriptsize}
\startdata
G10.29-0.13 &  13.97(0.01) &   1.18(0.01) &  1.58(+0.86,-1.13) & 1.94(0.04) &  4.76 & 22.1(0.6) & 0.096(0.002) &  4.52(0.05) & 21.19(0.17) & 1.19(0.04) & 0.61(0.05) & Y &   Y  \\
G10.34-0.14 &  12.02(0.02) &   1.07(0.02) &  1.29(+0.92,-1.23) & 1.56(0.06) &  4.28 & 18.2(1.0) & 0.047(0.002) &  3.95(0.07) & 28.23(0.38) & 0.94(0.05) & 0.76(0.07) & Y &   Y  \\
G11.11-0.11 &  29.79(0.02) &   0.68(0.01) & 12.67(+0.48,-0.41)\tablenotemark{f} & 1.79(0.07) &  5.15 & 11.9(0.6) & 0.150(0.006) &  4.45(0.09) & 14.15(0.27) & 0.65(0.06) & 0.32(0.07) & Y &   N  \\
G11.92-0.61 &  36.11(0.01) &   1.09(0.01) &  3.48(+0.44,-0.52) & 3.02(0.07) &  6.40 & 31.7(0.8) & 0.087(0.001) &  4.79(0.04) & 26.27(0.19) & 1.83(0.06) & 1.14(0.07) & Y &   Y  \\
G12.02-0.21 &  -3.15(0.06) &   1.04(0.05) &  5.30(+0.20,-0.20)\tablenotemark{g} & 0.46(0.06) &  5.18 & 13.2(2.1) & 0.042(0.004) &  3.16(0.15) & $<$12.85 & $<$0.19 & $<$0.22 & N &   N  \\
G12.20-0.03 &  51.16(0.06) &   1.24(0.05) & 11.70(+0.31,-0.27)\tablenotemark{f} & 0.55(0.06) &  3.76 & 13.5(2.3) & 0.032(0.004) &  3.27(0.13) & 19.56(0.76) & 0.28(0.05) & $<$0.25 & N &   Y  \\
G12.42+0.50 &  18.13(0.03) &   1.16(0.04) &  1.87(+0.69,-0.85) & 0.81(0.06) &  1.19 &  6.0(1.6) & 0.058(0.016) &  4.30(0.42) & 29.31(0.60) & 0.55(0.05) & 0.37(0.06) & N &   Y  \\
G12.68-0.18 &  55.69(0.01) &   1.18(0.01) &  4.46(+0.30,-0.34) & 1.45(0.05) &  7.48 & 32.4(0.9) & 0.051(0.001) &  3.87(0.04) & 24.95(0.21) & 0.99(0.05) & 0.66(0.05) & Y &   Y  \\
\enddata
\end{deluxetable}
\clearpage
\end{landscape}

\begin{deluxetable*}{lccccc}
\tablewidth{0pt}
\tablecaption{\ammonia\/ Properties: Two Component Fits\tablenotemark{a} \label{nh3_prop_tab_2comp}}
\tablehead{
\colhead{Source Name} &
\colhead{T$_{kin}$} &
\colhead{\vlsr} &
\colhead{$\sigma_{\rm v}$} &
\colhead{T$_{ex}$} & 
\colhead{N(\ammonia)} \\
\colhead{} & 
\colhead{(K)} & 
\colhead{(\kms)} &
\colhead{(\kms)} & 
\colhead{(K)} & 
\colhead{(cm$^{-2}$)} \\
\colhead{} & 
\colhead{} & 
\colhead{} &
\colhead{} & 
\colhead{} & 
\colhead{$\times$10$^{14}$} 
}
\tablecolumns{6}
%\setlength{\tabcolsep}{0.06in}
%\tabletypesize{\scriptsize}
\startdata
    G10.29-0.13 & 15.93(0.32) &  14.16(0.02) & 0.89(0.02) &  4.19(0.06) & 15.6(0.7) \\
                & 32.91(0.98) &  13.31(0.06) & 1.73(0.04) &  3.23(0.12) & 30.2(2.8) \\
    G10.34-0.14 & 15.62(0.78) &  12.05(0.03) & 0.70(0.04) &  4.04(0.17) &  6.5(1.1) \\
                & 41.51(2.42) &  12.05(0.05) & 1.43(0.05) &  3.32(0.23) & 21.3(2.7) \\
    G11.11-0.11 & 13.90(0.58) &  29.40(0.04) & 0.45(0.03) &  3.70(0.25) & 10.6(1.1) \\
                & 14.07(0.64) &  30.33(0.14) & 0.65(0.06) &  3.72(0.22) &  8.8(1.2) \\
    G11.92-0.61 & 13.55(0.20) &  36.10(0.01) & 0.73(0.01) &  5.16(0.07) & 19.3(0.6) \\
                & 54.06(1.38) &  35.89(0.06) & 3.43(0.07) & 54.06(0.00) &  3.4(0.1) \\
\enddata
\tablenotetext{a}{Uncertainties estimated from the model optimization are given in parentheses: values of 0.00 indicate cases where the model is poorly constrained (see \S\ref{nh3_modeling}).}
\end{deluxetable*}

\newpage
\begin{deluxetable*}{lcc}
\tablewidth{0pt}
\tablecaption{EGO Subsamples with Statistically Significant Differences in \ammonia\/ Properties \label{stat_sig_diff_table}}
\tablehead{ 
\colhead{Subsample} & 
\colhead{Property} &
\colhead{K-S significance}
} 
\tablecolumns{3}
%\tabletypesize{\small}
%\setlength{\tabcolsep}{0.05in}
\startdata
Likely/Possible & \ammonia(1,1) peak (T$_{\rm MB}$) &      8.4E-03 \\
Likely/Possible & \ammonia(2,2) peak (T$_{\rm MB}$) &      7.6E-03 \\
\\
IRDC/no IRDC & \ammonia(1,1) peak (T$_{\rm MB}$) &     1.5E-05 \\
IRDC/no IRDC & $\sigma_{\rm v}$ &                         8.2E-04 \\
IRDC/no IRDC & $\eta_{ff}$ &                       1.7E-03  \\ 
\\
\water\/ maser detections/nondetections &  $\sigma_{\rm v}$ &   5.5E-08  \\
\water\/ maser detections/nondetections & N(\ammonia) &   1.6E-04 \\
\water\/ maser detections/nondetections &  T$_{kin}$ &     4.7E-03\\    
\\          
Class I/Class I ND & \ammonia(1,1) peak (T$_{\rm MB}$) & 3.0E-03 \\
Class I/Class I ND & $\sigma_{\rm v}$ &                     5.1E-03 \\
Class I/Class I ND & N(\ammonia) &    4.5E-03 \\
Class I/Class I ND & T$_{kin}$ &                    5.1E-03 \\
\\
Class II/Class II ND & \ammonia(1,1) peak (T$_{\rm MB}$) &            3.2E-03 \\
Class II/Class II ND &   $\sigma_{\rm v}$ &                              4.0E-03 \\
Class II/Class II ND &  N(\ammonia) &    2.7E-06 \\
Class II/Class II ND &    T$_{kin}$ &                           6.8E-03 \\
Class II/Class II ND & \ammonia(2,2) peak (T$_{\rm MB}$) & 3.9E-03 \\ 
Class II/Class II ND & \ammonia(3,3) peak (T$_{\rm MB}$) & 3.5E-03 \\ 
\\
Both Class I \& II/Neither &  \ammonia(1,1) peak (T$_{\rm MB}$) &   8.9E-04 \\
Both Class I \& II/Neither &   $\sigma_{\rm v}$ &         1.6E-03\\
Both Class I \& II/Neither & N(\ammonia) &  3.9E-05\\

\enddata
\end{deluxetable*}

\newpage

\begin{deluxetable*}{lcccccccccc}
\tablewidth{0pt}
\tablecaption{\water\/ Maser Properties: Detections \label{water_detect_tab}}
\tablehead{ 
\colhead{Source Name} &
\colhead{$\sigma$} &
\colhead{V$_{\rm min}$} &
\colhead{V$_{\rm max}$} &
\colhead{V$_{\rm peak}$} &
\colhead{V$_{\rm range}$} &
\colhead{V$_{ NH_{3}}$} &
\colhead{S$_{\rm peak}$} &
\colhead{$\int$S$_{\nu}dV$} &
\colhead{L$_{\rm iso}$(\water)} & 
\colhead{Notes\tablenotemark{a}} \\
\colhead{} &
\colhead{Jy} &
\colhead{\kms} &
\colhead{\kms} &
\colhead{\kms} &
\colhead{\kms} &
\colhead{\kms} &
\colhead{Jy} &
\colhead{Jy \kms} &
\colhead{\lsun} & 
\colhead{}
}
\tablecolumns{11}
%\tabletypesize{\scriptsize}
\startdata
    G10.29-0.13  &  0.08   &     5.3   &    16.1   &   12.3   &    10.8   &    14.0  &     0.7   &     1.548   &      8.89E-08 & BE11 \\
    G10.34-0.14  &  0.11   &   -13.1   &    49.8   &   19.8   &    62.8   &    12.0  &    11.1   &    64.044   &      2.45E-06 & BE11 \\
    G11.92-0.61  &  0.14   &    18.2   &    43.8   &   39.8   &    25.6   &    36.1  &    53.1   &   145.844   &      4.06E-05 & HC96,BE11 \\
    G12.20-0.03  &  0.11   &    47.3   &    47.6   &   47.3   &     0.3   &    51.2  &     0.5   &     0.242   &      7.62E-07 & BE11  \\
    G12.42+0.50  &  0.11   &     4.5   &    11.2   &    5.8   &     6.7   &    18.1  &     2.9   &     7.828   &      6.30E-07 & \nodata \\
    G12.68-0.18  &  0.11   &     8.3   &   109.7   &   59.5   &   101.4   &    55.7  &   629.0   &  1381.320   &      6.32E-04 & BE11 \\
    G12.91-0.03  &  0.11   &    13.2   &    16.9   &   16.7   &     3.8   &    56.8  &     0.7   &     1.327   &      6.13E-07 & \nodata \\
    G12.91-0.26  &  0.14   &    40.4   &    49.3   &   40.7   &     8.9   &    37.1  &     1.0   &     1.010   &      2.70E-07 & \tablenotemark{*} \\
\enddata
\tablenotetext{a}{References are for previously reported \water\/ masers with accurate positions from interferometric observations that fall within the polygonal aperture for the EGO published by \citetalias{egocat} (\S\ref{water_dis}).  B02: \citet{Beuther02maser} B11: \citet{Bart11}. BE11: \citet{BreenEllingsen11} CG11: \citet{CaswellGreen11} FC99: \citet{FC99} HC96: \citet{HC96} S10: \citet{Sato10} W06: \citet{Wang06} \nodata indicates no \water\/ maser reference meeting these criteria was found in a SIMBAD search.}
\tablenotetext{*}{``Outflow-only'' source in \citetalias{egocat}.  The published polygonal aperture does not include the ``central'' source, which is associated with \water\/ masers \citep[G12.91-0.26; G34.26+0.15; G37.55+0.20:][respectively]{BreenEllingsen11,FC99,Beuther02maser}.}
\end{deluxetable*}

\clearpage
\newpage

\begin{deluxetable*}{lcc}
\tablewidth{0pt}
\tablecaption{\water\/ Maser Nondetections: Limits \label{water_nondetect_tab}}
\tablehead{ 
\colhead{Source Name} &
\colhead{$\sigma$} &
\colhead{L$_{\rm iso}$(\water)\tablenotemark{a}} \\
\colhead{} &
\colhead{Jy} &
\colhead{\lsun} 
}
\tablecolumns{3}
%\tabletypesize{\scriptsize}
\startdata
    G11.11-0.11  &  0.14   & $<$       1.09E-06   \\
    G12.02-0.21  &  0.10   & $<$       1.41E-07   \\
    G16.58-0.08  &  0.10   & $<$       5.04E-08   \\
    G16.61-0.24  &  0.10   & $<$       5.68E-08   \\
    G19.36-0.03  &  0.16   & $<$       3.40E-08   \\
    G21.24+0.19  &  0.10   & $<$       1.89E-08   \\
    G24.11-0.17  &  0.09   & $<$       9.87E-08   \\
    G24.11-0.18  &  0.10   & $<$       1.09E-07   \\
    G28.28-0.36  &  0.17   & $<$       8.07E-08   \\
    G28.83-0.25  &  0.17   & $<$       2.01E-07   \\
    G28.85-0.23  &  0.12   & $<$       1.73E-07   \\
    G29.84-0.47  &  0.14   & $<$       1.09E-07   \\
    G29.89-0.77  &  0.16   & $<$       1.74E-07   \\
    G29.91-0.81  &  0.17   & $<$       1.87E-07   \\
    G29.96-0.79  &  0.06   & $<$       6.52E-08   \\
    G35.04-0.47  &  0.12   & $<$       5.82E-08   \\
    G35.83-0.20  &  0.13   & $<$       2.10E-08   \\
    G36.01-0.20  &  0.10   & $<$       1.50E-07   \\
    G40.28-0.27  &  0.08   & $<$       8.94E-08   \\
    G40.60-0.72  &  0.10   & $<$       9.26E-08   \\
    G44.01-0.03  &  0.10   & $<$       1.16E-07   \\
    G48.66-0.30  &  0.16   & $<$       4.68E-08   \\
    G49.27-0.32  &  0.15   & $<$       2.18E-07   \\
    G54.11-0.04  &  0.16   & $<$       1.09E-07   \\
    G54.11-0.05  &  0.11   & $<$       7.35E-08   \\
    G56.13+0.22  &  0.06   & $<$       6.80E-08   \\
    G57.61+0.02  &  0.11   & $<$       1.11E-07   \\
    G58.09-0.34  &  0.06   & $<$       1.69E-09   \\
    G58.79+0.63  &  0.07   & $<$       6.48E-08   \\
    G62.70-0.51  &  0.06   & $<$       4.15E-08   \\
\enddata
\tablenotetext{a}{All limits are 4$\sigma$.}
\end{deluxetable*}

\clearpage
\LongTables
\begin{landscape}
\tabletypesize{\scriptsize}

\begin{deluxetable*}{lccccccccc}
\tablewidth{0pt}
\tablecaption{Properties of Associated BGPS Dust Clumps\tablenotemark{a}  \label{bgps_prop_tab}}
\tablehead{ 
\colhead{} &
\colhead{} &
\colhead{} &
\colhead{} &
\colhead{} &
\colhead{} &
\multicolumn{2}{c}{Single Component\tablenotemark{b}} &
\multicolumn{2}{c}{Two Component\tablenotemark{c}} \\
\colhead{Source Name} &
\colhead{BGPS Cat. ID} &
\colhead{S$_{1.1 mm}$\tablenotemark{d}} & 
\colhead{Distance\tablenotemark{e}} &
\colhead{Radius} & 
\colhead{Radius} & 
\colhead{Clump Mass} & 
\colhead{Log[n$_{H_{2}}$]} &
\colhead{Clump Mass} & 
\colhead{Log[n$_{H_{2}}$]} \\
\colhead{} &
\colhead{} &
\colhead{(Jy)} & 
\colhead{(kpc)} & 
\colhead{(\pp)} & 
\colhead{(pc)} & 
\colhead{(\msun)} & 
\colhead{(Log[cm$^{-3}$])} &
\colhead{(\msun)} & 
\colhead{(Log[cm$^{-3}$])} 
}
\tablecolumns{10}
%\tabletypesize{\scriptsize}
\startdata
    G10.29-0.13 & 1474 &  9.4(0.7) &   1.58(+0.86,-1.13) &  68.1 &    0.52 &   427 $^{+ 774 }_{- 398}$ &  4.1$^{+     6.2}_{- 4.1}$ &   585 &   4.2 \\
    G10.34-0.14 & 1483 &  8.6(0.6) &   1.29(+0.92,-1.23) &  84.9 &    0.43 &   179 $^{+ 441 }_{- 179}$ &  3.7$^{+     8.2}_{- 3.7}$ &   361 &   4.0 \\
    G11.11-0.11 & 1589 &  2.3(0.2) &  12.67(+0.48,-0.41)\tablenotemark{f} &  56.7 &    4.18 & 11740 $^{+3341 }_{-2679}$ &  3.1$^{+     2.7}_{- 2.6}$ &  7967 &   2.9 \\
    G12.02-0.21 & 1668 &  1.0(0.1) &   5.30(+0.20,-0.20)\tablenotemark{g} &  38.0 &    1.75 & $>$ 1024 $^{+ 361 }_{- 292}$ & $>$ 3.7$^{+     3.4}_{- 3.2}$ & \nodata & \nodata \\
    G12.20-0.03 & 1682 &  2.1(0.2) &  11.70(+0.31,-0.27)\tablenotemark{f} &  52.8 &    3.86 &  5732 $^{+1674 }_{-1373}$ &  3.0$^{+     2.5}_{- 2.4}$ & \nodata & \nodata \\
    G12.42+0.50 & 1710 &  7.6(0.5) &   1.87(+0.69,-0.85) &  50.6 &    0.62 &   318 $^{+ 384 }_{- 239}$ &  4.1$^{+     5.2}_{- 4.1}$ & \nodata & \nodata \\
    G12.68-0.18 & 1747 & 11.8(0.8) &   4.46(+0.30,-0.34) &  86.2 &    1.47 &  3454 $^{+1163 }_{- 978}$ &  3.4$^{+     3.2}_{- 3.0}$ &  6447 &   3.6 \\
    G12.91-0.03 & 1809 &  3.6(0.3) &   4.48(+0.30,-0.34) &  65.0 &    1.48 &  1129 $^{+ 392 }_{- 327}$ &  3.2$^{+     3.1}_{- 2.8}$ &  1715 &   3.4 \\
\enddata
\end{deluxetable*}
\clearpage
\end{landscape}

\clearpage
\LongTables
\begin{landscape}
\tabletypesize{\scriptsize}

\begin{deluxetable*}{lccccccc}
%\rotate
\tablewidth{0pt}
\tablecaption{\water\/ Maser and \ammonia\/ Surveys of MYSO Samples: Detection Rate Comparison \label{lit_detect_comparison_tab}}
\tablehead{ 
\colhead{Sample} & 
\colhead{Reference\tablenotemark{a}} &
\multicolumn{3}{c}{\underline{\water\/ Maser}} &
\multicolumn{3}{c}{\underline{\ammonia(1,1)}} \\
\colhead{} &
\colhead{} &
\colhead{Sensitivity} & 
\colhead{Resolution\tablenotemark{b}} & 
\colhead{Detection Rate} & 
\colhead{Sensitivity} & 
\colhead{Resolution\tablenotemark{b}} & 
\colhead{Detection Rate} 
}
\tablecolumns{8}
%\tabletypesize{\scriptsize}
\startdata
EGOs (overall)\tablenotemark{c} & this work & $\sigma \sim$ 110 mJy & SD (73\pp) & 68\% & $\sigma \sim$ 50 mK & SD (73\pp) & 97\% \\
EGOs (IRDC)\tablenotemark{c} &               &                      &            & 62\% &                    &              & 100\% \\
EGOs (Class I \meth)\tablenotemark{c} &    &                         &            & 90\% &                   &             & 100\% \\
EGOs (Class II \meth)\tablenotemark{c} &    &                         &            & 86\% &                   &             & 96\% \\
EGOs (Class I \& II \meth)\tablenotemark{c} &    &                         &            &95\% &                &              & 100\% \\
EGOs (Neither \meth)\tablenotemark{c} &    &                         &            &    33\% &                  &              & 93\% \\
\hline
Dust Clumps (overall)\tablenotemark{d} & BE11/H10 & $\sigma \sim$30-40 mJy  & Int. (\q 2\pp) & 44\% & variable & SD (52\pp) & 56\%  \\
Dust Clumps (mm only)\tablenotemark{d} & BE11/H10 &                                      &                & 23\% &          &         &     53\%          \\
\enddata
\tablenotetext{a}{A96: \citet{Anglada96}. B02: \citet{Beuther02maser}. BCEP10: \citet{Breen10water}. B11: \citet{Bart11}.  BE11: \citet{BreenEllingsen11}. Ch09: \citet{Chambers09}. C90: \citet{Churchwell90}. D11: \citet{nh3-innergal}. H10: \citet{Hill10} P12: \citet{Pandian12} Sr02: \citet{Sridharan02}. Sz05: \citet{Sz05}. U09: \citet{Urquhart09}.  U11: \citet{Urquhart11}}
\tablenotetext{b}{Int. indicates interferometric observations, and estimated positional accuracy is given.  SD indicates single dish observations, and the FWHP beam size is given.}
\tablenotetext{c}{See Table~\ref{detect_rates_table} for additional statistics on EGO subsamples.  'Class I': all EGOs with a Class I maser detection in \citetalias{Chen11}, regardless of Class II association (or lack of Class II information).  'Class II': all EGOs listed as Class II maser detections in \citetalias{Chen11}, regardless of Class I association.  'Class I \& II \meth': EGOs listed as both Class I and Class II detections in \citetalias{Chen11}.  'Neither \meth': EGOs listed as both Class I nondetections and Class II nondetections in \citetalias{Chen11}.}

\tablenotetext{d}{BE11: \water\/ maser survey; H10: \ammonia\/ survey.  These surveys both target subsets of the \citet{Hill05} dust clump sample, and divide their observed dust clumps into the same general categories.  The studies are, however, independent, and BE11 do not include the \ammonia\/ results of H10 in their analysis.}
\end{deluxetable*}

\clearpage
\end{landscape}

\end{document}